\documentclass[12pt]{article}
\pdfoutput=1
\usepackage{jheppub}
\usepackage{comment}
\usepackage{todonotes}
\usepackage{setspace}

\usepackage{amsmath, amsfonts, amsthm, amssymb}  
\usepackage{gensymb}
\usepackage{siunitx}
\usepackage{enumerate}
\usepackage{enumitem}
\usepackage{hyperref}
\usepackage[all]{xy}
\usepackage{wrapfig}
\usepackage{fancyvrb}
\usepackage[T1]{fontenc}
\usepackage{listings}
\usepackage{float}
\usepackage{graphicx}
\usepackage{caption}
\usepackage{subcaption}
\usepackage{physics}
\usepackage{booktabs}

\usepackage{centernot}
\usepackage{mathtools}

\theoremstyle{remark}

\theoremstyle{definition}

\newcommand\arash[1]{\textcolor{blue}{(Arash: #1)}}

\usetikzlibrary{arrows,arrows.meta,shapes,shapes.misc,patterns,decorations.pathmorphing,decorations.pathreplacing,positioning,chains,fit}
\pgfdeclarepatternformonly{coarsedots}{\pgfqpoint{-1pt}{-1pt}}{\pgfqpoint{5pt}{5pt}}{\pgfqpoint{12pt}{12pt}}{
    \pgfpathcircle{\pgfqpoint{.5pt}{.5pt}}{.5pt}
    \pgfusepath{fill}}
\tikzset{gauge/.style={rounded rectangle, draw=black!100, thick, minimum size=5mm},  gaugeD/.style={rounded rectangle, draw=black!100,double,thick,minimum size=5mm},  empty/.style={rounded rectangle, draw=white!100, thick, minimum size=5mm}, flavor/.style={rectangle, draw=black!100, thick, minimum size=5mm},flavorD/.style={rectangle, draw=black!100, double,thick, minimum size=5mm}}

\author[a]{Arash Ardehali,}
\author[b]{Dongmin Gang,}
\author[a]{Neville Joshua Rajappa,}
\author[c]{Matteo Sacchi}

\affiliation{${}^a\,$C.N. Yang Institute for Theoretical Physics, Stony Brook University,\\ ${}\,$\ Stony Brook, NY 11794, USA}

\affiliation{${}^b\,$Department of Physics and Astronomy \& Center for Theoretical Physics,\\
${}\,$\ Seoul National University, 1 Gwanak-ro, Seoul 08826, Korea}

\affiliation{${}^c\,$Simons Center for Geometry and Physics,
Stony Brook University,\\ ${}\,$\ Stony Brook, NY 11794-3636, USA\\}

\emailAdd{a.a.ardehali@gmail.com}
\emailAdd{arima275@snu.ac.kr}
\emailAdd{nevillejoshua.rajappa@stonybrook.edu}
\emailAdd{msacchi@scgp.stonybrook.edu}

\title{3d SUSY enhancement and non-semisimple TQFTs from four dimensions}

\date{}

\begin{document}

\abstract{It has been recently shown that the celebrated SCFT$_4$/VOA$_2$ correspondence can be bridged via three-dimensional field theories arising from a specific R-symmetry twisted circle reduction. We apply this twisted reduction to the $(A_1,A_{n})$ and $(A_1,D_{n})$ families of 4d $\mathcal{N}=2$ Argyres-Douglas SCFTs using their $\mathcal{N}=1$ Agarwal-Maruyoshi-Song Lagrangians. From $(A_1,A_{2n})$ we derive the Gang-Kim-Stubbs family of 3d $\mathcal{N}=2$ gauge theories with SUSY enhancement to $\mathcal{N}=4$ in the infrared, generalizing a recent derivation made in the special cases $n=1,2$. Topological twists of these theories are known to yield \emph{semisimple} TQFTs supporting \emph{rational} VOAs on holomorphic boundaries. From $(A_1,A_{2n-1})$, $(A_1,D_{2n+1})$, and $(A_1,D_{2n})$, we obtain three new infinite families of 3d $\mathcal{N}=2$ abelian gauge theories, all with monopole superpotentials, flowing to $\mathcal{N}=4$ SCFTs without Coulomb branch, but with the same non-trivial Higgs branch as the four-dimensional parent. Their topological A-twist yields \emph{non-semisimple} TQFTs related to \emph{logarithmic} VOAs such as $\widehat{\mathfrak{su}}(2)_{-4/3}\,$.
}

\maketitle

\section{Introduction and summary}\label{sec:intro}

Renormalization group (RG)  is a foundational concept in quantum field theory, describing how physical systems change as we vary the energy scale. There are several ways of triggering an RG flow from a given ultraviolet (UV) conformal field theory (CFT). Conventional textbook approaches are deformations using relevant local operators and gauging flavor symmetries. Another mechanism  involves dimensional reduction along a compact manifold, connecting fixed points in different spacetime dimensions. To enrich the dimensional reduction procedure or make it more controllable, we often turn on non-trivial background gauge fields coupled to a global symmetry $F$ along the internal direction, which we refer to as $F$-twisted dimensional reduction.  Under RG flow,  the infrared (IR) fixed point may exhibit more or fewer symmetries than the UV theory.  IR symmetry enhancement occurs when symmetry-breaking terms become irrelevant in the IR limit, whereas IR theories may exhibit reduced (faithfully acting) symmetries when some UV symmetries act trivially on IR observables.
The enhancement of supersymmetry, in particular, has garnered significant attention for two main reasons: 
(i) it provides examples of (potentially experimentally realizable) emergent supersymmetry arising from non-supersymmetric systems,\footnote{The first historical instance of this in continuum field theory might be the Landau-Ginzburg formulation \cite{Zamolodchikov:1986db} of the diagonal tri-critical Ising CFT \cite{Friedan:1984rv}.} and 
(ii) it offers access to exotic classes of superconformal field theories (SCFTs) that cannot be realized through conventional means with all symmetries manifest (see \emph{e.g.}~\cite{Maruyoshi:2016tqk, Gang:2018huc}).

In this paper, as a continuation of the previous work in \cite{ArabiArdehali:2024ysy}, we explore the interplay between twisted $S^1$ reductions and IR supersymmetry enhancements in 4d/3d supersymmetric gauge theories. 
\begin{align}
\begin{split}
\textrm{4d $\mathcal{N}=1$  theories } &\xrightarrow{ \;\;\textrm{RG}\;\; } \textrm{ 4d $\mathcal{N}=2$ AD SCFTs} 
\\
&\quad \; \Big\downarrow\; \textrm{$U(1)_r$-twisted reduction}
\\
 \textrm{3d $\mathcal{N}=2$  theories } &\xrightarrow{ \;\;\textrm{RG}\;\; } \textrm{ 3d  $\mathcal{N}=4$ SCFTs or unitary TQFTs} 
\end{split}
\end{align}
Specifically, we consider the 4d $\mathcal{N}=1$ gauge theories \cite{Maruyoshi:2016aim,Agarwal:2016pjo} that flow to 4d $\mathcal{N}=2$ Argyres-Douglas (AD) SCFTs of type $(A_1,A_n)$ and $(A_1,D_n)$ \cite{Cecotti:2010fi,Xie:2012hs,Wang:2015mra} in the IR with supersymmetry enhancement and  analyze their twisted $S^1$-reduction using the $U(1)_r$ factor of the $SU(2)_R\times U(1)_r$ 4d $\mathcal{N}=2$ R-symmetry.\footnote{In this paper we will only refer to the Lie algebra of the symmetries and be cavalier on their global structure, unless otherwise stated.}
Unlike conventional Lagrangian theories, AD theories contain local operators with fractional $r$-charge, making  the  $U(1)_r$-twisted reduction non-trivial. If the $r$-charge is quantized as $\frac{1} N \mathbb{Z}$, there are $N$ inequivalent $U(1)_r$-twisted reductions labeled by $\gamma=0, \ldots, N-1$, or equivalently, the $N$-th roots of unity $e^{\frac{2\pi i \gamma}N}$.  An intriguing aspect of  this twisted  $S^1$ reduction is that one can identify the 2d vertex operator algebras (VOAs) in \cite{Beem:2013sza} associated with 4d $\mathcal{N}=2$ theories as the boundary VOAs of A-twisted 3d theories \cite{Costello:2018fnz}, obtained from the minimal ($\gamma=1$) twisted reduction of the 4d theories \cite{Dedushenko:2023cvd}. This opens a pathway to associate more general VOAs with 4d $\mathcal{N}=2$ theories by varying $\gamma$ and boundary conditions.  

Under the  $U(1)_r$-twisted reduction with $\gamma\geq 1$, some Coulomb branches are typically lifted, and the UV 4d AD theories flow to IR 3d theories with simpler Coulomb branches. When the 4d AD theory has no Higgs branch and the $U(1)_r$-twisted reduction lifts all the Coulomb branches, we expect that the resulting 3d theories flow to either rank-0 SCFTs or (unitary) topological field theories (TQFTs),  both of which have trivial Coulomb and Higgs branches. This scenario occurs for the $U(1)_r$-twisted reduction of $(A_1, A_{2n})$ AD theories with $\gamma$  coprime to the $N$, which is $(2n+3)$ in this case.   In the TQFT cases, extended supersymmetry is absent, as it acts trivially on all IR observables, which are non-local operators. The TQFT supports a unitary rational VOA at the boundary. In the the rank-0 SCFT cases, A-twisting results in a (semisimple) non-unitary TQFT that supports a non-unitary rational VOA at the boundary. Interestingly, unitary and non-unitary TQFTs obtained through  the $U(1)_r$-twisted reduction (and subsequent topological twisting) of a 4d AD theory for different values of $\gamma$ are interconnected by a Galois conjugation operation \cite{Dedushenko:2018bpp,ArabiArdehali:2024ysy}. Thus, the $U(1)_r$-twisted reductions provide a physical realization of the Galois conjugation. 

When the 4d theory has a non-trivial Higgs branch instead, such as for $(A_1,A_{2n-1})$ and $(A_1,D_n)$, the resulting 3d theory flows to an $\mathcal{N}=4$ SCFT with the same Higgs branch. In such cases, the A-twisted theory is a non-semisimple, non-unitary TQFT that supports a non-rational boundary VOA.  

\subsection*{The reduction technique}

To identify 3d uv theories that flow to the IR fixed point of the twisted reduction of 4d AD theories, we utilize the 4d UV Lagrangian theory proposed by Agarwal, Maruyoshi, and Song \cite{Maruyoshi:2016aim,Agarwal:2016pjo}. This 4d UV gauge theory enables us to compute the fully generalized superconformal index of the AD theories. By analyzing the appropriate Cardy limit of the 4d superconformal index, one can derive a Coulomb branch integral expression for the partition function on a squashed 3-sphere of the 3d gauge theories. From this expression, one can extract the effective field theory (EFT) in 3d.

The reduction procedure has been streamlined in \cite{ArabiArdehali:2024ysy} for theories with a Lagrangian description, involving a reductive gauge group $G$ of rank $r_G$, and manifesting 4d $\mathcal{N}=1$ supersymmetry.
\begin{itemize}
    \item In the first step, one looks for stationary loci of a piecewise quadratic function $Q^\gamma\in C^1(\mathfrak{h}_\text{cl})$ (determined by the gauge and $U(1)_r$ charges of the matter fields---see \eqref{eq:Qht}) on the moduli-space $\mathfrak{h}_\text{cl}=(-\frac{1}{2},\frac{1}{2}]^{r_G}$ of the gauge holonomies around the circle. If these loci are isolated, they correspond to the 3d vacua. If the stationary loci are extended, however, they signal gauge invariant BPS monopoles in the 3d theory. These monopole operators parametrize the  perturbatively unlifted 3d $\mathcal{N}=2$ Coulomb branch.
    \item In the second step, one examines the slopes of various piecewise linear functions $L_F^\gamma\in C^0(\mathfrak{h}_\text{cl})$ (determined by the global $F$ as well as the gauge and $U(1)_r$ charges of the various fields---see \emph{e.g.}~\eqref{eq:Lht}, \eqref{eq:Lf}), along the flat directions of $Q^\gamma$. These slopes give the global (possibly R-) symmetry charges of the corresponding BPS monopoles. If one finds monopoles of R-charge $2$ and all other global charges $0,$ they ought to be included in the superpotential of the 3d theory as non-perturbatively generated corrections.\footnote{That monopole superpotentials can be generated in the circle reduction of 4d $\mathcal{N}=1$ theories to 3d $\mathcal{N}=2$ has been discussed in detail in \cite{Aharony:2013dha,Aharony:2013kma}, as well as \cite{ArabiArdehali:2019zac} where a holonomy saddle scenario similar to the ones in the present work was treated. One way to understand this is that the monopole superpotentials break in 3d certain abelian symmetries that were anomalous in 4d (equivalently, the condition of exact marginality of the monopole in 3d is equivalent to the one of the anomaly cancellation in 4d). See also \cite{Nardoni:2024sos} for a recent discussion on the fate of these anomalies in the compactifications of 4d theories to 3d and 2d.} These would then further shrink the 3d $\mathcal{N}=2$ Coulomb branch, from the stationary loci of $Q^\gamma$ to a smaller subset. In all the examples treated in \cite{ArabiArdehali:2024ysy} and the present paper, these smaller subsets are isolated. Therefore the 3d EFTs do not have Coulomb branches of vacua.

    \item In the third and final step, the massless field content supported on the Coulomb branch of vacua obtained as above, and the various (mixed) Chern-Simons couplings of the 3d EFT, are determined using again the $Q^\gamma$ and $L_F^\gamma$ functions. This is reviewed in the next section. In fact, the $U(1)_r$-twist would generically imply that all KK modes of the matter fields are massive and integrated out, however this can be compensated by going on a non-generic point of the classical 3d $\mathcal{N}=2$ Coulomb branch where some fields remain massless. The price to pay is that the 4d gauge symmetry is broken to a subgroup in 3d, which is typically the Cartan (however see Appendix \ref{app:higher_sheets} for a non-abelian example).
\end{itemize}

The resulting 3d EFT has only manifest 3d $\mathcal{N}=2$ supersymmetry, but it  either undergoes non-trivial IR supersymmetry enhancement and flow to an $\mathcal{N}=4$ SCFT,  or  dynamically generates a mass gap and flows to a unitary TQFT. Thus, through the twisted reduction of 4d $\mathcal{N}=2$ SCFTs using their $\mathcal{N}=1$ Lagrangians, one can systematically generate 3d $\mathcal{N}=2$ gauge theories that  exhibit either non-trivial IR SUSY enhancement or a dynamically generated mass gap.

\subsection*{The new 3d SUSY enhancing families}

In this paper, we consider the 3d $\mathcal{N}=2$ SUSY enhancing gauge theories that arise from the $U(1)_r$-twisted circle reduction of the 4d $\mathcal{N}=1$ Lagrangians for the 4d $\mathcal{N}=2$ SCFTs of type $(A_1,A_n)$ and $(A_1,D_n)$. We focus on $\gamma=1$. The 3d theories resulting from $\gamma=-1$ are obtained via simple conjugations of those arising from $\gamma=1$, as explained in Subsection~\ref{subsec:conj}. An example with $\gamma=2$ is treated in Appendix~\ref{app:higher_sheets}.

From the $(A_1,A_{2n})$ family, we obtain the Gang-Kim-Stubb $\mathcal{T}_n$ theories \cite{Gang:2023rei}, generalizing the derivation made in the special cases $n=1,2$ in \cite{ArabiArdehali:2024ysy}.
From the remaining three families, we obtain abelian SUSY enhancing gauge theories somewhat similar to the Gang-Kim-Stubb family, as follows.

From $(A_1,A_{2n-1}),$ we obtain a 3d $\mathcal{N}=2$ $U(1)^{n-1}$ gauge theory, with matter content as in Table~\ref{tab:masslessA1A2n-1}, the matrix of gauge-gauge CS couplings
\begin{align}
    K_{ij} &=\begin{cases}
    \hspace{1.5cm} 1,\qquad&\text{$n=2,$}\\
    \begin{pmatrix}
        2 & \frac{3}{2} & 1 & \dots & 1 & \frac{1}{2} \\
        \frac{3}{2} & 2 & \frac{3}{2} & \ddots & \vdots & \vdots \\
        1 & \frac{3}{2} & \ddots &\ddots & 1 & \vdots \\
        \vdots & \ddots & \ddots & \ddots & \frac{3}{2} & \frac{1}{2} \\
        1 & \dots & 1 & \frac{3}{2} & 2 & 1 \\
        \frac{1}{2} & \dots & \dots & \frac{1}{2} & 1 & 1 \\
    \end{pmatrix},\qquad&\text{$n\ge3,$}
    \end{cases}
\end{align}
and superpotential
\begin{equation}
    \mathcal{W}_V=V^{}_{1, -1, 0, \dots, 0}+V^{}_{0, 1, -1,0, \dots, 0}+ \dots+ V^{}_{0, \dots, 0, 1}\qquad\qquad(\text{$n-1$ terms}).
\end{equation}
Here $V_{m_1,\dots,m^{}_{r_G}}$ denotes the BPS monopole with magnetic gauge fluxes $m_1,\dots,m^{}_{r_G}$.

From $(A_1,D_{2n+1}),$ we obtain a 3d $\mathcal{N}=2$ $U(1)^{n}$ gauge theory, with matter content as in Table~\ref{tab:masslessA1D2n+1}, the matrix of gauge-gauge CS couplings
\begin{align}
  K_{ij} &=\begin{cases}
  \hspace{1.5cm}1,\qquad&\text{$n=1,$}\\
  \begin{pmatrix}
    \frac{3}{2} & \frac{1}{2} & 0 & \dots & 0  \\
    \frac{1}{2} & 1 & \ddots & \ddots & \vdots  \\
    0 & \ddots & \ddots & \ddots & 0  \\
    \vdots & \ddots & \ddots & 1 & \frac{1}{2}  \\
    0 & \dots & 0 & \frac{1}{2} & \frac{1}{2}  \\
  \end{pmatrix},\qquad&\text{$n\ge2,$}
    \end{cases} \label{eq:A1Dodd_Kij}
\end{align}
monopole superpotential
\begin{equation}
    \mathcal{W}^{}_V=V_{1, -1, 0, \dots, 0}+V_{0, 1, -1, 0, \dots, 0}+V_{0, \dots, 0, 1, -1}+V_{0, \dots, 0, 1}\qquad\qquad(\text{$n$ terms}),
\end{equation}
and matter superpotential as in \eqref{eq:W_Phi_D2n+1}.\\

From $(A_1,D_{2n}),$ we obtain a 3d $\mathcal{N}=2$ $U(1)^{n-1}$ gauge theory, with matter content as in Table~\ref{tab:masslessA1D2n}, the matrix of gauge-gauge CS couplings
\begin{align}
    K_{ij} &=\begin{cases}
    \hspace{1.5cm}0,\qquad&\text{$n=2$,}\\
    \begin{pmatrix}
        1 & \frac{1}{2} & 0 & \dots & 0 \\
        \frac{1}{2} & \ddots & \ddots & \ddots & \vdots \\
        0 & \ddots & \ddots & \ddots & 0 \\
        \vdots & \ddots & \ddots & \ddots & \frac{1}{2} \\
        0 & \dots & 0 & \frac{1}{2} & 1 \\
    \end{pmatrix},\qquad&\text{$n\ge3$,}
        \end{cases}\label{eq:A1Deven_Kij}
\end{align}
monopole superpotential
\begin{equation}
    \mathcal{W}^{}_V=V_{1, -1, \dots, 0}+V_{0, 1, -1, \dots, 0}+V_{0, \dots, 0, 1, -1}+V_{0, \dots, 0, 1}+V_{-1,0, \dots, 0}\qquad\quad(\text{$n$ terms}),
\end{equation}
and matter superpotential as in \eqref{eq:W_Phi_D2n}.

\subsection*{Outline of the paper}

We begin in Section~\ref{sec:generalities} by reviewing the $U(1)_r$-twisted 4d$\,\to\, $3d reduction, given a 4d $\mathcal{N}=1$ gauge theory description of the 4d $\mathcal{N}=2$ SCFT.

In Section~\ref{sec:AD_reduction} we apply the reduction procedure with $\gamma=\pm1$ to the $(A_1,A_n)$ and $(A_1,D_n)$ Argyres-Douglas theories. From the $(A_1,A_{2n})$ family we recover the Gang-Kim-Stubbs theories. From the remaining three families $(A_1,A_{2n-1})$, $(A_1,D_{2n+1})$ and $(A_1,D_{2n})$, three new infinite families of SUSY enhancing 3d gauge theories with monopole superpotentials are derived. The first two members of each new family are treated in detail: in most cases their extra SUSY current operators are determined, and in all cases their Higgs and Coulomb branch Hilbert series are matched with the expectations from the 4d reduction (namely, trivial 3d Coulomb branch and the 3d Higgs branch identical to the 4d one), using appropriate limits of the 3d superconformal index \cite{Razamat:2014pta}.
The simplest new theory is the one arising from $(A_1,A_3)$, which we discuss in detail in Subsection \ref{sec:A1A3EFT}. For this we present a dual description with manifest $\mathcal{N}=3$ supersymmetry, obtained by gauging the $SU(2)$ symmetry on the Coulomb branch of the $T[SU(2)]$ theory \cite{Intriligator:1996ex,Gaiotto:2008ak} with Chern-Simons level $k=3$, besides a topological sector.

In Section~\ref{sec:TQFT}, the TQFT arising from A-twisting the simplest new SUSY enhancing theory, namely the one arising from twisted reduction of $(A_1,A_3)$, is studied. In particular, its modular data are  determined using Bethe roots and BPS surgery techniques, and matched with those expected for the corresponding $\widehat{\mathfrak{su}}(2)_{-4/3}$ boundary VOA.

Section~\ref{sec:discussion} outlines some future directions. It also discusses an observation on the possibility of systematically obtaining unitary TQFTs from $\mathcal{N}=2$ preserving superpotential deformations of 3d $\mathcal{N}=4$ SCFTs with trivial Coulomb branch but non-trivial Higgs branch, such as those constructed in this work.

Appendix~\ref{app:higher_sheets} exemplifies the exploration of the higher sheets $|\gamma|>1$ of Argyres-Douglas theories, finding specifically a 3d $\mathcal{N}=2\to\mathcal{N}=4$ SUSY enhancing non-abelian gauge theory with monopole superpotential, no Coulomb branch, and Higgs branch $\mathbb{C}^2/\mathbb{Z}_2,$ from the $\gamma=2$ twisted reduction of $(A_1,D_5).$

\vspace{.5cm}
\noindent\textbf{Note added.} While this manuscript was being finalized, the work \cite{Gaiotto:2024ioj} appeared, which uses different methods to obtain similar (though different looking) 3d theories from the 4d $(A_1,A_n)$ and $(A_1,D_\text{odd})$ families.

\section{$U(1)_r$-twisted 4d$\ \to\ $3d reduction}\label{sec:generalities}

This section contains a short outline of the R-twisted circle reduction procedure \cite{ArabiArdehali:2024ysy} that will be applied to the $(A_1,A_n)$ and $(A_1,D_n)$ Argyres-Douglas theories \cite{Argyres:1995jj,Eguchi:1996ds,Gaiotto:2010jf,Cecotti:2010fi} in the next section. We start with the $\mathcal{N}=2$ index of a given Argyres-Douglas theory as per \cite{Maruyoshi:2016aim,Agarwal:2016pjo}, which we denote by $\mathcal{I}_t(p, q, t)$ \cite{Kinney:2005ej,Romelsberger:2005eg,Gadde:2011uv}:
\begin{equation}
    \mathcal{I}_t(p,q,t):=\mathrm{Tr}^{}_{S^3}(-1)^{2j_1+2j_2}\, p^{j_1+j_2+r}q^{j_1-j_2+r}t^{R-r}.
\end{equation}
Here $R,r$ are the generators of the Cartan of the 4d $\mathcal{N}=2$ R-symmetry $SU(2)_R\times U(1)_r$, and $j_1,j_2$ are the Lorentz spins. We then twist $p \to p \,e^{2 \pi i \gamma}$, and suppress all other chemical potentials by setting $p = q$ and $t = q^{1+s}$. We thus consider
\begin{equation}\label{eq:Isgamma}
\mathcal{I}_s^\gamma(q) \equiv \mathcal{I}_t(q\, e^{2 \pi i \gamma}, q, q^{1+s})=\mathrm{Tr}^{}_{S^3}(-1)^{2r}\, q^{2j_1+R_s}\,,
\end{equation}
where
\begin{equation}
    R_s:=r+R-s\,(r-R)\,.\label{eq:Rs_def}
\end{equation}
Our focus in this paper will be on $\gamma=1$, corresponding to the minimal $U(1)_r$ twist around the circle \cite{ArabiArdehali:2024ysy,Dedushenko:2023cvd}, to which we refer to as the \emph{second sheet}.

The value $s=\tfrac{1}{3}$ corresponds to the unrefined $\mathcal{N}=1$ index
\begin{equation}
    R^{}_{s=1/3}=\frac{4}{3}R+\frac{2}{3}r=R^{}_{\mathcal{N}=1}.
\end{equation}
This value of $s$ was the focus of \cite{ArabiArdehali:2024ysy}, however in the present paper we will be mostly interested in $s=0,\pm1$ for the reasons that we will next explain.

To make contact with the 3d $\mathcal{N}=4$ $SU(2)_C\times SU(2)_H$ R-symmetry in the infrared, we first note that the 4d $SU(2)_R$ is identified with the 3d $SU(2)_H$ (up to a normalization due to conventions). In terms of the Cartan generators
\begin{equation}
    R=\frac{H}{2}\,.
\end{equation}
This is because our $U(1)_r$-twisted reductions leave the Higgs branch (which both the UV $SU(2)_R$ and the IR $SU(2)_H$ act on) intact, and since both the $SU(2)_R$ and the $SU(2)_H$ are non-abelian, there is no room for any mixing along the RG flow across dimensions.

We also identify the UV $U(1)_r$ with the Cartan of the IR $SU(2)_C$ (again up to a normalization)
\begin{equation}\label{eq:U1rSU2}
    r=\frac{C}{2}\,.
\end{equation}
This identification is clear for Lagrangian theories, but it is more subtle for non-Lagrangian theories and especially for the AD theories we are considering. Indeed, AD theories contain Coulomb branch operators with fractional $U(1)_r$ charges (which are not just half-integers), while in 3d $\mathcal{N}=4$ theories we can only have Coulomb branch operators with integer charges under the Cartan $C$ due to the non-abelian nature of the $SU(2)_C$ symmetry. This fact is clearly not compatible with the identification \eqref{eq:U1rSU2}. The explanation is that, since the $U(1)_r$ symmetry is abelian, it can mix with other abelian symmetries (including part of the 4d superconformal group) along the RG flow before becoming the Cartan of the $SU(2)_C$. Consequently, in reductions where Coulomb branch operators with fractional (but not just half-integer) $U(1)_r$ charges survive, the identification \eqref{eq:U1rSU2} is not true (see \emph{e.g.}~\cite{Buican:2015hsa,Dedushenko:2019mnd}). On the other hand, in the twisted reductions we consider in this work, the Coulomb branch operators with fractional $U(1)_r$ charge decouple (the Coulomb branches are completely lifted in fact), so at least there is no need for the $U(1)_r$ to mix with anything to become the Cartan of $SU(2)_C$. We thus proceed with the assumption of no mixing, and corroborate this assumption by providing various consistency checks (as in Section~3.1.2 of \cite{Buican:2015hsa}).

The preceding two paragraphs justify the $s=-1$ and $s=1$ columns of the following table, averaging which gives the $s=0$ column:
\begin{center}
\begin{tabular}{|c|c|c|c|}
  \hline
  $s$ &$-1$ & $0$ & $1$ \\
  \hline
  $R_s$&$C=2r$ &$\frac{C+H}{2}=r+R$ &$H=2R$\\
  \hline
  {\small 3d $\mathcal{N}=2$ data}&B-twist&$\mathcal{N}=4$ SCFT &A-twist\\
  \hline
\end{tabular}
\end{center}
Note that $s=0$ (or $t=q$) corresponds to the Schur index \cite{Gadde:2011ik,Gadde:2011uv}. With this choice the Cardy limit $q\to1$ of the second-sheet index yields the 3d $\mathcal{N}=2$ R-charges of the $\mathcal{N}=4$ SCFT arising from the circle reduction, namely
\begin{equation}
    R_{s=0}=r+R=\frac{C+H}{2}\,.
\end{equation}

For a general 4d $\mathcal{N}=1$ gauge theory with a $U(1)_{R_s}$ R-symmetry and semi-simple gauge group $G$ of rank $r_G$, the indices of our interest take the schematic form \cite{Dolan:2008qi,ArabiArdehali:2024ysy}:
\begin{align}
  \mathcal{I}_s^\gamma(q) &= \frac{(q; q)^{2r_G}}{|W|} \int_{\mathfrak{h}_\text{cl}}   \prod_{i=1}^{r_G}\mathrm{d}^{r_G}x\,\frac{\prod_{\chi} \prod_{\rho^\chi} \Gamma_e(\boldsymbol{z}^{\rho^\chi} q^{R_s^\chi} e^{2 \pi i\, r^\chi \, \gamma})}{\prod_{\alpha_+} \Gamma_e(\boldsymbol{z}^{\alpha_+}) \Gamma_e (\boldsymbol{z}^{-\alpha_+})}\,,\label{eq:Is}
\end{align} 
where $(z; q)$ is the $q$-Pochhammer symbol and $\Gamma_e(z) \equiv \Gamma(z; q, q)$ is the elliptic Gamma function, while $\mathfrak{h}_\text{cl}=(-1/2,1/2]^{r_G}$ is the moduli space of the gauge holonomies $z_j=e^{2\pi i x_j}$ around the compactification circle and $|W|$ is the dimension of the Weyl group of $G$. Finally, the product over $\chi$ is on the set of chiral superfields in the theory, with $\rho^\chi$ the weights of the representation of $G$ under which they transform, $R_s^\chi$ their $U(1)_{R_s}$ charge and $r^\chi$ their $U(1)_r$ charge, and $\alpha_+$ are the positive roots of $G$. 

Upon taking the Cardy $q=e^{-\beta}\to1$ (or high-temperature $\beta\to0$) limit, the index takes the following form \cite{ArabiArdehali:2015ybk,ArabiArdehali:2024ysy}:
\begin{align}
  \mathcal{I}_s^\gamma(q) &\approx e^{-\frac{\pi^2}{3\beta} \Tr R_s} \left( \frac{2\pi}{\beta} \right)^{\dim \mathfrak{h}_\text{qu}} e^{- \frac{4\pi^2}{\beta} L_{R_s}^\gamma(\boldsymbol{x}^*) + i \frac{8\pi^3}{\beta^2} Q^\gamma(\boldsymbol{x}^*)},\label{eq:index_Cardy}
\end{align} 
where
\begin{align}
  Q^\gamma(\boldsymbol{x}) &\equiv \frac{1}{12} \sum_{\chi} \sum_{\rho^\chi} \kappa \left( \rho^\chi \cdot \boldsymbol{x} + r^\chi \cdot \gamma \right),\label{eq:Qht} \\
  L_{R_s}^\gamma(\boldsymbol{x}) &\equiv \frac{1}{2} \sum_{\chi} (1 - R_s^\chi) \sum_{\rho^\chi} \vartheta (\rho^\chi \cdot \boldsymbol{x} + r^\chi \cdot \gamma) - \sum_{\alpha_+} \vartheta (\alpha_+ \cdot \boldsymbol{x})\,,\label{eq:Lht} 
\end{align}
with
\begin{equation}
    \begin{split}
        \kappa(x) &\equiv \{ x \} (1 - \{ x \}) (1 - 2 \{ x \}),\\
        \vartheta(x) &\equiv \{ x \} (1 - \{ x \}).
    \end{split}
\end{equation}
In \eqref{eq:index_Cardy} by $\boldsymbol{x}^\ast$ we mean any point on the subset $\mathfrak{h}_\text{qu}\subset\mathfrak{h}_\text{cl}$ of the stationary locus of $Q^\gamma$ where $L_{R_s}^\gamma$ is minimized. These $\boldsymbol{x}^\ast$ are (roughly speaking \cite{Ardehali:2021irq}) saddle-points of the integral \eqref{eq:Is} in the $q\to1$ limit, and correspond to the vacua of the circle-compactified gauge theory.

As explained in the discussion section of \cite{ArabiArdehali:2024ysy}, single-valuedness of the Schur index of the Argyres-Dougals theories of our interest \cite{Cordova:2015nma}, together with the fact that they satisfy Di~Pietro-Komargodski type asymptotics (see \emph{e.g.}~\cite{Buican:2015ina}) on the first sheet, imply that the saddle-point values $Q^\gamma(\boldsymbol{x}^\ast)$ and $L_{R_s}^\gamma(\boldsymbol{x}^\ast)$ are $0$ for any $\gamma$, and in particular for $\gamma=1$.  Therefore all we need is some $\boldsymbol{x}^* \in \{ \boldsymbol{x} \in  \ \mathfrak{h}_\text{cl}| \ L_{R_s}^\gamma(\boldsymbol{x}) = Q^\gamma(\boldsymbol{x}) = 0 \}$. In the following sections, we shall propose such a saddle point for each of the four families of Argyres-Douglas theories.

From the knowledge of the saddle point $\boldsymbol{x}^\ast$ and of the functions $Q^\gamma$ and $L_{R_s}^\gamma$, we obtain the EFT data for the 3d theories arising from the twisted reduction as follows.
\begin{itemize}
    \item \textbf{Gauge symmetry and field content.} Reducing the theory with the $U(1)_r$-twist would naively integrate out all the matter fields, since the mass of the KK modes is shifted such that none of them stays massless. However, this is true at generic points of the 3d Coulomb branch, while at special points some fields might becomes massless if the gauge holonomy $\boldsymbol{x}$ compensates the mass induced by the twist. This is indeed the case at the saddle point $\boldsymbol{x}^\ast$ we focus on. The price to pay is that, except for specific points like the origin, the gauge group is partially broken on the Coulomb branch. The breaking can be at most to the Cartan subgroup, which will be the preserved gauge symmetry in most of the examples of this paper (except for Appendix \ref{app:higher_sheets}). In order to determine the massless field content of the EFT we simply have to look at which arguments of the $\vartheta$ functions in $L_{R_s}^\gamma(\boldsymbol{x}^*)$ vanish. The light vector fields will tell us what is the preserved gauge symmetry, while the light chiral fields will tell us what is the matter content.
    \item\textbf{Chern-Simons couplings.} The quadratic function $Q^\gamma(\boldsymbol{x})$ encodes the gauge-gauge CS interactions of the low energy theory at the point $\boldsymbol{x}$ of the Coulomb branch. The bare CS levels of the EFT of our interest are then given by the Hessian at the saddle point\footnote{Since we evaluate exactly at $\boldsymbol{x}^*$, we need not worry about ``inner'' versus ``outer'' patches. See footnote~7 in \cite{ArabiArdehali:2024ysy}.} $K_{ij} = \partial_i \partial_j Q^\gamma (\boldsymbol{x}^*)$.
    \item\textbf{Superpotential interactions.} The superpotential of the 3d EFT consists of two pieces, the one involving only matter fields and the monopole superpotential. The first one can be determined by considering the interactions that are compatible with the charge assignment of the massless matter fields and it typically descends from the 4d superpotential. The monopole superpotential can instead be determined from the behaviour of the functions $Q^\gamma$ and $L_{R_s}^\gamma$ around the saddle point. Indeed, the saddle point $\boldsymbol{x}^*$ generically has multiple directions along which $Q^{\gamma}$ remains constant under small perturbations from $\boldsymbol{x}^*$. Since away from $\boldsymbol{x}^*$ all fields are massive, the second derivative of $Q^{\gamma}$ is computing the effective CS level which is related to the charge under the gauge symmetry of the monopole, so these flat directions of $Q^{\gamma}$ are associated with gauge invariant monopoles. The function $L_{R_s}^{\gamma}$ is instead bilinear in $\boldsymbol{x}$ and in the parameters for the global and R symmetries (compactly encoded in the trial R-symmetry $U(1)_{R_s}$) and so it represents BF couplings between the gauge symmetry and these other symmetries. Similarly to the gauge-gauge CS couplings, the bare levels in the EFT of these BF couplings are given by the slope of $L_{R_s}^{\gamma}$ (as a function of $\boldsymbol{x}$) at the saddle point $\boldsymbol{x}^*$, while the effective levels are obtained considering small perturbations around $\boldsymbol{x}^*$. Hence, from the slope of $L_{R_s}^{\gamma}$ along the flat directions of $Q^{\gamma}$ at $\boldsymbol{x}^*$ we can determine the $R_s$-charge of the gauge invariant monopoles. In particular, if the slope (or some integer multiple of it associated to monopoles of higher flux) is 2, it means that a monopole term is generated in the superpotential of the EFT.
\end{itemize}

Once the 3d $\mathcal{N}=2$ EFT data is determined as above, we can compute the gauge and $R_s$ charges of the monopole operators via\footnote{A few comments are in order regarding these equations. First, we point out that we use an orientation-reversed convention here compared to \cite{ArabiArdehali:2024ysy}. Second, these equations apply to abelian symmetries; for non-abelian ones they can be applied to each factor of their Cartan subgroups. Finally, similar equations apply also if the symmetry is a global symmetry---see \emph{e.g.}~\eqref{eq:A(m)}.  \label{fn:parity_conv}} (see \emph{e.g.}~\cite{Borokhov:2002ib,Borokhov:2002cg,Borokhov:2003yu,Gaiotto:2008ak,Benna:2009xd,Bashkirov:2010kz,Cremonesi:2013lqa,Closset:2016arn,Pasquetti:2019uop})
\begin{equation}
    c_i(\boldsymbol{m})=\sum_j k^{}_{ij}\,{m}_j-\frac{1}{2}\sum_{\chi}\sum_{\rho^\chi}\rho^\chi_i|\rho^\chi(\boldsymbol{m})|,\label{eq:c(m)}
\end{equation}
\begin{equation}
    R_s(\boldsymbol{m})=\sum_j k^{}_{g_jR_s}\,{m}_j-\frac{1}{2}\sum_{\chi}(R_s^{\chi}-1)\sum_{\rho^\chi}|\rho^\chi(\boldsymbol{m})|-\frac{1}{2}\sum_{\alpha} |\alpha(\boldsymbol{m})|,\label{eq:r(m)}
\end{equation}
and perform various consistency checks, using in particular the 3d supersymmetric index (which coincides with the superconformal index when computed with the R-symmetry $R_{s=0}$) \cite{Bhattacharya:2008zy,Kim:2009wb,Imamura:2011su,Kapustin:2011jm,Dimofte:2011py,Aharony:2013kma}:
\begin{equation}
\begin{split}
    &\mathcal{I}^{}(q):=\mathrm{Tr}^{}_{S^2}(-1)^{2j^{}_3} q^{{R_s}/2+j^{}_3}\\
    &\qquad =\sum_{\boldsymbol{m}}\frac{q^{\frac{R_s(\boldsymbol{m})}{2}}}{|W_{\boldsymbol{m}}|}\oint \prod_{i=1}^{r_G}\frac{\mathrm{d}z_i}{2\pi i z_i}(-1)^{c_i(\boldsymbol{m})m_i} z_i^{c^{}_{i}(\boldsymbol{m})} \prod_{\alpha_+} (1-q^{|\alpha_+(\boldsymbol{m})|/2}\boldsymbol{z}^{\pm\alpha_+})\\
&\qquad\qquad \prod_{\chi}\prod_{\rho^\chi}\frac{(\boldsymbol{z}^{-\rho^\chi}q^{|\rho^\chi(\boldsymbol{m})|/2+1-R_s^\chi/2};q)}{(\boldsymbol{z}^{\rho^\chi}q^{|\rho^\chi(\boldsymbol{m})|/2+R_s^\chi/2};q)}\,,\label{eq:3dIndexDef}
    \end{split}
\end{equation}
where the gauge fluxes $\boldsymbol{m}$ live in the co-weight lattice of the gauge group (for us they will mostly be just integers since the gauge group is abelian) and $|W_{\boldsymbol{m}}|$ is the dimension of the Weyl group of the residual gauge symmetry in the monopole background.
Additional fugacities for flavor and topological symmetries are straightforward to introduce.

\subsection*{Flavor charges of the monopoles}

The combination 
\begin{equation}
    A:=r-R=\frac{C-H}{2}\,,\label{eq:Adef}
\end{equation}
corresponds to a $U(1)_A$ subgroup of the 4d $\mathcal{N}=2$ (resp.~3d $\mathcal{N}=4$) R-symmetry that is a flavor symmetry usually called axial symmetry from the 4d $\mathcal{N}=1$ (resp.~3d $\mathcal{N}=2$) perspective. Notice that this appears as the coefficient of $-s$ in \eqref{eq:Rs_def}. From \eqref{eq:Lht} we have
\begin{equation}
    L_{R_s}^\gamma(\boldsymbol{x})=L_{R_0}^\gamma(\boldsymbol{x})-s\cdot L_{A}^\gamma(\boldsymbol{x})\,,
\end{equation}
with
\begin{equation}
    L_{A}^\gamma(\boldsymbol{x})\equiv -\frac{1}{2} \sum_{\chi} A_\chi \sum_{\rho^\chi} \vartheta (\rho^\chi \cdot \boldsymbol{x} + r^\chi \cdot \gamma).\label{eq:Lf}
\end{equation}
Here $A_\chi$ is the $U(1)_A$ charge of the 4d chiral multiplet $\chi.$ Analogously to the discussion in \cite{ArabiArdehali:2024ysy}, we can use the following counterpart of \eqref{eq:r(m)}
\begin{equation}
    A(\boldsymbol{m})=\sum_j k^{}_{g_jA}\,{m}_j-\frac{1}{2}\sum_{\chi}A_{\chi}\sum_{\rho^\chi\in L_\ast}|\rho^\chi(\boldsymbol{m})|,\label{eq:A(m)}
\end{equation}
to deduce
\begin{equation}
    \sum_j\partial_j L_{A}^\gamma\cdot m_j\xrightarrow{\text{\ patch-wise\ }}A(\boldsymbol{m}),
\end{equation}
where by $A(\boldsymbol{m})$ we mean the $U(1)_A$ charge of the 3d monopole with GNO flux $\boldsymbol{m}.$

Note in particular that the slope $\partial_x$ of $L_{R_s}^\gamma$ being $2$ irrespective of $s$ implies that the slope of $L_{A}^\gamma$ is zero, meaning the corresponding monopole is invariant under $U(1)_A$. This is a necessary condition for the UV $U(1)_A$ symmetry to allow the corresponding monopole in the superpotential of the 3d EFT.

Piece-wise linear functions similar to \eqref{eq:A(m)} can be defined for other 4d $U(1)$ flavor symmetries as well, and their flatness along the directions of superpotential monopoles must be checked.

\subsection*{Notation}

All the 3d SUSY enhancing theories derived in this work have abelian gauge groups of the form $U(1)^{r_G}$, for some rank $r_G$ (except for Appendix \ref{app:higher_sheets} where we discuss a non-abelian example). We indicate the gauge charges of matter chiral multiplets in these theories as subscripts:
\begin{equation}
    \Phi^{}_{q^{}_1,\dots,q^{}_{r_G}}\,.
\end{equation}
The gauge charges of monopole operators will be kept implicit, but their flux quantum numbers will be indicated as subscripts:
\begin{equation}
    V^{}_{m^{}_1,\dots,m_{r_G}}\,.
\end{equation}

To compare with \cite{Gang:2023rei} note that
\begin{equation}
    s_\text{here}=-\nu_\text{there}\,,
\end{equation}
while for the 3d axial $U(1)_A$ symmetry \eqref{eq:Adef} we use the fugacity
\begin{equation}
    y^{}_\text{here}=\eta^{}_\text{there}\, .
\end{equation}

\section{3d SUSY enhancement from 4d SUSY enhancement}\label{sec:AD_reduction}

In this section, we apply the technology we have just reviewed to derive the 3d $\mathcal{N}=2$ EFTs coming from the $U(1)_r$-twisted compactification of the 4d $\mathcal{N}=1$ Lagrangians of \cite{Maruyoshi:2016aim,Agarwal:2016pjo} for the AD theories of type $(A_1,A_n)$ and $(A_1,D_n)$. We will then provide evidence that these flow to the 3d $\mathcal{N}=4$ SCFTs associated to the twisted compactification of the AD theories. For the entirety of this section, $R$ will denote the three-dimensional $R_{s=0}$, not the Cartan of the four-dimensional $SU(2)_R$.

\subsection{Gang-Kim-Stubbs from $(A_1,A_{2n})$}


The $\mathcal{N}=2$ index for the $(A_1,A_{2n})$ theory reads
\begin{align}
  \mathcal{I}_{t}^{(A_1, A_{2n})} &= \frac{\big((p;p)(q;q)\big)^n}{2^n n!} \prod_{i=1}^n \frac{\Gamma_e \left( \left( \frac{pq}{t} \right)^{\frac{2(n+1+i)}{2n+3}} \right)}{\Gamma_e\left( \left( \frac{pq}{t} \right)^{\frac{2i}{2n+3}} \right)} \Gamma_e \left( \left( \frac{pq}{t} \right)^{\frac{1}{2n+3}} \right)^n \nonumber\\
  &\hspace{.5cm}\times\int_{\mathfrak{h}_\text{cl}}\! \mathrm{d}^{n}x\  \prod_{i=1}^{n} \Gamma_e \left( z_i^\pm \left( \frac{pq}{t} \right)^{\frac{n+1}{2n+3}} t^{\frac{1}{2}} \right) \Gamma_e \left( z_i^\pm \left( \frac{pq}{t} \right)^{-\frac{n}{2n+3}} t^{\frac{1}{2}} \right) \nonumber\\
  & \hspace{-1cm}\cdot \prod_{1 \leq i < j \leq n} \frac{\Gamma_e \left( \left( z_i z_j \right)^\pm (\frac{pq}{t})^{\frac{1}{2n+3}} \right) \Gamma_e \left( \left( \frac{z_i}{z_j} \right)^\pm \left( \frac{pq}{t} \right)^{\frac{1}{2n+3}} \right)}{\Gamma_e \left( \left( \frac{z_i}{z_j} \right)^\pm \right)} \ \prod_{i=1}^{n} \Gamma_e \left( z_i^{\pm 2} \left( \frac{pq}{t} \right)^{\frac{1}{2n+3}} \right)
\end{align}
where $z_1, \dots, z_n$ are the fugacities corresponding to the $Sp(n)$ gauge group. We twist $p \to p e^{2 \pi i}$ and then set $p = q$ and $t = q^{1+s}$, which gives $\mathcal{I}_s^{\gamma=1}$ as in \eqref{eq:Isgamma}. Recall $z_i = e^{2 \pi i x_i}$. Then, we know that the leading asymptotics in this limit is controlled by the piecewise quadratic function 
\begin{eqnarray}
  &12 Q^{\gamma=1}(\boldsymbol{x}) = n \kappa \left( \frac{1}{2n+3} \right) + \sum_{j=1}^n \Big( \kappa(\alpha_j) - \kappa(\beta_j) + \kappa \left( 2x_j + \frac{1}{2n+3} \right) + \kappa \left( -2x_j + \frac{1}{2n+3} \right) \Big)\nonumber \\
  & + \sum_{j=1}^n \Big( \kappa \left( x_j + \frac{n+1}{2n+3} \right) + \kappa \left( -x_j + \frac{n+1}{2n+3} \right) + \kappa \left( x_j - \frac{n}{2n+3} \right) + \kappa \left( -x_j - \frac{n}{2n+3} \right) \Big)\nonumber \\
  & + \sum_{1 \leq i < j \leq n} \bigg( \kappa \left( x_i + x_j + \frac{1}{2n+3} \right) + \kappa \left( x_i - x_j + \frac{1}{2n+3} \right)\nonumber \\
  &\hspace{4cm} + \kappa \left( -x_i + x_j + \frac{1}{2n+3} \right) + \kappa \left( -x_i - x_j + \frac{1}{2n+3} \right) \bigg),
\end{eqnarray} 
and the piecewise-linear function
\begin{eqnarray}
  & L_{R_s}^{\gamma=1}(\boldsymbol{x}) = \frac{n}{2} \left( 1 - \frac{1 - s}{2n+3} \right) \vartheta \left( \frac{1}{2n+3} \right) + \sum_{j=1}^n \left( \left( \frac{1-(1-s)\alpha_j}{2} \right) \vartheta(\alpha_j) - \left( \frac{1-(1-s)\beta_j}{2} \right) \vartheta(\beta_j) \right) \nonumber\\
  & + \frac{1-s}{4(2n+3)} \sum_{j=1}^n \left( \vartheta \left( x_j + \frac{n+1}{2n+3} \right) + \vartheta \left( -x_j + \frac{n+1}{2n+3} \right) \right) \nonumber\\
  & + \frac{(4n+3)(1-s)}{4(2n+3)} \sum_{j=1}^n \left( \vartheta \left( x_j - \frac{n}{2n+3} \right) + \vartheta \left( -x_j - \frac{n}{2n+3} \right) \right) \nonumber\\
  & + \frac{1}{2} \left( 1 - \frac{1-s}{2n+3} \right) \bigg( \sum_{j=1}^n \Big( \vartheta \left( 2x_j + \frac{1}{2n+3} \right) + \vartheta \left( -2x_j + \frac{1}{2n+3} \right) \Big) \nonumber\\
  & \hspace{3cm} + \sum_{1 \leq i < j \leq n} \Big( \vartheta \left( x_i + x_j + \frac{1}{2n+3} \right) + \vartheta \left( x_i - x_j + \frac{1}{2n+3} \right) \nonumber\\
  & \hspace{5cm} + \vartheta \left( -x_i + x_j + \frac{1}{2n+3} \right) + \vartheta \left( -x_i - x_j + \frac{1}{2n+3} \right) \Big) \bigg) \nonumber\\
  & - \sum_{j=1}^n \vartheta(2x_j) - \sum_{1 \leq i < j \leq n} \Big( \vartheta (x_i + x_j) + \vartheta \left( x_i - x_j \right) \Big),
\end{eqnarray}
where we have defined $\alpha_j = \frac{2(n+1+j)}{2n+3}$ and $\beta_j = \frac{2j}{2n+3}$. 

The saddle point we propose for this theory is
\begin{equation}
  \hspace{5cm}x^*_j = \frac{j}{2n+3}\,,\hspace{3cm} j=1,\dots, n\,.
\end{equation}
This corresponds to a 3d $\mathcal{N}=2$ $U(1)^{n}$  gauge theory.

To show that both of the functions above vanish at this point is as simple as plugging the saddle point into the functions. The resulting sums can then be evaluated somewhat straightforwardly using computational software (specifically Mathematica).

We then propose that $Q^{\gamma=1}$ has $n-1$ flat directions near this saddle point
\begin{align}
  (\delta x_1,\dots,\delta x_n)\,=\, (1,-1, 0, \dots, 0),\ (0, 1, -1, 0, \dots, 0),\ \dots,\ (0, \dots, 0, 1, -1)\,.
\end{align}
We show that these are flat directions of $Q^{\gamma=1}$ by first considering the terms in $Q^{\gamma=1}$ that depend on $x_k$ and $x_{k+1}$ for $1 \leq k \leq n-1$. For example, in the present case this would be
\begin{eqnarray}
  &\hspace{-4cm}Q^{\gamma=1}(x_k, x_{k+1}) = \sum_{j=k, k+1} \Big( \kappa \left( 2x_j + \frac{1}{2n+3} \right) + \kappa \left( -2x_j + \frac{1}{2n+3} \right) \Big)\nonumber \\
  & \hspace{1cm}+ \sum_{j=k, k+1} \Big( \kappa \left( x_j + \frac{n+1}{2n+3} \right) + \kappa \left( -x_j + \frac{n+1}{2n+3} \right) + \kappa \left( x_j - \frac{n}{2n+3} \right) + \kappa \left( -x_j - \frac{n}{2n+3} \right) \Big) \nonumber\\
  & \hspace{-4cm}+ \sum_{\substack{j = k, k+1 \\ i \neq j}} \bigg( \kappa \left( x_i + x_j + \frac{1}{2n+3} \right) + \kappa \left( x_i - x_j + \frac{1}{2n+3} \right) \\
  & \hspace{2cm} + \kappa \left( -x_i + x_j + \frac{1}{2n+3} \right) + \kappa \left( -x_i - x_j + \frac{1}{2n+3} \right) \bigg) - \kappa \left( x_k - x_{k+1} + \frac{1}{2n+3} \right),\nonumber
\end{eqnarray} 
We then make the replacement $(x_k, x_{k+1}) \to (x_k^* + x, x_{k+1}^* - x)$, and evaluate the resulting expression on Mathematica\footnote{In order to help Mathematica evaluate the function, include the assumption that the perturbation $x$ is small enough that the arguments of the various terms at the saddle point do not cross an integer when we perturb around it. For instance, in the above setting, imposing that $|x| < \frac{1}{53(2n+3)}$ works. Of course, this would break down if one the arguments is exactly an integer at the saddle point. To mitigate this, we must also include the assumption that $x > 0$
to ensure that the perturbation always pushes the argument to one side of the integer.}. Doing so will show that the $x$-dependence of the above function cancels out, proving that these are indeed flat directions of $Q^{\gamma=1}$. The same strategy can also be used to show that $L_{R_s}^{\gamma=1}$ has slope $+2$ when perturbing along these directions from $x^*$. This implies the monopole superpotential
\begin{equation}
    \hspace{1.5cm}\mathcal{W}_V=V_{1,-1, \dots, 0}+\ V_{0, 1, -1, \dots, 0}+\ \dots+\ V_{0, \dots, 1, -1} \qquad\qquad(\text{$n-1$ terms}).\label{eq:WV_A1A2n}
\end{equation}
Note that the monopoles $V$ in the above equation have $n$ subscripts, as many as the rank of the $U(1)^{n}$ gauge group. These indeed indicate the value of the magnetic flux under each of the gauge $U(1)$'s.

For the matrix of CS couplings, we first need to derive the corresponding expression, i.e., $K_{ij} = \partial_i \partial_j Q^\gamma(\boldsymbol{x}^*)$. By defining 
\begin{align}
  \overline{B}_1(x) \equiv 
  \begin{dcases}
    \{ x \} - \frac{1}{2},\qquad &x \notin \mathbb{Z} ;\\
    0,\qquad &x \in \mathbb{Z},
  \end{dcases}
\end{align} 
and noting that this is just the second derivative of $\kappa(x)$ up to a multiplicative factor:
\begin{equation}
    \kappa''(x)=12\overline{B}_1(x),
\end{equation}
we see that the off-diagonal elements of the CS coupling matrix are given by 
\begin{align}
  K_{ij} &= \overline{B}_1 \left( x_i^* + x_j^* + \frac{1}{2n+3} \right) - \overline{B}_1 \left( x_i^* - x_j^* + \frac{1}{2n+3} \right)\nonumber\\
  &\qquad- \overline{B}_1 \left( -x_i^* + x_j^* + \frac{1}{2n+3} \right) + \overline{B}_1 \left( -x_i^* - x_j^* + \frac{1}{2n+3} \right)
\end{align} 
for $i \neq j$, and the diagonal elements are given by
\begin{align}
  &K_{ii} = 4 \overline{B}_1 \Big( 2x_i^* + \frac{1}{2n+3} \Big) + 4 \overline{B}_1 \Big( -2x_i^* + \frac{1}{2n+3} \Big) \nonumber\\
  &+ \overline{B}_1 \Big( x_i^* + \frac{n+1}{2n+3} \Big) + \overline{B}_1 \Big(\!\! -x_i^* + \frac{n+1}{2n+3} \Big) + \overline{B}_1 \Big( x_i^* - \frac{n}{2n+3} \Big) + \overline{B}_1 \Big(\!\! -x_i^* - \frac{n}{2n+3} \Big) \nonumber\\
  &+ \sum_{j \neq i} \bigg( \overline{B}_1 \Big( x_i^* + x_j^* + \frac{1}{2n+3} \Big) + \overline{B}_1 \Big( x_i^* - x_j^* + \frac{1}{2n+3} \Big) \\
  &\qquad\qquad+ \overline{B}_1 \Big( -x_i^* + x_j^* + \frac{1}{2n+3} \Big) + \overline{B}_1 \Big( -x_i^* - x_j^* + \frac{1}{2n+3} \Big) \bigg),\nonumber 
\end{align} 
Just as earlier, these are sums that can be evaluated straightforwardly on Mathematica. The result takes the form of an $n \times n$ matrix, reading 
\begin{equation}
    k_{gg}=\tfrac{3}{2}\,, \qquad \text{for $n=1$,}
\end{equation}
\begin{align}
  K_{ij} &= 
  \begin{pmatrix}
    \frac{3}{2} & \frac{1}{2} & 0 & \dots & 0  \\
    \frac{1}{2} & 1 & \ddots & \ddots & \vdots  \\
    0 & \ddots & \ddots & \ddots & 0  \\
    \vdots & \ddots & \ddots & \ddots & \frac{1}{2}  \\
    0 & \dots & 0 & \frac{1}{2} & 1  \\
  \end{pmatrix},\qquad\text{for $n\ge2.$}
\end{align} 
That is, the diagonal elements are all 1 except for the first one, which is $K_{11} = \frac{3}{2}$, the lower and upper diagonals are all $\frac{1}{2}$, and every other element vanishes. 

Finally, we look at the massless field content of the theory. As mentioned earlier, this is determined by which arguments of the $\vartheta$ functions vanish in $L_{R_s}^{\gamma=1}(x^*)$. At this point, we also switch to new gauge coordinates $y_i = x_i - x_{i + 1}, 1 \leq i \leq n - 1$ and $y_n = x_n$ to make contact with the results of Gang-Kim-Stubbs \cite{Gang:2023rei}. Note that this only affects the gauge charges of the field content and not the R-charges. See Table~\ref{tab:masslessA1A2n}.
\begin{table}[t]
  \centering
  \begin{tabular}{c|c|c|c}
      \toprule
      {\small$\#$ of chirals} & $R_s=R_0-s\,A$ & {\small Gauge charge(s)} & {\footnotesize Gauge charge(s) (GKS coords)} \\
    \midrule
      1 & $\frac{3}{4n+6} + \frac{4n+3}{4n+6} s$ & $e_n^{(n)}$ & $e_n^{(n)}$ \\
      $n-1$ &  $\frac{1-s}{2n+3}$ & $\{ e^{(n)}_i - e_{i+1}^{(n)} \ |${\footnotesize$\, 1 \leq i \leq n-1$}$\}$ & $\{ e_i^{(n)} \ |${\footnotesize$\, 1 \leq i \leq n-1$}$\}$ \\
      \bottomrule
  \end{tabular}
  \caption{Massless field content of the 3d EFT descending from the $\gamma=1$ reduction of the $(A_1, A_{2n})$ Argyres-Douglas theory. We use the notation $e_i^{(m)}$ to mean the $i$th vector in the standard basis of $\mathbb{R}^m$, in order to present results succinctly.}
  \label{tab:masslessA1A2n}
\end{table}

It can be checked that no piece of the 4d matter superpotential \cite{Maruyoshi:2016aim} survives the $\gamma=1$ reduction.

\textbf{Remark}. Since we have one fewer monopole superpotential terms (namely $n-1$) than the rank  (namely $n$), the superpotential \eqref{eq:WV_A1A2n} \emph{under-determines} the vector $k_{jR_s}$ of mixed gauge-$R_s$ CS couplings. Table~\ref{tab:masslessA1A2n} needs therefore to be accompanied with at least one entry of $k_{jR}$ as a function of $s$, to provide a complete description. For the case of $(A_1,A_2)$, \emph{i.e.}~the Gang-Yamazaki theory \cite{Gang:2018huc}, the additional required data is found from averaging $\partial_x L_{R_s}^{\gamma=1}(x)$ on the two sides of the holonomy saddle as in \cite{ArabiArdehali:2024ysy} to be $k_{gR_s}=\frac{s-1}{20}$. For $(A_1,A_{2n})$ with $n>1$, \emph{i.e.}~the Gang-Kim-Stubbs $\mathcal{T}_{n>1}$ theories, we have similarly computed $k_{g_nR_s}=\frac{-1+4n(-2+s)-5s}{12+8n}.$\footnote{As a consistency check, we note that for $n=2$ and $s=\tfrac{1}{3}$ we get $k_{g_2R_{1/3}}=-\tfrac{4}{7},$ matching the result reported below Eq.~(4.2) in \cite{ArabiArdehali:2024ysy} (modulo the different sign convention, see footnote~\ref{fn:parity_conv}).}

\subsection{New family from $(A_1, A_{2n-1})$ with $n \geq 2$}

The $\mathcal{N}=2$ index for the $(A_1, A_{2n - 1})$ theory, after turning off all flavor fugacities for simplicity, is given by
\begin{align}
    \mathcal{I}^{(A_1, A_{2n-1})}_{t} &= \frac{\big((p;p)(q;q)\big)^{n-1}}{n!} \prod_{i=1}^{n-1} \frac{\Gamma_e \left( \left( \frac{pq}{t} \right)^{\frac{2n+1-i}{n+1}} \right)}{\Gamma_e \left( \left( \frac{pq}{t} \right)^{\frac{i+1}{n+1}} \right)} \Gamma_e \left( \left( \frac{pq}{t} \right)^{\frac{1}{n+1}} \right)^{n-1}\nonumber\\
    &\quad
    \int_{\mathfrak{h}_\text{cl}}\! \mathrm{d}^{n-1}x\  \prod_{i \neq j} \frac{\Gamma_e \left( \frac{z_i}{z_j} \left( \frac{pq}{t} \right)^{\frac{1}{n+1}} \right)}{\Gamma_e \left( \frac{z_i}{z_j} \right)} \prod_{i=1}^n \Gamma_e \left( (z_i)^\pm \left( \frac{pq}{t} \right)^{-\frac{n-1}{2n+2}} t^{\frac{1}{2}} \right), 
\end{align} 
where $z_1, \dots, z_n$ are the fugacities corresponding to the $SU(n)$ gauge group. In particular, this imposes the condition that $\prod_{i=1}^n z_i = 1$, or equivalently, $\sum_{i=1}^n x_i = 0$, which we will henceforth refer to as the $SU(n)$ condition. Following the same process as earlier, we get the piecewise quadratic and linear functions from $\mathcal{I}^{\gamma=1}_s$. Defining
\begin{equation}
    \alpha_i = \frac{2n + 1 - i}{n+1},\qquad
    \beta_i = \frac{i + 1}{n + 1}, 
\end{equation}
we have that
\begin{eqnarray}
    &\hspace{-3cm}12 Q^{\gamma=1}(\mathbf{x}) = (n-1) \kappa \left( \frac{1}{n+1} \right)+ \sum_{j=1}^{n-1} \Big( \kappa(\alpha_j) - \kappa(\beta_j) \Big) \\
    & \hspace{.5cm}+ \sum_{j=1}^n \Big( \kappa \left( x_j - \frac{n-1}{2n+2} \right) + \kappa \left( - x_j - \frac{n-1}{2n+2} \right) \Big) \nonumber\\
    & \hspace{3cm}+ \sum_{1 \leq i < j \leq n} \Big( \kappa \left( x_i - x_j + \frac{1}{n+1} \right) + \kappa \left( - x_i + x_j + \frac{1}{n+1} \right) \Big),\nonumber
\end{eqnarray}
and 
\begin{eqnarray}
    &L_{R_s}^{\gamma=1}(\mathbf{x}) = \frac{n-1}{2} \left( 1 - \frac{1-s}{n + 1} \right) \vartheta \left( \frac{1}{n+1} \right) + \sum_{j = 1}^{n-1} \bigg( \left( \frac{1 - (1-s)\alpha_j}{2} \right) \vartheta \left( \alpha_j \right) - \left( \frac{1 - (1-s) \beta_j}{2} \right) \vartheta (\beta_j)\bigg) \nonumber\\
    &+ \sum_{j = 1}^{n}\frac{n(1-s)}{2(n+1)} \Big( \vartheta \left( x_i - \frac{n-1}{2n+2} \right) + \vartheta \left( -x_i - \frac{n-1}{2n+2} \right) \Big)  \\
    & + \sum_{1 \leq i < j \leq n} \bigg( \frac{1}{2} \left( 1 - \frac{1-s}{n+1} \right) \Big( \vartheta \left( x_i - x_j + \frac{1}{n + 1} \right) + \vartheta \left( - x_i + x_j + \frac{1}{n + 1} \right) \Big) - \vartheta \left( x_i - x_j \right) \bigg), \nonumber
\end{eqnarray}
We propose that the saddle point for this family of theories is at
\begin{equation}
  \hspace{5cm}x_j^* = \frac{j}{n+1} - \frac{1}{2}\,,\hspace{3cm} j=1,\dots, n-1\,.
\end{equation} 
This corresponds to a 3d $\mathcal{N}=2$ $U(1)^{n-1}$  gauge theory.

The theory possesses $n-1$ gauge invariant monopoles corresponding to the following $n-1$ flat directions of $Q^{\gamma=1}(\boldsymbol{x})$:
\begin{align}
  (\delta x_1,\dots,\delta x_n)\,=\,(1, -1, 0, \dots, 0),\  (0, 1, -1,  \dots, 0),\  \dots,\  (0, \dots, 0, 1, -1)\,.
\end{align}
These can all be shown via evaluation on software using the strategies outlined in the previous case. Checking that $L^{\gamma=1}_{R_s}$ has slope $2$ along these directions, we get the monopole superpotential
\begin{equation}
    \mathcal{W}^{}_V=V^{}_{1, -1, 0, \dots, 0}+V^{}_{0, 1, -1,0, \dots, 0}+ \dots+ V^{}_{0, \dots, 0, 1}\qquad\qquad(\text{$n-1$ terms}).
\end{equation}
Note that the monopoles $V$ in the above equation have $n-1$ subscripts, as many as the rank of the $U(1)^{n-1}$ gauge group. We are of course allowed to use $n-1$ subscripts due to the redundancy of one of our $n$ coordinates coming from the $SU(n)$ condition. 

The matrix of Chern-Simons couplings can also be derived similarly to the previous case, i.e., as the Hessian of $Q^{\gamma=1}$ evaluated at $\boldsymbol{x}^*$. However, note that the $SU(n)$ condition must be imposed (for example by setting $x_n = - \sum_{i=1}^{n-1} x_i$), which just makes the expressions slightly more involved. The resulting CS couplings form an $(n-1) \times (n-1)$ matrix, reading
\begin{equation}
    k_{gg}=1\,,\qquad\text{for $n=2,$}
\end{equation}
\begin{align}
    K_{ij} &=
    \begin{pmatrix}
        2 & \frac{3}{2} & 1 & \dots & 1 & \frac{1}{2} \\
        \frac{3}{2} & 2 & \frac{3}{2} & \ddots & \vdots & \vdots \\
        1 & \frac{3}{2} & \ddots &\ddots & 1 & \vdots \\
        \vdots & \ddots & \ddots & \ddots & \frac{3}{2} & \frac{1}{2} \\
        1 & \dots & 1 & \frac{3}{2} & 2 & 1 \\
        \frac{1}{2} & \dots & \dots & \frac{1}{2} & 1 & 1 \\
    \end{pmatrix},\qquad\text{for $n\ge3.$}
\end{align} 
That is, the main diagonal elements are all 2 except for $K_{nn} = 1$, the upper and lower diagonals are all $\frac{3}{2}$ except for $K_{n-1,n}=K_{n,n-1}=1$, and the off-diagonals are all $1$ except for $K_{in} = K_{ni} = \frac{1}{2}$. 

Finally, we tabulate the massless field content of this family of theories below: 
\begin{table}[H]
  \centering
  \begin{tabular}{c|c|c}
      \toprule
      {\small $\#$ of chirals} & $R_s=R_0-s\,A$ & {\small Gauge charge(s)} \\
    \midrule
      2 & $\frac{1}{n+1} + \frac{n}{n+1} s$ & $
      (-1, \dots, -1),\ (-1,  0, \dots, 0)$ \\
      $n-1$ & $\frac{1-s}{n+1}$ & $\{ e^{(n-1)}_i - e_{i+1}^{(n-1)} \ |${\footnotesize$\, 1 \leq i \leq n-2$}$\}$, $\ \sum_{i=1}^{n-2} e_i^{(n-1)} + 2e_{n-1}^{(n-1)}$ \\
      \bottomrule
  \end{tabular}
  \caption{Massless field content of the 3d EFT descending from the $\gamma=1$ reduction of the $(A_1, A_{2n-1})$ Argyres-Douglas theory.}
  \label{tab:masslessA1A2n-1}
\end{table}

It can be checked that no piece of the 4d matter superpotential \cite{Maruyoshi:2016aim} survives the $\gamma=1$ reduction.

\subsubsection{$(A_1, A_3)$ in more detail}\label{sec:A1A3EFT}

As explained in Section~\ref{sec:generalities}, the $\mathcal{N}=2$ data of the 3d $\mathcal{N}=4$ SCFT, in particular the R-charge assignment, can be obtained by reducing the 4d Schur index, corresponding to $s=0$.

\begin{figure}[h]
\centering
\includegraphics[scale =0.5]{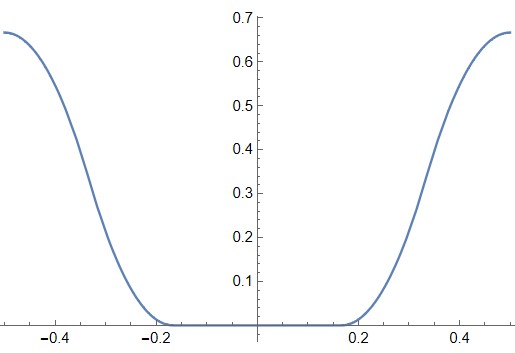}
\caption{The plot of $12Q^{\gamma=1}(x)$ versus $x$ for $(A_1,A_3)$. The interior region between $x=\pm1/6$ is flat and the value of $Q^{\gamma=1}$ there is zero.}
\label{fig:QhtA1A3}
\end{figure}

\begin{figure}[h]
\centering
\includegraphics[scale =0.55]{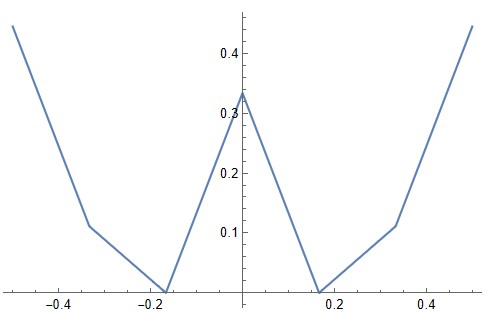}
\caption{The plot of $L^{\gamma=1}_{R_0}(x)$ versus $x$ for $(A_1,A_3)$. The minima at $x=\pm1/6$ are exactly zero. The interior slopes are $2.$}
\label{fig:LhtA1A3}
\end{figure}

The functions $Q^{\gamma=1}$ and $L_{R_0}^{\gamma=1}$ for this case are plotted in Figures~\ref{fig:QhtA1A3} and \ref{fig:LhtA1A3}. The holonomy saddle is at $x=-1/6$ (and its Weyl image). The flatness of $Q^{\gamma=1}$ to the right implies that the monopole $V_1$ is gauge invariant, and the slope of $L_{R_0}^{\gamma=1}$ in that direction implies it also has R-charge $2$. 

The EFT data is found easily using the methods of Section~\ref{sec:generalities}. We have a 3d $\mathcal{N}=2$ $U(1)_1$ gauge theory with two chiral multiplets of charge $-1$ and one of charge $2$. We denote these fields respectively by $\Phi_{-1}^i$ and $\Phi_2$, where the subscript indicates the gauge charge and the superscript $i=1,2$ is an $SU(2)_f$ flavor index. The SCFT R-charges are found by setting $s=0$ in Table~\ref{tab:masslessA1A2n-1} to be
\begin{equation}\label{eq:A1A3Rsc}
     R[\Phi_{-1}]=\frac{1}{3}\,,\qquad R[\Phi_2]=\frac{1}{3}\,.
\end{equation}
The superpotential consists only of the monopole term that we have already mentioned
\begin{equation}
    \mathcal{W}= V_{1}\label{eq:WV_A1A3}
\end{equation}
and this fixes the BF coupling between the gauge and the R-symmetry at level $k_{gR}=\tfrac{2}{3}$. 
One can indeed check that this monopole is gauge invariant given the above field content and CS level $k_{gg}=1$, and that it has R-charge 2 provided that $k_{gR}=\tfrac{2}{3}$
and that the R-charges of the fields are properly constrained. 
There is no superpotential term involving the matter fields.

On top of the R-symmetry and of the $SU(2)_f$ flavor symmetry, the theory also possesses an additional $U(1)_A$ global symmetry as indicated by the following counting. Each of the two fields $\Phi_{-1}$ and $\Phi_2$ can be acted upon by two independent abelian symmetries. Moreover, the $U(1)$ gauge group supplements an additional $U(1)_J$ topological symmetry. However, one combination of these three symmetries is broken by the monopole superpotential and another one can be reabsorbed by a gauge rotation. As we shall see, the residual $U(1)_A$ is identified with the axial symmetry of the IR $\mathcal{N}=4$ SCFT. A convenient way to parametrize $U(1)_A$ is such that $\Phi_{-1}$ is uncharged and $\Phi_2$ has charge $-1$. Moreover, the monopole $V_1$ can be made invariant under $U(1)_A$ and thus compatible with the superpotential by introducing a BF coupling between this symmetry and the gauge symmetry at level $k_{gA}=-1$. (We could instead derive $k_{gA}=-1$ from a 4d calculation as in \cite{ArabiArdehali:2024ysy}.) It then follows that $V_{m}$ has $U(1)_A$ charge $0$ for $m>0$, and $-2m$ for $m<0$.\footnote{This is equivalent to saying that the symmetry acting on the matter fields with such charges and the topological symmetry which acts on the monopole $V_m$ with charge $m$ are broken to their off-diagonal combination.}

Performing $F$-extremization \cite{Jafferis:2010un}, we have checked that the $U(1)_A$ symmetry does not mix with the trial $U(1)_R$ R-symmetry in \eqref{eq:A1A3Rsc} and so, as expected from our general discussion of Section \ref{sec:generalities}, $U(1)_R$ is the low energy superconformal R-symmetry. 
The way the matter fields and the bare monopoles of the IR SCFT transform under the gauge, global and superconformal R symmetries is summarized in the following table:

\begin{table}[H]
  \centering
  \begin{tabular}{c|cccc}
      \toprule
       & $U(1)_{g}$ & $SU(2)_f$ & $U(1)_A$ & $U(1)_R$ \\
    \midrule
      $\Phi_{-1}$ & $-1$ & $\bf 2$ & 0 & $\frac{1}{3}$ \\
      $\Phi_2$ & 2 & $\bf 1$ & $-1$ & $\frac{1}{3}$ \\ \hline
      $V_{m<0}$ & $2m$ & $\bf 1$ & $-2m$ & $-\frac{2}{3}m$ \\
      $V_{m>0}$ & 0 & $\bf 1$ & 0 & $2m$ \\
      \bottomrule
  \end{tabular}
  \caption{Transformation properties of the matter fields and of the bare monopoles under the gauge, global and R symmetries in the $(A_1,A_3)$ 3d EFT. Note that for convenience we have added $-\frac{2}{3}$ of the gauge charge to the $U(1)_A$ charges of Table~\ref{tab:masslessA1A2n-1}.} 
  \label{tab:A1A33dEFTcharges}
\end{table}

\subsubsection*{SUSY enhancement and moduli space}

We claim that this 3d $\mathcal{N}=2$ theory flows at low energies to the $\mathcal{N}=4$ SCFT that can be obtained by circle compactification of $(A_1,A_3)$ with a $U(1)_r$ twist. We will provide strong evidence by studying the SUSY enhancement and the moduli space using the superconformal index as a tool.

The index of the theory computed with the charge assignment in Table \ref{tab:A1A33dEFTcharges} as a power series in $q$ is to the first few orders
\begin{align}\label{eq:indA1A33dEFT}
    \mathcal{I}&=1+\chi^{SU(2)_f}_{[2]}(f)\frac{q^\frac{1}{2}}{y}+\left(\frac{1}{y^2}\chi^{SU(2)_f}_{[4]}(f)-1-\chi^{SU(2)_f}_{[2]}(f)\right)q\nonumber\\
    &+\left(y+\frac{1}{y}+\frac{1}{y^3}\chi^{SU(2)_f}_{[6]}(f)-\frac{1}{y}\chi^{SU(2)_f}_{[4]}(f)\right)q^{\frac{3}{2}}+\cdots\,,
\end{align}
where $\chi^{SU(2)_f}_{[2j]}(f)$ denotes the character of the representation with Dynkin label $[2j]$ of $SU(2)_f$ written in terms of the fugacity $f$, while $y$ is the fugacity for $U(1)_A$. We can see that this index is consistent with the claim that the theory exhibits SUSY enhancement to $\mathcal{N}=4$, where $SU(2)_f$ corresponds to the flavor symmetry of $(A_1,A_3)$ and $U(1)_A$ combines with the $\mathcal{N}=2$ R-symmetry $U(1)_R$ to give the full $SU(2)_H\times SU(2)_C$ $\mathcal{N}=4$ R-symmetry.

Assuming that the theory has extended supersymmetry, the $\mathcal{N}=4$ index can be obtained by setting $y=1$
\begin{align}
    \lim_{y\to1}\mathcal{I}&=1+\chi^{SU(2)_f}_{[2]}(f)q^\frac{1}{2}+\left(\chi^{SU(2)_f}_{[4]}(f){\color{red}-1}-\chi^{SU(2)_f}_{[2]}(f)\right)q\nonumber\\
    &+\left(2+\chi^{SU(2)_f}_{[6]}(f)-\chi^{SU(2)_f}_{[4]}(f)\right)q^{\frac{3}{2}}+\cdots\,.
\end{align}
In an $\mathcal{N}=4$ index, the terms at order $q^\frac{1}{2}$ correspond to the moment map operators in the adjoint representation of the (non-R) global symmetry. In accordance with this, in our index we find the character of the adjoint representation of $SU(2)_f$ at this order. Moreover, at order $q$ contribute with a positive sign the $\mathcal{N}=2$ exactly marginal operators, while with a minus sign we can find the conserved currents for the flavor symmetry and a component of the $\mathcal{N}=4$ stress-energy tensor multiplet (see e.g.~\cite{Beem:2012yn,Razamat:2016gzx,Evtikhiev:2017heo,Garozzo:2019ejm,Beratto:2020qyk,Comi:2023lfm}). Again we can then see that our index is consistent with it being an $\mathcal{N}=4$ index, where in particular we highlighted in red the contribution of the $\mathcal{N}=4$ stress-energy tensor.

Another consistency check of the SUSY enhancement can be performed considering again the $\mathcal{N}=2$ index \eqref{eq:indA1A33dEFT} and looking at the positive terms at order $q^\frac{3}{2}$. These can either correspond to conformal primary operators (CPOs) or to extra-SUSY currents \cite{Gang:2018huc,Gang:2023rei}. We will show momentarily that the terms $y^{-3}\chi^{SU(2)_f}_{[6]}(f)$ correspond to chiral operators of the Higgs branch. On the other hand, the terms $y+y^{-1}$ are two extra-SUSY currents that indicate an enhancement of supersymmetry from $\mathcal{N}=2$ to $\mathcal{N}=4$. These contributions to the index can be identified with the following gauge invariant bosonic operators in the $\mathcal{N}=2$ theory which have R-charge and spin 1:\footnote{The spin of a monopole operator $V_{\boldsymbol{m}}$ is $|\sum_i c_i(\boldsymbol{m})m_i|/2\,.$}
\begin{equation}
   y\,q^\frac{3}{2}:\,\, V_{-1}\Phi_2\,,\qquad \frac{q^\frac{3}{2}}{y}:\, \epsilon_{ij}\Phi_{-1}^i\left(\partial_\mu\Phi_{-1}^j\right)\Phi_2\,.
\end{equation}

We next study the moduli space of the $\mathcal{N}=4$ fixed point theory, focusing on its Higgs branch and Coulomb branch. Their Hilbert series can be obtained from suitable limits of the superconformal index in terms of the R-symmetry fugacities $q$ and $y$ \cite{Razamat:2014pta}. In order to obtain the Higgs branch Hilbert series we consider the limit
\begin{equation}
    x=q^\frac{1}{2}y\to0\,,\qquad t=\frac{q^\frac{1}{2}}{y}\,\,\,\,\text{fixed}\,,
\end{equation}
where $x$ and $t$ can be understood as fucacities for $SU(2)_H$ and $SU(2)_C$, respectively. Using an expansion in $t$ we can check that in such limit our index reduces to
\begin{equation}
    \lim_{x\to 0}\mathcal{I}=\sum_{j=0}^\infty\chi^{SU(2)_f}_{[2j]}(f)t^j=\frac{1-t^2}{(1-t)(1-f^2t)(1-f^{-2}t)}\,.
\end{equation}
As expected this coincides with the Hilbert series of $\mathbb{C}^2/\mathbb{Z}_2$, which is the Higgs branch of $(A_1,A_3)$ (see e.g.~\cite{Xie:2012hs,Closset:2020afy,Giacomelli:2020ryy,Xie:2021ewm}). In particular, the terms $\chi^{SU(2)_f}_{[6]}(f)t^3$ correspond to the aforementioned CPOs. In fact, we can easily identify the generators of the Higgs branch chiral ring in terms of the fields of the microscopic $\mathcal{N}=2$ theory
\begin{equation}
    x=(\Phi_{-1}^1)^2\Phi_{2}\,,\qquad y=(\Phi_{-1}^2)^2\Phi_{2}\,,\qquad z=\Phi_{-1}^1\Phi_{-1}^2\Phi_{2}\,,
\end{equation}
which trivially satisfy the defining relation of $\mathbb{C}^2/\mathbb{Z}_2$
\begin{equation}
    x\cdot y =z^2\,.
\end{equation}

The Hilbert series of the Coulomb branch is extracted from the index by taking the opposite limit
\begin{equation}
    t=\frac{q^\frac{1}{2}}{y}\to0\,,\qquad x=q^\frac{1}{2}y\,\,\,\,\text{fixed}\,.
\end{equation}
We find that the Coulomb branch of the theory is trivial
\begin{equation}
    \lim_{t\to 0}\mathcal{I}=1\,,
\end{equation}
which is consistent with the fact that the one-dimensional Coulomb branch of $(A_1,A_3)$ is completely lifted by the $U(1)_r$ twist.

\subsubsection*{Dual description with manifest $\mathcal{N}=3$ supersymmetry}\label{subsec:A1A3_dual_TSU2/SU2}

We propose the following IR duality:
\begin{align}
\big( (A_1, A_3) \textrm{ 3d EFT} \big) \otimes U(1)_2 \sim \frac{T[SU(2)]}{SU(2)^{(1)}_{k=3}}\;.
\end{align}
Here, the $T[SU(2)]$ theory represents the 3d $\mathcal{N}=4$ SQED with $N_f=2$ flavors \cite{Intriligator:1996ex,Gaiotto:2008ak}.  This theory has both Coulomb and Higgs branches, each of which is isomorphic to $\mathbb{C}^2/\mathbb{Z}_2$, on which the $SU(2)^{(1)}$ and $SU(2)^{(2)}$ flavor symmetries act respectively.  The theory on the right-hand side is obtained by gauging the $SU(2)^{(1)}$ symmetry with $\mathcal{N}=3$ Chern-Simons terms with level $k=3$. The $\mathcal{N}=3$ theory is expected to flow to an $\mathcal{N}=4$ SCFT with a SUSY enhancement, due to the nilpotent property of the moment maps of the $SU(2)^{(1)}$ \cite{Gaiotto:2008ak}. The enhancement can be explicitly confirmed using the superconformal index \cite{Evtikhiev:2017heo}. Gauging lifts all the Coulomb branch, leaving the theory with only the Higgs branch $\mathbb{C}^2/\mathbb{Z}_2$ and $SU(2)$ flavor symmetry, which matches that of the $(A_1, A_3)$ EFT. This duality can also be verified through computations of the superconformal index, in particular one can detect the presence of the topological sector by turning on background magnetic fluxes for the $SU(2)$ flavor symmetry \cite{Beratto:2021xmn}. The round 3-sphere partition functions $Z_{S^3}$ for the theories appearing in the duality can be computed using localization, yielding
\begin{align}
\begin{split}
&\big|Z_{S^3} [\textrm{$(A_1,A_3)$ EFT}]\big| = \frac{1}{3 \sqrt{3}}\;,\quad 
\\
&\bigg|Z_{S^3} \left[\frac{T[SU(2)]}{SU(2)^{(1)}_{k=3}}\right]\bigg| =\bigg{|} \int_{-\infty}^{\infty} \frac{dX}{2\pi} (2 \sinh^2 X) e^{\frac{3 X^2}{2\pi i }}  \frac{\sin (\frac{XY}\pi )}{2\sinh X \sinh Y}\big|^{}_{Y \rightarrow 0} \bigg{|}= \frac{1}{3 \sqrt{6}}\;.
\end{split}
\end{align}
These results further support the duality, alongside the fact that $Z_{S^3}[U(1)_2] = \frac{1}{\sqrt{2}}$. The theories on both sides possess an anomalous $\mathbb{Z}_2$ 1-form symmetry, which arises either from the  decoupled topological sector $U(1)_2=  SU(2)_1$ or from the dynamical $SU(2)_3$ gauge field.

\subsubsection{$(A_1,A_5)$ in more detail}

The saddle at $(x_1,x_2)=(-1/4,0)$ yields a 3d $\mathcal{N}=2$ $U(1)\times U(1)$ gauge theory with
\begin{equation}
    K_{ij}=\begin{pmatrix}
        2&1\\
        1&1
    \end{pmatrix},
\end{equation}
and the field content as in Table~\ref{tab:A1A53dEFTcharges}.
\begin{table}[t]
  \centering
  \begin{tabular}{c|cccc}
      \toprule
       & $U(1)_{g_1}\times U(1)_{g_2}$ & $U(1)_f$ & $U(1)_A$ & $U(1)_R$ \\
    \midrule
      $\Phi_{-1,0}$ & $(-1,0)$ & $\frac{2}{3}$ & 0 & $\frac{1}{4}$ \\
      $\Phi_{-1,-1}$ & $(-1,-1)$ & 0 & 0 & $\frac{1}{4}$ \\
      $\Phi_{1,-1}$ & $(1,-1)$ & $-\frac{1}{3}$& $-\frac{1}{2}$ & $\frac{1}{4}$ \\
      $\Phi_{1,2}$ & $(1,2)$ & $-\frac{1}{3}$ & $-\frac{1}{2}$ & $\frac{1}{4}$ \\
      \bottomrule
  \end{tabular}
  \caption{Transformation properties of the matter fields under the gauge, global and R symmetries in the $(A_1,A_5)$ 3d EFT. Note that for convenience we have added $-\frac{3}{4}g_1$ to the $U(1)_A$ charges of Table~\ref{tab:masslessA1A2n-1}.}
  \label{tab:A1A53dEFTcharges}
\end{table}

The global symmetries and the charge assignments are determined by the superpotential interaction
\begin{equation}
\mathcal{W}=V_{1,-1}+V_{0,1}\,.
\end{equation}
In particular, the monopoles involved in the superpotential have R-charge 2 and are uncharged under all other global symmetries provided that the following BF couplings are included:
\begin{align}
    &k_{g_1 f}=-\frac{2}{3}\,,\quad k_{g_1 A}=-\frac{3}{2}\,,\quad k_{g_1 R}=1\,,\nonumber\\
    &k_{g_2 f}=-\frac{1}{2}\,,\quad k_{g_2 A}=-\frac{3}{2}\,,\quad k_{g_2 R}=\frac{1}{2}\,,
\end{align}
where the labels $g_1$ and $g_2$ distinguish the two $U(1)$ gauge factors.
The R-symmetry we specified in Table \ref{tab:A1A53dEFTcharges} (corresponding to $s=0$ in Table \ref{tab:masslessA1A2n-1}) turns out to be the IR superconformal one, as can be checked via $F$-extremization.

\subsubsection*{SUSY enhancement and moduli space}

The index of the theory computed with the charge assignment in Table \ref{tab:A1A53dEFTcharges} is
\begin{align}\label{eq:indA1A53dEFT}
    \mathcal{I}&=1+\frac{q^\frac{1}{2}}{y}+\left(f+\frac{1}{f}\right)\frac{q^\frac{3}{4}}{y^\frac{3}{2}}+\left(\frac{1}{y^2}{\color{red}-1}-1\right)q\nonumber\\
    &+\left(f+\frac{1}{f}\right)\left(\frac{1}{y^\frac{5}{2}}-\frac{1}{y^\frac{1}{2}}\right)q^\frac{5}{4}+\left(y+\frac{1}{y^3}\left(1+f^2+\frac{1}{f^2}\right)\right)q^{\frac{3}{2}}+\cdots\,.
\end{align}

Once again this index provides strong evidence that the 3d $\mathcal{N}=2$ theory flows at low energies to the $\mathcal{N}=4$ SCFT that correspond to the $U(1)_r$-twisted compactication of $(A_1,A_5)$. In particular, the $U(1)_f$ symmetry is identified with the flavor symmetry of $(A_1,A_5)$, while the $U(1)_A$ symmetry becomes as usual part of the $\mathcal{N}=4$ R-symmetry. Indeed, at order $y^{-1}q^\frac{1}{2}$ we can find the contribution of the $U(1)_f$ moment map, while at order $q$ the negative contribution (in red) that is not coming from the flavor symmetry conserved current corresponds to the $\mathcal{N}=4$ stress-energy tensor.

Moreover, it is useful to focus on the order $q^\frac{3}{2}$. The contribution $y^{-3}(1+f^2+f^{-2})$ to this order, as we will see momentarily, corresponds to a chiaral operator of the Higgs branch of the IR $\mathcal{N}=4$ theory. Instead, we claim that the contribution $y$ comes from one of the two extra-SUSY current, while the second one $y^{-1}$ is not explicitly visible due to accidental cancellations. In fact, we have two of them which can be identified with the following gauge invariant bosonic operators in the $\mathcal{N}=2$ theory which have R-charge and spin 1:
\begin{equation}
   y\,q^\frac{3}{2}:\,\, V_{-1,0}\Phi_{1,-1}\Phi_{1,2}\,,\qquad \frac{q^\frac{3}{2}}{y}:\, \Phi_{-1,0}\left(\partial_\mu\Phi_{-1,-1}\right)\Phi_{1,-1}\Phi_{1,2}\,.
\end{equation}
Note that the contribution from the second operator seems to cancel against a fermionic operator in the index \eqref{eq:indA1A53dEFT}. Also, 
we could have applied the $\partial$ on any of the other fields; we have written a representative of the four possibilities, and leave clarification of the other options to future work.

We finally consider the Higgs and Coulomb branch limits of the index. In the Higgs limit we find
\begin{align}\label{eq:HBHSA1A53dEFT}
    \lim_{x\to 0}\mathcal{I}&=1+t+\left(f+\frac{1}{f}\right)t^\frac{3}{2}+t^2+\left(f+\frac{1}{f}\right)t^\frac{5}{2}+\left(1+f^2+\frac{1}{f^2}\right)t^3+\cdots\,.
\end{align}
In this expansion, we point out the terms $+\left(1+f^2+f^{-2}\right)t^3$ corresponding to the aforementioned CPOs. Moreover, we would like to compare this with the Hilbert series of the Higgs branch of $(A_1,A_5)$, which is $\mathbb{C}^2/\mathbb{Z}_3$ (see e.g.~\cite{Xie:2012hs,Closset:2020afy,Giacomelli:2020ryy,Xie:2021ewm})
\begin{equation}
    \mathrm{HS}[(A_1,A_5)]=\frac{1-t^3}{(1-t)(1-f\,t^\frac{3}{2})(1-f^{-1}t^\frac{3}{2})}\,.
\end{equation}
We verified in a power series in $t$ that the Higgs limit of our index \eqref{eq:HBHSA1A53dEFT} coincides with this Hilbert series.  In fact, we can easily identify the generators of the Higgs branch chiral ring in terms of the fields of the microscopic $\mathcal{N}=2$ theory
\begin{equation}
    x=\Phi_{-1,0}^3\Phi^2_{1,-1}\Phi_{1,2}\,,\quad y=\Phi_{-1,-1}^3\Phi_{1,-1}\Phi^2_{1,2}\,,\quad z=\Phi_{-1,0}\Phi_{-1,-1}\Phi_{1,-1}\Phi_{1,2}\,,
\end{equation}
which trivially satisfy the defining relation of $\mathbb{C}^2/\mathbb{Z}_3$
\begin{equation}
    x\cdot y =z^3\,.
\end{equation}

In the Coulomb limit we instead find
\begin{equation}
    \lim_{t\to 0}\mathcal{I}=1\,,
\end{equation}
which is again consistent with the fact that the two-dimensional Coulomb branch of $(A_1,A_5)$ is completely lifted by the $U(1)_r$ twist.

\subsection{New family from $(A_1, D_{2n+1})$ with $n \geq 1$}

The $\mathcal{N}=2$ index for the theory $(A_1, D_{2n+1})$, after turning off all flavor fugacities, is given by \cite{Agarwal:2016pjo}:
\begin{align}
  \mathcal{I}_{t}^{(A_1, D_{2n+1})} &= \frac{\big((p;p)(q;q)\big)^n}{2^n n!} \prod_{i=1}^{n} \frac{\Gamma_e \left( \left( \frac{pq}{t} \right)^{\frac{2(n+i)}{2n+1}} \right)}{\Gamma_e\left( \left( \frac{pq}{t} \right)^{\frac{2i}{2n+1}} \right)} \Gamma_e \left( \left( \frac{pq}{t} \right)^{\frac{1}{2n+1}} \right)^n\nonumber\\
  &\times \int_{\mathfrak{h}_\text{cl}}\! \mathrm{d}^{n}x\   \prod_{i=1}^{n} \Gamma_e \left( z_i^\pm \left( \frac{pq}{t} \right)^{\frac{n}{2n+1}} t^{\frac{1}{2}} \right)^3 \Gamma_e \left( z_i^\pm \left( \frac{pq}{t} \right)^{-\frac{n}{2n+1}} t^{\frac{1}{2}} \right) \\
  & \cdot \prod_{1 \leq i < j \leq n} \frac{\Gamma_e \left( \left( z_i z_j \right)^\pm (\frac{pq}{t})^{\frac{1}{2n+1}} \right) \Gamma_e \left( \left( \frac{z_i}{z_j} \right)^\pm \left( \frac{pq}{t} \right)^{\frac{1}{2n+1}} \right)}{\Gamma_e \left( \left( \frac{z_i}{z_j} \right)^\pm \right)} \prod_{i=1}^{n} \frac{\Gamma_e \left( z_i^{\pm 2} \left( \frac{pq}{t} \right)^{\frac{1}{2n+1}} \right)}{\Gamma_e(z_i^{\pm 2})},\nonumber
\end{align}  
where $z_1, \dots, z_n$ are the fugacities corresponding to the $Sp(n)$ gauge group. Following the same process as earlier, we get the piecewise quadratic and linear functions from $\mathcal{I}^{\gamma=1}_s$. Defining \begin{equation}
\alpha_i= \frac{2(n + i)}{2n+1},\qquad
    \beta_i = \frac{2i}{2n + 1},
    \end{equation}
    we have
\begin{eqnarray}
  & 12Q^{\gamma=1}(\boldsymbol{x}) = n \kappa \left( \frac{1}{2n+1} \right) + \sum_{j=1}^n \Big( \kappa(\alpha_j) - \kappa(\beta_j) + \kappa \left( 2x_j + \frac{1}{2n+1} \right) + \kappa \left( - 2x_j + \frac{1}{2n+1} \right) \Big) \nonumber\\
  & + \sum_{j=1}^n \Big( 3 \kappa \left( x_j + \frac{n}{2n+1} \right) + 3 \kappa \left( -x_j + \frac{n}{2n+1} \right) + \kappa \left( x_j - \frac{n}{2n+1} \right) + \kappa \left( -x_j - \frac{n}{2n+1} \right) \Big) \nonumber\\
  & \hspace{-2cm}+ \sum_{1 \leq i < j \leq n} \Big( \kappa \left( x_i + x_j + \frac{1}{2n+1} \right) + \kappa \left( x_i - x_j + \frac{1}{2n+1} \right) \\
  & \hspace{4.5cm} + \kappa \left( -x_i + x_j + \frac{1}{2n+1} \right) + \kappa \left( -x_i - x_j + \frac{1}{2n+1} \right) \Big),\nonumber
\end{eqnarray} 
and 
\begin{eqnarray}
  &L_{R_s}^{\gamma=1}(\boldsymbol{x}) = \frac{n}{2} \left( 1 - \frac{1-s}{2n+1} \right) \vartheta \left( \frac{1}{2n+1} \right) + \sum_{j=1}^{n} \left( \left( \frac{1 - (1-s) \alpha_j}{2} \right) \vartheta(\alpha_j) - \left( \frac{1 - (1-s) \beta_j}{2} \right) \vartheta(\beta_j) \right) \nonumber\\
  & + \frac{1-s}{4(2n+1)} \sum_{j=1}^n \Big( 3 \vartheta \left( x_j + \frac{n}{2n+1} \right) + 3 \vartheta \left( -x_j + \frac{n}{2n+1} \right) \Big)\nonumber \\
  & + \frac{(4n+1)(1-s)}{4(2n+1)} \sum_{j=1}^n \Big( \vartheta\left( x_j - \frac{n}{2n+1} \right) + \vartheta \left( -x_j - \frac{n}{2n+1} \right) \Big) \\
  & \hspace{-3cm}+ \frac{1}{2} \left( 1 - \frac{1-s}{2n+1} \right) \bigg( \sum_{j=1}^n \Big( \vartheta \left( 2x_j + \frac{1}{2n+1} \right) + \vartheta \left( -2x_j + \frac{1}{2n+1} \right) \Big) \nonumber\\
  & \hspace{1.5cm} + \sum_{1 \leq i < j \leq n} \Big( \vartheta \left( x_i + x_j + \frac{1}{2n+1} \right) + \vartheta \left( x_i - x_j + \frac{1}{2n+1} \right) \nonumber\\
  & \hspace{5cm} + \vartheta \left( -x_i + x_j + \frac{1}{2n+1} \right) + \vartheta \left( -x_i - x_j + \frac{1}{2n+1} \right) \Big) \bigg) \nonumber\\
  & - \sum_{j=1}^n \vartheta \left( 2x_j \right) - \sum_{1 \leq i < j \leq n} \Big( \vartheta \left( x_i + x_j \right) + \vartheta \left( x_i - x_j \right) \Big).\nonumber 
\end{eqnarray} 

We propose that the saddle point for this family of theories is 
\begin{equation}
  \hspace{5cm} x_j^* = \frac{j}{2n+1}\,,\hspace{3cm} j=1,\dots, n\,.
\end{equation} 
This corresponds to a 3d $\mathcal{N}=2$ $U(1)^{n}$  gauge theory.

There are $n$ gauge invariant monopoles corresponding to the flat directions:
\begin{equation}
  (\delta x_1,\dots,\delta x_n)\,=\,(1, -1, \dots, 0), \,\dots\,,\ (0, \dots, 1, -1),\ (0, \dots, 0, 1).
\end{equation}
of $Q^{\gamma=1}(\boldsymbol{x})$. Just as before, these can all be shown via evaluation on software. The slope of $L^{\gamma=1}_{R_s}$ being 2 along these directions implies the monopole superpotential
\begin{equation}
    \mathcal{W}^{}_V=V_{1, -1, 0, \dots, 0}+V_{0, 1, -1, 0, \dots, 0}+V_{0, \dots, 0, 1, -1}+V_{0, \dots, 0, 1}\qquad\qquad(\text{$n$ terms}).\label{eq:WV_A1Dodd}
\end{equation}

The Chern-Simons coupling is given by an $n \times n$ matrix 
\begin{equation}
    k_{gg}=1,\qquad\text{for $n=1,$}
\end{equation}
\begin{eqnarray}
  K_{ij} &= 
  \begin{pmatrix}
    \frac{3}{2} & \frac{1}{2} & 0 & \dots & 0  \\
    \frac{1}{2} & 1 & \ddots & \ddots & \vdots  \\
    0 & \ddots & \ddots & \ddots & 0  \\
    \vdots & \ddots & \ddots & 1 & \frac{1}{2}  \\
    0 & \dots & 0 & \frac{1}{2} & \frac{1}{2}  \\
  \end{pmatrix},\qquad\text{for $n\ge2.$}
\end{eqnarray} 

\noindent That is, the main diagonal elements are all 1 except for $K_{11} = \frac{3}{2}$ and $K_{nn} = \frac{1}{2}$, the upper and lower diagonals are all $\frac{1}{2}$, and all other elements vanish. 

Finally, we tabulate the massless field content of this family below, with the last column indicating a potentially useful change of gauge coordinates analogously to \cite{Gang:2023rei}: 
\begin{table}[H]
  \centering
  \makebox[\linewidth][c]{
  \begin{tabular}{c|c|c|c}
      \toprule
      {\small$\#$ of chirals} & $R_s=R_0-s\,A$ & {\small Gauge charge(s)} & {\footnotesize Gauge charge(s) (GKS type)} \\
    \midrule
      3 & $\frac{4n+1}{4n+2} + \frac{s}{4n+2}$ & $-e_n^{(n)}$ & $-e_n^{(n)}$ \\
      1 & $\frac{1}{4n+2} + \frac{4n+1}{4n+2} s$ & $e_n^{(n)}$ & $e_n^{(n)}$ \\
      $n$ & $\frac{1-s}{2n+1}$ & $\{ e^{(n)}_i - e_{i+1}^{(n)} \ |${\footnotesize$\, 1 \leq i \leq n-1$}$\},\ 2e_n^{(n)}$ & $e_1^{(n)},\dots,e_{n-1}^{(n)},\ 2e_n^{(n)}$ \\
      \bottomrule
  \end{tabular}}
  \caption{Massless field content of the 3d EFT descending from the $\gamma=1$ reduction of the $(A_1, D_{2n+1})$ Argyres-Douglas theory.}
  \label{tab:masslessA1D2n+1}
\end{table}

Denoting the $SU(2)_f$ triplet of chiral multiplets on the first row of Table~\ref{tab:masslessA1D2n+1} by $\Phi^{ab}_{0,\dots,0,-1}$, and the last chiral multiplet on the last row by $\Phi_{0,\dots,0,2},$ the 4d superpotential \cite{Agarwal:2016pjo} yields the 3d matter superpotential
\begin{equation}
    \mathcal{W}_{\Phi}=\epsilon_{ab}\,\epsilon_{cd}\,\Phi^{ac}_{0,\dots,0,-1}\Phi^{bd}_{0,\dots,0,-1}\Phi_{0,\dots,0,2}\,.\label{eq:W_Phi_D2n+1}
\end{equation}
The full superpotential of the 3d theory is $\mathcal{W}=\mathcal{W}_V+\mathcal{W}_\Phi\,.$

\subsubsection{$(A_1,D_3)$ in more detail}\label{subsec:A1D3}

In this case we find a 3d $\mathcal{N}=2$ $U(1)_1$ gauge theory with three chiral multiplets of charge $-1$, one of charge $2$, and another of charge $1$. For convenience, we collect the charge $-1$ fields in a $2\times 2$ traceless matrix which we denote by $\Phi_{-1}^{ab}$ for $a,b=1,2$. This is because the superpotential of the theory, which we will present momentarily, explicitly breaks the $SU(3)$ under which these three fields would form a triplet to its $SU(2)_f$ subgroup. The fields $\Phi_{-1}^{ab}$ then transform in the adjoint of $SU(2)_f$. We denote the remaining fields by $\Phi_2$ and $\Phi_{-1}$. We summarize how the matter fields and the monopoles of the theory transform under the gauge, global and R-symmetry in Table~\ref{tab:A1D33dEFTcharges}.
\begin{table}[t]
  \centering
  \begin{tabular}{c|cccc}
      \toprule
       & $U(1)_{g}$ & $SU(2)_f$ & $U(1)_A$ & $U(1)_R$ \\
    \midrule
      $\Phi_{-1}$ & $-1$ & $\bf 3$ & $0$ & $\frac{5}{6}$ \\
      $\Phi_1$ & 1 & $\bf 1$ & $-1$ & $\frac{1}{6}$ \\
      $\Phi_2$ & 2 & $\bf 1$ & $0$ & $\frac{1}{3}$ \\
      \hline
      $V_{m<0}$ & $2m$ & $\bf 1$ & $-m$ & $-\frac{2}{3}m$ \\
      $V_{m>0}$ & 0 & $\bf 1$ & $0$ & $2m$ \\
      \bottomrule
  \end{tabular}
  \caption{Transformation properties of the matter fields and of the bare monopoles under the gauge, global and R symmetries in the $(A_1,D_3)$ 3d EFT. Note that for convenience we have added $-\frac{1}{6}$ of the gauge charge to the $U(1)_A$ charges of Table~\ref{tab:masslessA1D2n+1}.}
  \label{tab:A1D33dEFTcharges}
\end{table}

The global symmetries and the charge assignment are determined by the superpotential interaction
\begin{equation}
\mathcal{W}^{}_V=V_1+\epsilon_{ab}\epsilon_{cd}\Phi_{-1}^{ac}\Phi_{-1}^{bd}\Phi_2\,.
\end{equation}
In particular, the monopoles involved in the superpotential have R-charge 2 and are uncharged under all other global symmetries provided that the following BF couplings are included:
\begin{equation}
    k_{gA}=-\frac{1}{2}\,,\qquad k_{gR}=\frac{2}{3}\,.
\end{equation}
The R-symmetry we specified in Table \ref{tab:A1D33dEFTcharges} (corresponding to $s=0$ in Table \ref{tab:masslessA1D2n+1}) turns out to be the IR superconformal one, as can be checked via $F$-extremization.

\subsubsection*{SUSY enhancement and moduli space}

The index of the theory computed with the charge assignment in Table \ref{tab:A1D33dEFTcharges} is
\begin{align}\label{eq:indA1D33dEFT}
    \mathcal{I}&=1+\chi^{SU(2)_f}_{[2]}(f)\frac{q^\frac{1}{2}}{y}+\left(\frac{1}{y^2}\chi^{SU(2)_f}_{[4]}(f)-1-\chi^{SU(2)_f}_{[2]}(f)\right)q\nonumber\\
    &+\left(y+\frac{1}{y}+\frac{1}{y^3}\chi^{SU(2)_f}_{[6]}(f)-\frac{1}{y}\chi^{SU(2)_f}_{[4]}(f)\right)q^{\frac{3}{2}}+\cdots\,.
\end{align}

Once again this index provides strong evidence that the 3d $\mathcal{N}=2$ theory flows at low energies to the $\mathcal{N}=4$ SCFT that correspond to the $U(1)_r$-twisted compactication of $(A_1,D_3)$. First of all, the symmetry $SU(2)_f$ corresponds to the flavor symmetry of $(A_1,D_3)$. Instead, the $U(1)_A$ symmetry becomes as usual part of the $\mathcal{N}=4$ R-symmetry.

Secondly, strong indications of the SUSY enhancement can be found by looking at the would be $\mathcal{N}=4$ index
\begin{align}
    \lim_{y\to 1}\mathcal{I}&=1+\chi^{SU(2)_f}_{[2]}(f)q^\frac{1}{2}+\left(\chi^{SU(2)_f}_{[4]}(f){\color{red}-1}-\chi^{SU(2)_f}_{[2]}(f)\right)q\nonumber\\
    &+\left(2+\chi^{SU(2)_f}_{[6]}(f)-\chi^{SU(2)_f}_{[4]}(f)\right)q^{\frac{3}{2}}+\cdots\,.
\end{align}
Indeed, at order $q^\frac{1}{2}$ we can find the contribution of the $SU(2)_f$ moment map, while at order $q$ the negative contribution (in red) that is not coming from the $SU(2)_f$ conserved current corresponds to the $\mathcal{N}=4$ stress-energy tensor.

Going back to the $\mathcal{N}=2$ index \eqref{eq:indA1D33dEFT}, we can again focus on the positive contributions at order $q^\frac{3}{2}$. We will see momentarily that the terms $y^{-3}\chi^{SU(2)_f}_{[6]}(f)$ correspond to chiral operators of the Higgs branch, while the terms $y+y^{-1}$ are the two extra-SUSY currents that indicate the SUSY enhancement from $\mathcal{N}=2$ to $\mathcal{N}=4$. These contributions to the index can be identified with the following gauge invariant bosonic operators in the $\mathcal{N}=2$ theory which have R-charge and spin 1:
\begin{equation}
   y\,q^\frac{3}{2}:\,\, V_{-1}\Phi_2\,,\qquad \frac{q^\frac{3}{2}}{y}:\,\epsilon_{ab}\Phi_1\partial_\mu\Phi_{-1}^{ab} \,.
\end{equation}
Note that the other option $\epsilon_{ab}(\partial_\mu\Phi_1)\Phi_{-1}^{ab}$ for the second operator differs from the one written above by a superconformal descendant $\epsilon_{ab}\partial_\mu(\Phi_1\Phi_{-1}^{ab})$.

We finally consider the Higgs and Coulomb branch limits of the index. In the Higgs limit we find the Hilbert series of $\mathbb{C}^2/\mathbb{Z}_2$
\begin{equation}
    \lim_{x\to 0}\mathcal{I}=\sum_{j=0}^\infty\chi^{SU(2)_f}_{[2j]}(f)t^j=\frac{1-t^2}{(1-t)(1-f^2t)(1-f^{-2}t)}\,.
\end{equation}
which is indeed the Higgs branch of $(A_1,D_3)$ (see e.g.~\cite{Carta:2021whq,Xie:2021ewm}). In particular, the terms $\chi^{SU(2)_f}_{[6]}(f)t^3$ correspond to the aforementioned CPOs.  In fact, we can easily identify the generators of the Higgs branch chiral ring in terms of the fields of the microscopic $\mathcal{N}=2$ theory
\begin{equation}
x=\Phi_1\Phi_{-1}^{12}\,,\qquad y=\Phi_1\Phi_{-1}^{21}\,,\qquad z=\Phi_1\Phi_{-1}^{11}\,,
\end{equation}
which satisfy the relation
\begin{equation}
    x\cdot y =z^2\,.
\end{equation}
as can be checked by using the F-term equation of $\Phi_2$. In the Coulomb limit we instead find
\begin{equation}
    \lim_{t\to 0}\mathcal{I}=1\,,
\end{equation}
which is again consistent with the fact that the one-dimensional Coulomb branch of $(A_1,D_3)$ is completely lifted by the $U(1)_r$ twist.

\subsubsection{$(A_1,D_5)$ in more detail}\label{subsec:A1D5}

In this case we find a 3d $\mathcal{N}=2$ $U(1)\times U(1)$ gauge theory with CS matrix
\begin{equation}
    K_{ij}=\begin{pmatrix}
        \frac{3}{2}&\frac{1}{2}\\
        \frac{1}{2}&\frac{1}{2}
    \end{pmatrix}\,,
\end{equation}
and the field content as in Table~\ref{tab:A1D53dEFTcharges}.
\begin{table}[t]
  \centering
  \begin{tabular}{c|cccc}
      \toprule
       & $U(1)_{g_1}\times U(1)_{g_2}$ & $SU(2)_f$ & $U(1)_A$ & $U(1)_R$ \\
    \midrule
      $\Phi_{0,-1}$ & $(0,-1)$ & $\bf 3$ & $-1$ & $\frac{9}{10}$ \\
      $\Phi_{0,1}$ & $(0,1)$ & $\bf 1$ & $0$ & $\frac{1}{10}$ \\
      $\Phi_{1,-1}$ & $(1,-1)$ & $\bf 1$ & 0 & $\frac{1}{5}$ \\
      $\Phi_{0,2}$ & $(0,2)$ & $\bf 1$ & $2$ & $\frac{1}{5}$ \\
      \bottomrule
  \end{tabular}
  \caption{Transformation properties of the matter fields under the gauge, global and R symmetries in the $(A_1,D_5)$ 3d EFT. Note that for convenience we have added $\frac{7}{10}g_1+\frac{9}{10}g_2$ to the $U(1)_A$ charges of Table~\ref{tab:masslessA1D2n+1}.}
  \label{tab:A1D53dEFTcharges}
\end{table}

The global symmetries and the charge assignments are determined by the superpotential interactions
\begin{equation}
\mathcal{W}=V_{1,-1}+V_{0,1}+\epsilon_{ab}\epsilon_{cd}\Phi_{0,-1}^{ac}\Phi_{0,-1}^{bd}\Phi_{0,2}\,.
\end{equation}
Notice in particular that, similarly to the case of $(A_1,D_3)$, the superpotential for the matter fields breaks the $SU(3)$ symmetry of the three fields with gauge charges $(0,-1)$ to its $SU(2)_f$ subgroup, under which they transform in the adjoint representation. Moreover, the monopoles involved in the superpotential have R-charge 2 and are uncharged under all other global symmetries provided that the following BF couplings are included:
\begin{equation}
    k_{g_1 A}=1\,,\quad k_{g_1 R}=0\,,\qquad k_{g_2 A}=\frac{1}{2}\,,\quad k_{g_2 R}=\frac{1}{5}\,,
\end{equation}
where the labels $g_1$ and $g_2$ distinguish the two $U(1)$ gauge factors.
The R-symmetry we specified in Table \ref{tab:A1D53dEFTcharges} (corresponding to $s=0$ in Table \ref{tab:masslessA1D2n+1}) turns out to be the IR superconformal one, as can be checked via $F$-extremization.

\subsubsection*{SUSY enhancement and moduli space}

The index of the theory computed with the charge assignment in Table \ref{tab:A1D53dEFTcharges} is
\begin{align}\label{eq:indA1D53dEFT}
    \mathcal{I}&=1+\chi^{SU(2)_f}_{[2]}(f)\frac{q^\frac{1}{2}}{y}+\left(\frac{1}{y^2}\chi^{SU(2)_f}_{[4]}(f)-1-\chi^{SU(2)_f}_{[2]}(f)\right)q\nonumber\\
    &+\left(y+\frac{1}{y}+\frac{1}{y^3}\chi^{SU(2)_f}_{[6]}(f)-\frac{1}{y}\left(\chi^{SU(2)_f}_{[4]}(f)+\chi^{SU(2)_f}_{[2]}(f)\right)\right)q^{\frac{3}{2}}+\cdots\,.
\end{align}

Once again this index provides strong evidence that the 3d $\mathcal{N}=2$ theory flows at low energies to the $\mathcal{N}=4$ SCFT that correspond to the $U(1)_r$-twisted compactication of $(A_1,D_5)$. First of all, the symmetry $SU(2)_f$ corresponds to the flavor symmetry of $(A_1,D_3)$. Instead, the $U(1)_A$ symmetry becomes as usual part of the $\mathcal{N}=4$ R-symmetry.

Secondly, strong indications of the SUSY enhancement can be found by looking at the would be $\mathcal{N}=4$ index
\begin{align}
    \lim_{y\to 1}\mathcal{I}&=1+\chi^{SU(2)_f}_{[2]}(f)q^\frac{1}{2}+\left(\chi^{SU(2)_f}_{[4]}(f){\color{red}-1}-\chi^{SU(2)_f}_{[2]}(f)\right)q\nonumber\\
    &+\left(2+\chi^{SU(2)_f}_{[6]}(f)-\chi^{SU(2)_f}_{[4]}(f)-\chi^{SU(2)_f}_{[2]}(f)\right)q^{\frac{3}{2}}+\cdots\,.
\end{align}
Indeed, at order $q^\frac{1}{2}$ we can find the contribution of the $SU(2)_f$ moment map, while at order $q$ the negative contribution (in red) that is not coming from the $SU(2)_f$ conserved current corresponds to the $\mathcal{N}=4$ stress-energy tensor.

Going back to the $\mathcal{N}=2$ index \eqref{eq:indA1D53dEFT}, we can again focus on the positive contributions at order $q^\frac{3}{2}$. We will see momentarily that the terms $y^{-3}\chi^{SU(2)_f}_{[6]}(f)$ correspond to chiral operators of the Higgs branch, while the terms $y+y^{-1}$ are the two extra-SUSY currents that indicate the SUSY enhancement from $\mathcal{N}=2$ to $\mathcal{N}=4$. These contributions to the index can be identified with the following gauge invariant bosonic operators in the $\mathcal{N}=2$ theory which have R-charge and spin 1:
\begin{equation}
y\,q^\frac{3}{2}:\,\, V_{-1,0}\left(\Phi_{1,-1}\right)^2\Phi_{0,2}\,,\qquad \frac{q^\frac{3}{2}}{y}:\,\epsilon_{ab}\Phi_{0,1}\partial_\mu\Phi_{0,-1}^{ab} \,.
\end{equation}
Note that the other option $\epsilon_{ab}(\partial_\mu\Phi_{0,1})\Phi_{0,-1}^{ab}$ for the second operator, differs from the one we wrote in the equation by a superconformal descendant $\epsilon_{ab}\partial_\mu(\Phi_{0,1}\Phi_{0,-1}^{ab})$.

We finally consider the Higgs and Coulomb branch limits of the index. These work very similarly to $(A_1,D_3)$. In the Higgs limit we find the Hilbert series of $\mathbb{C}^2/\mathbb{Z}_2$
\begin{equation}
    \lim_{x\to 0}\mathcal{I}=\sum_{j=0}^\infty\chi^{SU(2)_f}_{[2j]}(f)t^j=\frac{1-t^2}{(1-t)(1-f^2t)(1-f^{-2}t)}\,.\label{eq:A1D5_HB_HS}
\end{equation}
which is indeed the Higgs branch of $(A_1,D_5)$ (see e.g.~\cite{Carta:2021whq,Xie:2021ewm}) and whose chiral ring contains the aforementioned CPOs contributing as $\chi^{SU(2)_f}_{[6]}(f)t^3$.  The generators of the Higgs branch chiral ring in terms of the fields of the microscopic $\mathcal{N}=2$ theory read\footnote{As for $(A_1,D_3)$ we collect the triplet of fields with gauge charges $(0,-1)$ in a $2\times 2$ traceless matrix $\Phi_{0,-1}^{ab}$, with $a,b=1,2$ being $SU(2)_f$ indices.}
\begin{equation}
x=\Phi_{0,1}\Phi_{0,-1}^{12}\,,\qquad y=\Phi_{0,1}\Phi_{0,-1}^{21}\,,\qquad z=\Phi_{0,1}\Phi_{0,-1}^{11}\,,
\end{equation}
which satisfy the relation
\begin{equation}
    x\cdot y =z^2\,.
\end{equation}
as can be checked by using the F-term equation of $\Phi_{0,2}$. In the Coulomb limit we instead find
\begin{equation}
    \lim_{t\to 0}\mathcal{I}=1\,,\label{eq:A1D5_CB_HS}
\end{equation}
which is again consistent with the fact that the two-dimensional Coulomb branch of $(A_1,D_5)$ is completely lifted by the $U(1)_r$ twist.

\subsection{New family from $(A_1, D_{2n})$ with $n \geq 2$}

The $\mathcal{N}=2$ index for the $(A_1, D_{2n})$ theory, after turning off all flavor fugacities, is given by \cite{Agarwal:2016pjo}:
\begin{align}
    \mathcal{I}_{t}^{(A_1, D_{2n})} &= \frac{\big((p;p)(q;q)\big)^{n-1}}{n!} \prod_{j=1}^{n-1} \frac{\Gamma_e \left( \left( \frac{pq}{t} \right)^{\frac{j+n}{n}} \right)}{\Gamma_e \left( \left( \frac{pq}{t} \right)^{\frac{j+1}{n}} \right)} \Gamma_e \left( \left( \frac{pq}{t} \right)^{\frac{1}{n}} \right)^{n-1} \\
    &\hspace{-.7cm} \times \int_{\mathfrak{h}_\text{cl}}\! \mathrm{d}^{n-1}x\  \prod_{i\neq j} \frac{\Gamma_e \left( \frac{z_i}{z_j} \left( \frac{pq}{t} \right)^{\frac{1}{n}} \right)}{\Gamma_e \left( \frac{z_i}{z_j} \right)} \prod_{i=1}^n \Gamma_e \left( (z_i)^\pm (\frac{pq}{t})^{\frac{n-1}{2n}} t^{\frac{1}{2}}  \right) \Gamma_e \left( (z_i)^\pm (\frac{pq}{t})^{\frac{1-n}{2n}} t^{\frac{1}{2}} \right)\nonumber,
\end{align} 
where $z_1, \dots, z_n$ are the fugacities corresponding to the $SU(n)$ gauge group (recall that they respect the $SU(n)$ condition $\sum_{i=1}^n x_i = 0$). Defining 
\begin{equation}
    \alpha_j = \frac{j + n}{n},\qquad 
    \beta_j = \frac{j + 1}{n},
\end{equation}
the associated piecewise quadratic and linear functions are given by
\begin{eqnarray}
    &\hspace{-5cm}12 Q^{\gamma=1}(\mathbf{x}) = (n-1) \kappa \left( \frac{1}{n} \right) + \sum_{j = 1}^{n-1} \big( \kappa \left( \alpha_j \right) - \kappa \left( \beta_j \right) \big)\nonumber \\
     &+ \sum_{j = 1}^n \Big( \kappa \left( x_j + \frac{n-1}{2n} \right) + \kappa \left( - x_j + \frac{n-1}{2n} \right) + \kappa \left( x_j - \frac{n-1}{2n} \right) + \kappa \left( -x_j - \frac{n-1}{2n} \right)\Big)\nonumber \\
    &  \hspace{-2cm}+ \sum_{1 \leq i < j \leq n} \Big( \kappa \left( x_i - x_j + \frac{1}{n} \right) + \kappa \left( - x_i + x_j + \frac{1}{n} \right) \Big),
    \end{eqnarray}
and 
\begin{eqnarray}
    &L^{\gamma=1}_{R_s}(\mathbf{x}) = \frac{n-1}{2} \left( 1 - \frac{1-s}{n} \right) \vartheta \left( \frac{1}{n} \right) + \sum_{j = 1}^{n-1} \left( \left( \frac{1 - (1-s) \alpha_j}{2} \right) \vartheta \left( \alpha_j \right) - \left( \frac{1 - (1-s)\beta_j}{2} \right) \vartheta (\beta_j)  \right) \nonumber\\
    & \hspace{-2cm}+ \sum_{j = 1}^n \bigg( \frac{1-s}{4n} \Big( \vartheta \left( x_j + \frac{n-1}{2n} \right) + \vartheta \left( - x_j + \frac{n-1}{2n} \right) \Big) \nonumber\\
    & \qquad \qquad + \frac{(2n-1)(1-s)}{4n} \Big( \vartheta \left( x_j - \frac{n-1}{2n} \right) + \vartheta \left( -x_j - \frac{n-1}{2n} \right) \Big) \bigg) \\
    & + \frac{1}{2} \left( 1 - \frac{1-s}{n} \right) \sum_{1 \leq i < j \leq n} \Big( \vartheta \left( x_i - x_j + \frac{1}{n} \right) + \vartheta \left( - x_i + x_j + \frac{1}{n} \right)\! \Big) - \sum_{1 \leq i < j \leq n} \vartheta(x_i - x_j).\nonumber
\end{eqnarray}

A saddle point for this family of theories is given by
\begin{equation}
  x_j^* = \frac{j}{n} - \frac{n+1}{2n}\,.
\end{equation}
This corresponds to a 3d $\mathcal{N}=2$ $U(1)^{n-1}$  gauge theory.

The theory also possesses $n$ gauge invariant monopoles corresponding to the following flat directions of $Q^{\gamma=1}(\boldsymbol{x})$:
\begin{align}
  (\delta x_1,\dots,\delta x_n)\,=\,(1, -1, \dots, 0),\ (0, 1, -1, \dots, 0), \dots, (0, \dots, 1, -1),\  (-1, 0, \dots, 0, 1).
\end{align}
As usual, these can all be shown via evaluation on software. The slope of $L^{\gamma=1}_{R_s}$ being 2 along these directions implies the monopole superpotential
\begin{equation}
    \mathcal{W}^{}_V=V_{1, -1, \dots, 0}+V_{0, 1, -1, \dots, 0}+V_{0, \dots, 0, 1, -1}+V_{0, \dots, 0, 1}+V_{-1,0, \dots, 0}\qquad\quad(\text{$n$ terms}).\label{eq:WV_A1D2n}
\end{equation}
Note that the monopoles $V$ in the above equation have $n-1$ subscripts, as many as the rank of the $U(1)^{n-1}$ gauge group.

When computing the CS couplings, similar care as in the case of $(A_1, A_{2n-1})$ has to be taken due to the $SU(n)$ constraint. Here we have an $(n-1) \times (n-1)$ matrix
\begin{equation}
    k_{gg}=0,\qquad\text{for $n=2,$}
\end{equation}
\begin{eqnarray}
    K_{ij} &=
    \begin{pmatrix}
        1 & \frac{1}{2} & 0 & \dots & 0 \\
        \frac{1}{2} & \ddots & \ddots & \ddots & \vdots \\
        0 & \ddots & \ddots & \ddots & 0 \\
        \vdots & \ddots & \ddots & \ddots & \frac{1}{2} \\
        0 & \dots & 0 & \frac{1}{2} & 1 \\
    \end{pmatrix},\qquad\text{for $n\ge3$.}
\end{eqnarray} 
That is, the main diagonal is all 1's, and the lower and upper diagonals are all $\frac{1}{2}$. 

The massless field content of the theory is tabulated in Table~\ref{tab:masslessA1D2n}.
\begin{table}[t]
  \centering
  \makebox[\linewidth][c]{
  \begin{tabular}{c|c|c}
      \toprule
      {\small$\#$ of chirals} & $R_s=R_0-s\,A$ &{\small Gauge charge(s)} \\
    \midrule
      2 & $1 - \frac{1 - s}{2n}$ & $ (1, 0, \dots, 0),\, (1, \dots, 1) $ \\
      2 & $s + \frac{1 - s}{2n}$ & $ (-1, 0, \dots, 0),\, (-1, \dots, -1) $ \\
      $n$ & $\frac{1-s}{n}$ & $\{ e^{(n-1)}_i - e_{i+1}^{(n-1)} \ |${\footnotesize$\, 1 \leq i \leq n-2$}$\}$,$\,(1, \dots, 1, 2)$,$\,(-2, -1, \dots, -1)$\\
      $1$ & $1+s$ & $(0, \dots, 0)$ \\
      \bottomrule
  \end{tabular}}
  \caption{Massless field content of the 3d EFT arising from the $\gamma=1$ reduction of the $(A_1, D_{2n})$ Argyres-Douglas theory. }
  \label{tab:masslessA1D2n}
\end{table}
Denoting the two chiral multiplets on the first row of Table~\ref{tab:masslessA1D2n} by $\Phi_{1,0,\dots,0},\,\Phi_{1,\dots,1}$, the $n$ chiral multiplets on the next-to-last row by $\Phi_{e_i-e_{i+1}},\,\Phi_{1,\dots,1,2},\,\Phi_{-2,-1,\dots,-1},$ and the chiral on the last row by $\beta,$ the 4d superpotential \cite{Agarwal:2016pjo,Benvenuti:2017bpg} yields the 3d matter superpotential
\begin{equation}
    \mathcal{W}_{\Phi}=\Phi_{1,0,\dots,0}\,\Phi_{1,\dots,1}\,\Phi_{-2,-1,\dots,-1}+\beta\,\Big(\prod_{i=1}^{n-2}\Phi_{e_i-e_{i+1}}\Big)\,\Phi_{1,\dots,1,2}\,\Phi_{-2,-1,\dots,-1}\,.\label{eq:W_Phi_D2n}
\end{equation}
The full superpotential of the 3d theory is $\mathcal{W}=\mathcal{W}_V+\mathcal{W}_\Phi\,.$\\

\textbf{Remark}. Since we have one more monopole superpotential terms (namely $n$) than the rank  (namely $n-1$), the superpotential \eqref{eq:WV_A1D2n} \emph{over-determines} the vector $k_{jR}$ of mixed gauge-$R$ CS couplings in this case.

\subsubsection{$(A_1,D_4)$ in more detail}

In this case we find a 3d $\mathcal{N}=2$ $U(1)$ gauge theory with the following content:
\begin{table}[H]
  \centering
  \begin{tabular}{c|ccccc}
      \toprule
       & $U(1)_{g}$ & $SU(2)_f$ & $U(1)_v$ & $U(1)_A$ & $U(1)_R$ \\
    \midrule
      $\Phi_1$ & $1$ & $\bf 1$ & 1 & 0 & $\frac{3}{4}$ \\
      $\tilde{\Phi}_1$ & $1$ & $\bf 1$ & $-1$ & 0 & $\frac{3}{4}$ \\
      $\Phi_{-1}$ & $-1$ & $\bf 2$ & 0 & $-1$ & $\frac{1}{4}$ \\
      $\Phi_{2}$ & $2$ & $\bf 1$ & 0 & $1$ & $\frac{1}{2}$ \\
      $\Phi_{-2}$ & $-2$ & $\bf 1$ & 0 & 0 & $\frac{1}{2}$ \\
      $\beta$ & 0 & $\bf 1$ & 0 & $-1$ & $1$ \\ \hline
      $V_{m<0}$ & 0 & $\bf 1$ & 0 & $0$ & $-2m$ \\
      $V_{m>0}$ & 0 & $\bf 1$ & 0 & $0$ & $2m$ \\
      \bottomrule
  \end{tabular}
  \caption{Transformation properties of the matter fields and of the bare monopoles under the gauge, global and R symmetries in the $(A_1,D_4)$ 3d EFT. Note that for convenience we have added $\frac{1}{4}$ of the gauge charge to the $U(1)_A$ charges of Table~\ref{tab:masslessA1D2n}.}
  \label{tab:A1D43dEFTcharges}
\end{table}
The global symmetries and the charge assignments are determined by the superpotential interactions
\begin{equation}
\mathcal{W}=V_{+1}+V_{-1}+\Phi_{1}\tilde{\Phi}_{1}\Phi_{-2}+\beta\Phi_2\Phi_{-2}\,.
\end{equation}

Notice in this case there is no pure CS interaction nor BF couplings of the abelian global symmetries with the gauge symmetry
\begin{equation}
    k_{gg}=k_{gv}=k_{gA}=k_{gR}=0\,.
\end{equation}
The R-symmetry we specified in Table \ref{tab:A1D43dEFTcharges} (corresponding to $s=0$ in Table \ref{tab:masslessA1D2n}) turns out to be the IR superconformal one, as can be checked via $F$-extremization.

\subsubsection*{SUSY enhancement and moduli space}

The index of the theory computed with the charge assignment in Table \ref{tab:A1D43dEFTcharges} is
\begin{align}\label{eq:indA1D43dEFT}
    \mathcal{I}&=1+\chi^{\mathfrak{su}(3)}_{[1,1]}(f,v)\frac{q^\frac{1}{2}}{y}+\left(\frac{1}{y^2}\chi^{SU(3)}_{[2,2]}(f,v)-1-\chi^{SU(3)}_{[1,1]}(f,v)\right)q\nonumber\\
    &+\left(\frac{1}{y}+\frac{1}{y^3}\chi^{SU(3)}_{[3,3]}(f,v)-\frac{1}{y}\left(\chi^{SU(3)}_{[2,2]}(f,v)+\chi^{SU(3)}_{[3,0]}(f,v)+\chi^{SU(3)}_{[0,3]}(f,v)\right)\right)q^{\frac{3}{2}}+\cdots\,,
\end{align}
where we wrote the index in terms of characters for the enhanced $SU(3)$ symmetry with the following embedding of the manifest $SU(2)_f\times U(1)_v$ symmetries of the UV $\mathcal{N}=2$ theory:
\begin{equation}
    \chi^{SU(3)}_{[1,1]}(f,v)=\chi^{SU(2)_f}_{[2]}(f)+\left(v+\frac{1}{v}\right)\chi^{SU(2)_f}_{[1]}(f)+1\,.
\end{equation}

Once again this index provides strong evidence that the 3d $\mathcal{N}=2$ theory flows at low energies to the $\mathcal{N}=4$ SCFT that correspond to the $U(1)_r$-twisted compactication of $(A_1,D_4)$. First of all, we can see the enhancement of the $SU(2)_f\times U(1)_v$ symmetry to the $SU(3)$ flavor symmetry of $(A_1,D_4)$, since the index rearranges into $SU(3)$ characters and at order $q$ with a negative sign we can see the contribution of the $SU(3)$ conserved current. Instead, the $U(1)_A$ symmetry becomes as usual part of the $\mathcal{N}=4$ R-symmetry.

Strong indications of the SUSY enhancement can be found by looking at the would be $\mathcal{N}=4$ index
\begin{align}
    \lim_{y\to 1}\mathcal{I}&=1+\chi^{\mathfrak{su}(3)}_{[1,1]}(f,v)q^\frac{1}{2}+\left(\chi^{SU(3)}_{[2,2]}(f,v){\color{red}-1}-\chi^{SU(3)}_{[1,1]}(f,v)\right)q\nonumber\\
    &+\left(\chi^{SU(3)}_{[3,3]}(f,v)-\left(\chi^{SU(3)}_{[2,2]}(f,v)+\chi^{SU(3)}_{[3,0]}(f,v)+\chi^{SU(3)}_{[0,3]}(f,v)\right)\right)q^{\frac{3}{2}}+\cdots\,.
\end{align}
Indeed, at order $q^\frac{1}{2}$ we can find the contribution of the $SU(3)$ moment map, while at order $q$ the negative contribution (in red) that is not coming from the $SU(3)$ conserved current corresponds to the $\mathcal{N}=4$ stress-energy tensor.

Going back to the $\mathcal{N}=2$ index \eqref{eq:indA1D43dEFT}, we can again focus on the positive contributions at order $q^\frac{3}{2}$. We will see momentarily that the terms $y^{-3}\chi^{SU(3)}_{[3,3]}(f,v)$ correspond to chiral operators of the Higgs branch, while the term $y^{-1}$ is one of the two extra-SUSY currents. We do not see explicitly the contribution $y$ of the second extra-SUSY current  due to an accidental cancellation in the index. In fact, we can explicitly construct the two extra-SUSY currents as gauge invariant operators in the $\mathcal{N}=2$ theory of R-charge and spin 1 as follows:
\begin{equation}
    y\,q^\frac{3}{2}:\,\left(\partial_\mu\Phi_2\right)\Phi_{-2}\sim\Phi_2\partial_\mu\Phi_{-2}\,,\qquad \frac{q^\frac{3}{2}}{y}:\,\epsilon_{ij}\Phi_{-1}^i\left(\partial_\mu\Phi_{-1}^j\right)\Phi_2\,,
\end{equation}
where the identification is due to the F-term equation of $\beta$ which reads $\Phi_2\Phi_{-2}\sim0$. The cancellation of the contribution of the first operator in the index is due to the presence of a fermionic operator which provides a similar index contribution but with opposite sign. We leave its determination to future work.

We finally consider the Higgs and Coulomb branch limits of the index. In the Higgs limit we find
\begin{equation}
    \lim_{x\to 0}\mathcal{I}=\sum_{j=0}^\infty\chi^{\mathfrak{su}(3)}_{[j,j]}(f,v)t^j\,,
\end{equation}
which, as expected, coincides with the Hilbert series of $\overline{\text{min}}_{SU(3)}$ \cite{Benvenuti:2010pq}, that is the Higgs branch of $(A_1,D_4)$ (see e.g.~\cite{Xie:2012hs,Buican:2015hsa,Xie:2016uqq,Beratto:2020wmn,Carta:2021whq,Xie:2021ewm}). In particular, the terms $\chi^{SU(3)}_{[3,3]}(f,v)t^3$ correspond to the aforementioned CPOs. In the Coulomb limit we instead find
\begin{equation}
    \lim_{t\to 0}\mathcal{I}=1\,,
\end{equation}
which is again consistent with the fact that the one-dimensional Coulomb branch of $(A_1,D_4)$ is completely lifted by the $U(1)_r$ twist.

\subsubsection{$(A_1,D_6)$ in more detail}

In this case we find a 3d $\mathcal{N}=2$ $U(1)\times U(1)$ gauge theory with CS matrix
\begin{equation}
    K_{ij}=\begin{pmatrix}
        1&\frac{1}{2}\\
        \frac{1}{2}&1
    \end{pmatrix},
\end{equation}
and the field content as in Table~\ref{tab:A1D63dEFTcharges}.
\begin{table}[t]
  \centering
  \begin{tabular}{c|ccccc}
      \toprule
       & $U(1)_{g_1}\times U(1)_{g_2}$ & $U(1)_f$ & $U(1)_v$ & $U(1)_A$ & $U(1)_R$ \\
    \midrule
      $\Phi_{1,0}$ & $(1,0)$ & 0 & 0 & 0 & $\frac{5}{6}$ \\
      $\Phi_{1,1}$ & $(1,1)$ & 0 & 0 & 0 & $\frac{5}{6}$ \\
      $\Phi_{-1,0}$ & $(-1,0)$ & $2$ & 0 & $-1$ & $\frac{1}{6}$ \\
      $\Phi_{-1,-1}$ & $(-1,-1)$ & $-2$ & 0 & $-1$ & $\frac{1}{6}$ \\
      $\Phi_{1,-1}$ & $(1,-1)$ & $-5$ & $1$ & $\frac{1}{2}$ & $\frac{1}{3}$ \\
      $\Phi_{1,2}$ & $(1,2)$ & $5$ & $-1$ & $\frac{1}{2}$ & $\frac{1}{3}$ \\
      $\Phi_{-2,-1}$ & $(-2,-1)$ & 0 & 0 & $0$ & $\frac{1}{3}$ \\
      $\beta$ & $(0,0)$ & 0 & 0 & $-1$ & $1$ \\
      \bottomrule
  \end{tabular}
  \caption{Transformation properties of the matter fields under the gauge, global and R symmetries in the $(A_1,D_6)$ 3d EFT. Note that for convenience we have added $\frac{1}{6}g_1$ to the $U(1)_A$ charges of Table~\ref{tab:masslessA1D2n}.}
  \label{tab:A1D63dEFTcharges}
\end{table}

The global symmetries and the charge assignments are determined by the superpotential interactions
\begin{equation}
\mathcal{W}=V_{1,-1}+V_{0,1}+V_{-1,0}+\Phi_{1,0}\Phi_{1,1}\Phi_{-2,-1}+\beta\Phi_{1,-1}\Phi_{1,2}\Phi_{-2,-1}\,.
\end{equation}
In particular, the monopoles involved in the superpotential have R-charge 2 and are uncharged under all other global symmetries provided that the following BF couplings are included:
\begin{align}
    &k_{g_1 f}=k_{g_1 v}=0\,,\quad k_{g_1 A}=1\,,\quad k_{g_1 R}=\frac{1}{3}\,,\nonumber\\
    &k_{g_2 f}=\frac{3}{2}\,,\quad k_{g_2 v}=-\frac{1}{2}\,,\quad k_{g_2 A}=\frac{1}{2}\,,\quad k_{g_2 R}=\frac{1}{6}\,,
\end{align}
where the labels $g_1$ and $g_2$ distinguish the two $U(1)$ gauge factors.
The R-symmetry we specified in Table \ref{tab:A1D63dEFTcharges} (corresponding to $s=0$ in Table \ref{tab:masslessA1D2n}) turns out to be the IR superconformal one, as can be checked via $F$-extremization.

\subsubsection*{SUSY enhancement and moduli space}

The index of the theory computed with the charge assignment in Table \ref{tab:A1D63dEFTcharges} is
\begin{eqnarray}\label{eq:indA1D63dEFT}
    &\hspace{-3cm}\mathcal{I}=1+\left(1+\chi^{SU(2)_f}_{[2]}(f)\right)\frac{q^\frac{1}{2}}{y}+\left(v+\frac{1}{v}\right)\chi^{SU(2)_f}_{[1]}(f)\frac{q^\frac{3}{4}}{y^\frac{3}{2}}\nonumber\\
    &+\left(\frac{1}{y^2}\left(\chi^{SU(2)_f}_{[4]}(f)+\chi^{SU(2)_f}_{[2]}(f)+1\right){\color{red}-1}-1-\chi^{SU(2)_f}_{[2]}(f)\right)q\nonumber\\
    &+\left(v+\frac{1}{v}\right)\left(\frac{1}{y^\frac{5}{2}}\left(\chi^{SU(2)_f}_{[3]}(f)+\chi^{SU(2)_f}_{[1]}(f)\right)-\frac{1}{y^\frac{1}{2}}\chi^{SU(2)_f}_{[1]}(f)\right)q^\frac{5}{4}\nonumber\\
    &+\left(\frac{1}{y^3}\left(\chi^{SU(2)_f}_{[6]}(f)+\chi^{SU(2)_f}_{[4]}(f)+\left(1+v^2+\frac{1}{v^2}\right)\chi^{SU(2)_f}_{[2]}(f)+1\right)\right.\nonumber\\
    &\left.-\frac{1}{y}\left(\chi^{SU(2)_f}_{[4]}(f)+\chi^{SU(2)_f}_{[2]}(f)\right)\right)q^{\frac{3}{2}}+\cdots\,.
\end{eqnarray}

Once again this index provides strong evidence that the 3d $\mathcal{N}=2$ theory flows at low energies to the $\mathcal{N}=4$ SCFT that correspond to the $U(1)_r$-twisted compactication of $(A_1,D_6)$. First of all, we can see the enhancement of the $U(1)_f\times U(1)_v$ symmetry to the $SU(2)_f\times U(1)_v$ flavor symmetry of $(A_1,D_6)$, since the index rearranges into $SU(2)_f$ characters and at order $q$ with a negative sign we can see the contribution of the $SU(2)_f\times U(1)_v$ conserved current. Instead, the $U(1)_A$ symmetry becomes as usual part of the $\mathcal{N}=4$ R-symmetry.

We also have strong indications of the SUSY enhancement. Indeed, at order $y^{-1}q^\frac{1}{2}$ we can find the contribution of the $SU(2)_f\times U(1)_v$ moment map, while at order $q$ the negative contribution (in red) that is not coming from the flavor symmetry conserved current corresponds to the $\mathcal{N}=4$ stress-energy tensor.

In this case we do not see explicitly in the $\mathcal{N}=2$ index \eqref{eq:indA1D43dEFT} the contribution of any of the two extra-SUSY currents due to cancellations. 
Some possible options are $\Phi_{1,-1}\partial\Phi_{1,2}\Phi_{-2,-1}$ and $\Phi_{-1,0}\partial\Phi_{-1,-1}\Phi_{1,-1}\Phi_{1,2}$, or other variants where the $\partial$ is applied on one of the other fields. We leave the clarification of this to future work as well.

We finally consider the Higgs and Coulomb branch limits of the index. In the Higgs limit we find
\begin{align}\label{eq:HBHSA1D63dEFT}
    \lim_{x\to 0}\mathcal{I}&=1+\left(1+\chi^{SU(2)_f}_{[2]}(f)\right)t+\left(v+\frac{1}{v}\right)\chi^{SU(2)_f}_{[1]}(f)t^\frac{3}{2}+\left(\chi^{SU(2)_f}_{[4]}(f)+\chi^{SU(2)_f}_{[2]}(f)+1\right)t^2\nonumber\\
    &+\left(v+\frac{1}{v}\right)\left(\chi^{SU(2)_f}_{[3]}(f)+\chi^{SU(2)_f}_{[1]}(f)\right)t^\frac{5}{2}\nonumber\\
    &+\left(\chi^{SU(2)_f}_{[6]}(f)+\chi^{SU(2)_f}_{[4]}(f)+\left(1+v^2+\frac{1}{v^2}\right)\chi^{SU(2)_f}_{[2]}(f)+\right)t^3+\cdots\,.
\end{align}
We would like to compare this with the Higgs branch Hilbert series of $(A_1,D_6)$, which in turn can be computed as the Coulomb branch Hilbert series of the following 3d $\mathcal{N}=4$ quiver describing the mirror dual of the (non-twisted) circle reduction (see e.g.~\cite{Carta:2021whq,Xie:2021ewm}):
\begin{equation}
    \begin{tikzpicture}[scale=1.2,every node/.style={scale=1.2},font=\scriptsize]
    \node[flavor] (f0) at (0,0) {$\,2\,$};
    \node[gauge] (g0) at (1,0) {$\,1\,$};
    \node[gauge] (g1) at (2,0) {$\,1\,$};
    \node[flavor] (f1) at (3,0) {$\,1\,$};
    \draw (f0)--(g0)--(g1)--(f1);
\end{tikzpicture}
\end{equation}
Such a Coulomb branch Hilbert series can be computed using the monopole formula \cite{Cremonesi:2013lqa}, which for this theory reads
\begin{equation}
    \mathrm{HS}\left[(A_1,D_6)\right]=\frac{1}{(1-t)^2}\sum_{m_1,m_2=-\infty}^\infty f^{-m_1+2m_2}v^{m_1}t^{\frac{1}{2}\left(2|m_1|+|m_1-m_2|+|m_2|\right)}\,,
\end{equation}
where we are parametrizing the fugacities for the topological symmetries of the two nodes as $v\,f^{-1}$ and $f^2$ respectively. We verified in a power series in $t$ that the Higgs limit of our index \eqref{eq:HBHSA1D63dEFT} coincides with this Hilbert series. In the Coulomb limit we instead find
\begin{equation}
    \lim_{t\to 0}\mathcal{I}=1\,,
\end{equation}
which is again consistent with the fact that the two-dimensional Coulomb branch of $(A_1,D_6)$ is completely lifted by the $U(1)_r$ twist.

\subsection{Conjugate theories from the opposite sheet}\label{subsec:conj}

Since the Agarwal-Maruyoshi-Song Lagrangians are non-chiral, it can be easily checked from the definitions in \eqref{eq:Qht} and \eqref{eq:Lht} that
\begin{equation}
    \begin{split}
        Q^{-\gamma}(\boldsymbol{x})&=-Q^{\gamma}(-\boldsymbol{x})=-Q^{\gamma}(\boldsymbol{x}),\\
        L_{R_s}^{-\gamma}(\boldsymbol{x})&=L_{R_s}^{\gamma}(-\boldsymbol{x})=L_{R_s}^{\gamma}(\boldsymbol{x}).\label{eq:gamma_to_minus_gamma}
    \end{split}
\end{equation}
The first equalities follow from $\kappa$ being odd and $\vartheta$ being even, while the second follow from the Weyl redundancy in the $SU(n)$ and $Sp(n)$ gauge groups.

We see from \eqref{eq:gamma_to_minus_gamma} that any saddle $\boldsymbol{x}^\ast$ for $\gamma$ implies a saddle $\boldsymbol{x}^\ast$ for $-\gamma$. The negative sign on the first line of \eqref{eq:gamma_to_minus_gamma} implies that $K_{ij}$ will be negated. It is easy to check that the gauge charges of the light chiral multiplets at the saddle must be negated as well.

Thus, starting from the theories derived above for $\gamma=1$, negating the gauge charges of the chirals as well as $K_{ij},$ we obtain the 3d theories arising from the opposite $U(1)_r$ twist with $\gamma=-1.$ The A-twisted version of these theories have MTCs that are conjugate to those arising from $\gamma=1.$

\section{$A$-twisting and TQFT modular data}\label{sec:TQFT}

All the 3d $\mathcal{N}=4$ SCFTs obtained above lack a Coulomb branch, and thus their A-twist yields TQFTs without local operators. In this section we outline the calculation of the TQFT modular data for the simplest case of $(A_1,A_3)$. See Appendix~C.2 of \cite{ArabiArdehali:2024ysy} for our conventions and a review of the techniques used in the present section.

The 3d $\mathcal{N}=2$ data appropriate for the A-twist of the 3d SCFT arising from $(A_1,A_3)$ reads
\begin{equation}
    {\text{\textbf{TQFT}}^{\gamma=1}_{(A_1,A_3)}}:\qquad\quad U(1)_{1}\ +\ \Phi_{-1}^{R_1=1}+\tilde{\Phi}_{-1}^{R_1=1}+\, \Phi_{+2}^{R_1=0}\ \ \text{and}\ \ k_{gR_1}=1\,.\label{eq:A-twisted_A1A3}
\end{equation}
The $R$-charges are obtained by setting $s=1$ in Table~\ref{tab:masslessA1A2n-1}. The monopole superpotential \eqref{eq:WV_A1A3} fixes $k_{gR_1}.$

Since the VOA of the 4d $(A_1,A_3)$ theory is $\widehat{\mathfrak{su}}(2)_{-4/3}$ \cite{Beem:2013sza,Buican:2015ina,Cordova:2015nma,Beem:2017ooy}, we expect our TQFT to reproduce the corresponding  modular data of the admissible characters
\begin{equation}
\begin{split}
    T_{\widehat{\mathfrak{su}}(2)_{-4/3}}&=\begin{pmatrix}
        e^{i\pi/2} & 0 & 0 \\
        0 & e^{-i\pi/6} & 0\\
        0 & 0 & e^{-i\pi/6}
    \end{pmatrix},\qquad S_{\widehat{\mathfrak{su}}(2)_{-4/3}}=-\frac{1}{\sqrt{3}}
    \begin{pmatrix}
    1 & -1 & 1 \\
    -1 & \omega^2 & -\omega \\
    1 & -\omega & \omega^2
    \end{pmatrix},
    \end{split}\label{eq:su2-4/3_S&T}
\end{equation}
where $\omega=e^{\frac{2\pi i}{3}}.$

To compute the $S$ and $T$ matrices of the TQFT, we need (\emph{cf.}~\cite{Closset:2019hyt})
\begin{equation}
    \begin{split}
        k_{gg}^+&=1+2\times\frac{1}{2}(-1)^2+\frac{1}{2}(2)^2=4\,,\\
        k_{gR_1}^+&=1+2\times\frac{1}{2}(1-1)\times (-1)+\frac{1}{2}(0-1)\times (2)=0\,.
    \end{split}
\end{equation}
The effective twisted superpotential and dilaton \cite{Nekrasov:2014xaa,Closset:2019hyt} (in the conventions of \cite{ArabiArdehali:2024ysy})
\begin{equation}
W(Z):=-4\pi^2\,\widetilde{W}(u)=\frac{1}{2}k^{+}_{jl}Z_j Z_l+\frac{1}{2}k^{+}_{jj}\cdot2\pi i Z_j+\sum_{\chi}\sum_{\rho^\chi}\mathrm{Li}_2\big(e^{ \rho^\chi (\boldsymbol{Z})}\big),
\end{equation}
\begin{equation}
    \Omega(Z):=2\pi i\,\widetilde{\Omega}(u)=k^{+}_{jR_1}Z_j+\sum_{\chi}(1-R_1^\chi)\sum_{\rho^\chi}\log\big(1-e^{\rho^\chi (\boldsymbol{Z})}\big)-\sum_{\alpha_+}\log\big(1-e^{\pm  \alpha_+(\boldsymbol{Z})}\big),
\end{equation}
read for our theory
\begin{equation}
    \begin{split}
        W(Z,a)&=\mathrm{Li}_2\big(a\,e^{-Z}\big)+\mathrm{Li}_2\big(a^{-1}\,e^{-Z}\big)+\mathrm{Li}_2\big(e^{2Z}\big)+2Z^2+4\pi iZ,\\
        \Omega(Z)&=\log (1-e^{2Z})\,.
    \end{split}
\end{equation}

The Bethe equation $\exp\big(\partial_Z W(Z,a)\big)=1$ reads in terms of $z=-e^{-Z}$:
\begin{equation}
    \frac{(a+z_\ast) (a z_\ast+1)}{a \left(z_\ast^2-1\right)^2}=1.
\end{equation}
We discard the root $z_\ast=0$ since it yields a divergent handle-gluing operator; see Eq.~\eqref{eq:Hdef} below. The Bethe equation can then be simplified to
\begin{equation}
    z_\ast^3=a+a^{-1}\!+3z_\ast\,.
\end{equation}
The three Bethe roots are at
\begin{equation}
    z_\ast\in\Big\{a^{1/3}+a^{-1/3},\ \omega a^{1/3}+\omega^2 a^{-1/3},\ \omega^2 a^{1/3}+\omega a^{-1/3}\Big\}\,.
\end{equation}

We next consider the handle-gluing and fibering operators (see \emph{e.g.}~\cite{Closset:2018ghr}):
\begin{equation}
    \mathcal{H}(Z):=e^{\Omega(Z)}\,\det \partial^{}_{Z_i}\!\partial^{}_{Z_j}\! W(Z),\label{eq:Hdef}
\end{equation}
\begin{equation}
    \mathcal{F}(Z):=e^{[W(Z)-Z_i\,\partial^{}_{Z_i}\! W(Z)]/(2\pi i)}.\label{eq:Fdef}
\end{equation}
In terms of these we have the map (see \emph{e.g.}~\cite{Gang:2023rei,Cho:2020ljj,Gang:2021hrd})\footnote{Instead of identifying $T^2$ with $\mathcal{F}^{-2}$, we could identify $T^2$ with $\mathcal{F}^{2}$ but either $i)$ consider the conjugate theory corresponding to $\gamma=-1$, or $ii)$ reverse parity by negating our $k_{gg}$ and $k_{gR}$. The result would again match the modular data of $\widehat{\mathfrak{su}}(2)_{-4/3}$, up to the phase/sign ambiguities. Alternatively, we can identify $T^2$ with $\mathcal{F}^{2}$ with our current data, but match with the MTC conjugate to that of $\widehat{\mathfrak{su}}(2)_{-4/3}$.}
\begin{equation}
    S^2_{0\alpha}\longleftrightarrow \mathcal{H}(Z^\ast_\alpha)^{-1},\qquad T^{2}_{\alpha\alpha}\longleftrightarrow\mathcal{F}(Z^\ast_\alpha)^{-2}\,,\label{eq:HandF_vs_SandT}
\end{equation}
up to $\alpha$-independent overall phases. This gives in the limit $a\to1$ that
\begin{equation}
\begin{split}
    T&=\begin{pmatrix}
        e^{i\pi/2} & 0 & 0 \\
        0 & e^{-i\pi/6} & 0\\
        0 & 0 & e^{-i\pi/6}
    \end{pmatrix},
    \end{split} \label{eq:surgery_T_A1A3}
\end{equation}
\begin{equation}
    S=-\frac{1}{\sqrt{3}}
    \begin{pmatrix}
    \pm1 & \pm1 & \pm1 \\
    \pm1 & \ast & \ast \\
    \pm1 & \ast & \ast
    \end{pmatrix},\label{eq:surgery_S_A1A3}
\end{equation}
up to overall phases, and with the asterisks undetermined by \eqref{eq:HandF_vs_SandT}. This partial determination is compatible with \eqref{eq:su2-4/3_S&T}. Note that in order to reproduce the expected $T$ matrix as in \eqref{eq:surgery_T_A1A3}, we have identified $a^{1/3}+a^{-1/3}$ as the Bethe root $z_{\ast\,0}$ corresponding to the TQFT unit object (or the VOA vacuum module).

We have also checked that in the limit $a\to1$ the $S^3$ partition function (see \emph{e.g.}~\cite{Closset:2018ghr})
\begin{equation}
    Z_{S^3}=\sum_{z^\ast}\mathcal{H}^{-1}(z^\ast)\mathcal{F}(z^\ast),
\end{equation}
coincides with
\begin{equation}
    S_{00}=\pm1/\sqrt{\mathcal{H}(z^\ast_0)},
\end{equation}
up to an overall phase, as expected from a TQFT.

To fill in the asterisks in \eqref{eq:surgery_S_A1A3}, we need to identify the bulk line operators corresponding to the VOA admissible modules. Guided by \cite{Cordova:2016uwk,Neitzke:2017cxz}, we make the boundary/bulk identification
\begin{equation}
    \begin{split}
        1\ \ &\longleftrightarrow\ L_0=1,\\
        \Phi_1\ &\longleftrightarrow\ L_1=\frac{2+a^{-1}z-z^2}{1-a^{-2}},\\
        \Phi_2\ &\longleftrightarrow\ L_2=\frac{2+a\,z-z^2}{1-a^{-2}}.
    \end{split}\label{eq:A1A3_admissible_lines}
\end{equation}
Using
\begin{equation}
    L_\alpha\big|_{z^\ast_0}=\langle L_\alpha\rangle^{}_{S^3}\overset{{\small\text{\cite{Witten:1988hf}}}}{=}\frac{S_{\alpha 0}}{S_{00}}    
    \,,\label{eq:identifyLines}
\end{equation}
in the $a\to1$ limit (where $z_\ast\to 2,-1,-1$), we can confirm that the line operators in \eqref{eq:A1A3_admissible_lines} are compatible with the first column of \eqref{eq:surgery_S_A1A3}. Finally, using (\emph{cf.}~(A.52) in \cite{Gang:2021hrd})
\begin{equation}
    L_\alpha\big|_{z^\ast_\beta}=\frac{S_{\alpha\beta}}{S_{0\beta}}\,,\label{eq:Smatrix_formula}
\end{equation}
we fill in the asterisks in \eqref{eq:surgery_S_A1A3}, compatibly with \eqref{eq:su2-4/3_S&T}, modulo the phase ambiguities.

Note that due to the $1-a^{-2}$ denominators in \eqref{eq:A1A3_admissible_lines}, we can not interpret $L_{1,2}$ as conventional combinations of gauge and flavor Wilson loops. A conservative approach would be to fix $a$, and interpret $L_{1,2}$ as linear combinations of gauge Wilson loops. We leave a more thorough analysis of these lines to future work.

\section{Discussion and future directions}\label{sec:discussion}

The most pressing open end of this work is reproducing the characters of the admissible highest weight modules of $\widehat{\mathfrak{su}}(2)_{-4/3}$ via half-index calculations in the TQFT of Section~\ref{sec:TQFT}. It would be very interesting to also construct line operators corresponding to the (semi-) relaxed highest-weight modules of $\widehat{\mathfrak{su}}(2)_{-4/3}$ \cite{Creutzig:2012sd,Creutzig:2013yca}, and produce their characters via decorated half-indices.
It appears that gauge and flavor Wilson lines would not suffice, and more intricate line operators are needed (\emph{cf.}~\cite{Dimofte:2019zzj,Creutzig:2021ext,Garner:2023pmt,Garner:2024yin}).

Besides the simplest new case of $(A_1,A_3),$ it would be interesting to study the TQFTs arising from the A-twist of the next-to-simplest 3d SCFTs on the second sheet of $(A_1,D_3)$ and $(A_1,D_4).$ The case of $(A_1,D_3)$ should be similar to that of $(A_1,A_3)$, since they are dual to each other, but the case of $(A_1,D_4)$ is expected to be quite different, and related to the VOA $\widehat{\mathfrak{su}}(3)_{-3/2}$ \cite{Buican:2015ina,Cordova:2015nma,Beem:2017ooy}. The 3d $\mathcal{N}=2$ data appropriate for the A-twist in the latter case can be found by setting $s=1$ in Table~\ref{tab:masslessA1D2n}. The result is displayed in Table~\ref{tab:A1D4TQFTcharges}.
\begin{table}[t]
  \centering
  \begin{tabular}{c|ccccc}
      \toprule
       & $U(1)_{g}$ & $SU(2)_f$ & $U(1)_v$ &  $U(1)_{R_{s=1}}$ \\
    \midrule
      $\Phi_1$ & $1$ & $\bf 1$ & 1 &  1 \\
      $\tilde{\Phi}_1$ & $1$ & $\bf 1$ & $-1$ & 1 \\
      $\Phi_{-1}$ & $-1$ & $\bf 2$ & 0 & 1 \\
      $\Phi_{2}$ & $2$ & $\bf 1$ & 0 & 0 \\
      $\Phi_{-2}$ & $-2$ & $\bf 1$ & 0 & 0 \\
      $\beta$ & 0 & $\bf 1$ & 0 &  $2$ \\
      \bottomrule
  \end{tabular}
  \caption{The 3d $\mathcal{N}=2$ data of the A-twist $\text{TQFT}^{\,\gamma=1}_{(A_1,D_4)}$. All CS levels vanish.}
  \label{tab:A1D4TQFTcharges}
\end{table}
Our preliminary analysis indicates that the corresponding TQFT has only a single Bethe root. Identifying the (four \cite{Creutzig:2021seo}) bulk line operators associated with the VOA admissible modules seems therefore more challenging for $(A_1,D_4)$ than in the $(A_1,A_3)$ case.

As in \cite{ArabiArdehali:2024ysy}, we can also study the higher sheets $\gamma>1$ of the Argyres-Douglas theories discussed above to obtain new TQFTs. The simplest cases of $(A_1,A_3)$, $(A_1,D_3),$ $(A_1,D_4)$, and $(A_1,D_6)$ do not have inequivalent higher sheets (modulo conjugation), but all the other cases do. A higher sheet of $(A_1,D_5)$ is treated in Appendix~\ref{app:higher_sheets}.

It would be interesting to apply the above $U(1)_r$-twisted reduction procedure to other non-Lagrangian $\mathcal{N}=2$ SCFTs with an $\mathcal{N}=1$ description as well. 

Finally, the following observation about the theories considered in the present paper deserves further investigations.

\subsection*{Two TQFTs associated with an $\mathcal{N}=4$ SCFT that has a trivial Coulomb branch but a non-trivial Higgs branch} 

The new 3d $\mathcal{N}=4$ theories we obtained from $U(1)_r$-twisted $S^1$ reduction of 4d AD theories with a non-trivial Higgs branch exhibit the intriguing property that we can associate two distinct TQFTs to them.  One TQFT is obtained through a topological A-twisting, while the other can be obtained from a specific $\mathcal{N}=2$ superpotential deformation. Indeed, all theories with a non-trivial Higgs branch have at least a $U(1)$ flavor symmetry and an associated moment map operator in the $\mathbf{3}$ of the $SU(2)_{H}$ part of the $\mathcal{N}=4$ R-symmetry $SO(4) \cong SU(2)_H \times SU(2)_C$.\footnote{When the flavor symmetry is non-abelian, the moment map transforms in its adjoint representation and the operator we are interested in is one of its singlet components. This breaks part of the symmetry, however this is not an issue since either way it ends up acting trivially in the low energy TQFT.} In terms of the $\mathcal{N}=2$ subalgebra, such a moment map contains a chiral primary operator with superconformal $U(1)_R$ charge $1$ and $U(1)_A$ charge $-1$ (equivalently, it contributes as $y^{-1}q^\frac{1}{2}$ to the index), which is the deformation that we want to turn on in the superpotential. The SUSY partition functions of the deformed theory can be derived from those of the undeformed theory by specializing the R-symmetry mixing parameter $s$ to  $1$, which is equivalent to setting the R-charge of the superpotential to be $2$, and turning off the real mass or fugacity for the $U(1)_A$ symmetry (this means setting $y^{-1}q^\frac{1}{2}=q$ at the level of the index fugacities). This is the same constraint as that of the A-twist and hence, by construction, this $\mathcal{N}=2$ deformation  leads to an IR theory with the same partition functions as in the A-twist. In particular, since the theory has trivial Coulomb branch, the superconformal index of the deformed theory is just 1 indicating that it flows to a unitary TQFT, which is the second TQFT we are considering.  

The A-twisted TQFT is a non-semisimple TQFT which supports a boundary VOA that is non-unitary and logarithmic. On the other hand, the unitary TQFT is semisimple and supports a rational VOA. Notably, as previously mentioned, the two TQFTs are expected to have the same partition functions on arbitrary SUSY backgrounds, including the twisted partition functions $Z_{g,p}$ on a circle bundle of degree $p$ over a genus $g$ Riemann surface. The partition functions can be expressed in terms of modular matrices of the boundary VOA as follows:
 \begin{align}
 Z_{g,p} = \sum_{\alpha} (S_{0\alpha})^{2-2g} (T_{\alpha \alpha})^p\;.\label{eq:Zgp}
 \end{align}
For the $(A_1,A_3)$ EFT, one can use the operator $\epsilon_{ab}\Phi^a_{-1}\Phi^b_{-1}\Phi_{+2}$ for the superpotential deformation. The resulting IR TQFT is expected to have the same $Z_{g,p}$ partition function (possibly modulo a phase factor) as the $\widehat{\mathfrak{su}}(2)_{-4/3}$ theory, which can be computed using the modular matrices given in \eqref{eq:su2-4/3_S&T}. Interestingly, the conjugate (or the time-reversed) TQFT of $\widehat{\mathfrak{su}}(3)_{1}$ fulfills this role and can be identified as the IR TQFT. The modular matrices of $\widehat{\mathfrak{su}}(3)_{1}$ are 
\begin{equation}
\begin{split}
    {T}_{\widehat{\mathfrak{su}}(3)_{1}}&=\begin{pmatrix}
        e^{-\frac{\pi i}{6}} & 0 & 0 \\
        0 & e^{\frac{\pi i}{2}} & 0\\
        0 & 0 & e^{\frac{\pi i}{2}}
    \end{pmatrix},\qquad {S}_{\widehat{\mathfrak{su}}(3)_{1}}=\frac{1}{\sqrt{3}}
    \begin{pmatrix}
    1 & 1 & 1 \\
    1 & \omega & \omega^2 \\
    1 & \omega^2 & \omega
    \end{pmatrix}.
    \end{split}
\end{equation}
${T}_{\widehat{\mathfrak{su}}(3)_{1}}$ and $T_{\widehat{\mathfrak{su}}(2)_{-4/3}}$ are indeed complex conjugate modulo an overall phase, while ${S}_{\widehat{\mathfrak{su}}(3)_{1}}$ and $S_{\widehat{\mathfrak{su}}(2)_{-4/3}}$ differ by complex conjugation up to an overall sign, as well as negating the simple object corresponding to the second row/column.\footnote{In the rational context of \cite{ArabiArdehali:2024ysy}, imposing positivity of the fusion coefficients could limit the sign flip possibilities of the simple objects in that setting.} The overall phase and sign differences, and the sign flip of the simple object, are all accommodated by the unresolved ambiguities in \eqref{eq:HandF_vs_SandT}, and manifest themselves as overall phase differences in the partition functions \eqref{eq:Zgp}.

In this way, we can relate unitary semisimple TQFTs to non-unitary logarithmic TQFTs that share the same $Z_{g,p}$ partition functions (up to overall phases), provided that the latter can be realized as A-twists of 3d $\mathcal{N}=4$ SCFTs with a trivial Coulomb branch but a non-trivial Higgs branch. 
This relationship may offer a simple criterion for determining whether a given non-unitary logarithmic TQFT can be realized as an A-twisted theory of such 3d $\mathcal{
N}=4$ SCFTs. It would also be interesting to investigate further this systematic way of constructing unitary TQFTs in 3d $\mathcal{N}=4$ SCFTs with trivial Coulomb branch but non-trivial Higgs branch, for example relating specific observables in the TQFTs to the geometric properties of the Higgs branch only.

\begin{acknowledgments}

We are grateful to C.~Beem, M.~Boisvert, C.~Closset, A.~Deb, M.~Dedushenko, N.~Garner, H.~Kim, M.~Litvinov, L.~Rastelli, and F.~Yan for helpful correspondences and conversations related to this project. The work of AA was supported in part by the NSF grant PHY-2210533 and the Simons Foundation grants 397411 (Simons Collaboration on the Nonperturbative Bootstrap) and 681267 (Simons Investigator Award). AA is also indebted to the Galileo Galilei Institute (GGI) in Florence for hospitality during part of this work, and to the organizers of the BPS Dynamics and Quantum Mathematics program at GGI where some of these results were presented. The work of DG is supported in part by the National Research Foundation of Korea grant  NRF-2022R1C1C1011979. DG also acknowledges support by the National Research Foundation of Korea (NRF) Grant No. RS-2024-00405629.

\end{acknowledgments}

\appendix

\section{3$^\text{rd}$ sheet of $(A_1,D_5)$: SUSY enhancement in a non-abelian theory}\label{app:higher_sheets}

In \cite{ArabiArdehali:2024ysy}, besides the second sheet $\gamma=1$ and its conjugate $\gamma=-1,$ higher sheets of $(A_1,A_2)$ and $(A_1,A_4)$ were studied as well. On some higher sheets (the third and sixth sheets of $(A_1,A_4)$) SUSY enhancements were observed, while on others (the third and fourth sheets of $(A_1,A_2)$, and the fourth and fifth sheets of $(A_1,A_4)$) gapped theories were encountered that presumably flow to unitary TQFTs.

The families $(A_1,A_\text{odd})$ and $(A_1,D_n)$ studied here have non-trivial Higgs branches, which are uncharged under $U(1)_r$ and thus survive the $U(1)_r$-twisted reductions on all sheets. Therefore the gapped scenario can not occur. We are left with the possibility of SUSY enhancement on all sheets, and we show in a concrete example below that this expectation pans out.
As mentioned in Section~\ref{sec:discussion}, the cases $(A_1,A_3)$, $(A_1,D_3)$, and $(A_1,D_4)$ do not have higher sheets besides $\gamma=\pm1.$ The next simplest cases to explore are hence $(A_1,A_5)$ and $(A_1,D_5)$.

The higher (namely, third $\gamma=2$ and fourth $\gamma=3=-2$ {\small mod 5}) sheets of $(A_1,A_5)$ are expected to be subtle, because $(A_1,A_5)$ has a Coulomb branch operator of dimension $\frac{3}{2}$ that would survive the $\gamma=\pm 2$ $U(1)_r$ twists, since $\pm2\times\frac{3}{2}\in\mathbb{Z}\,.$ An unlifted Coulomb branch is thus expected in the 3d EFT, whose analysis we leave to future work.

The higher sheets of $(A_1,D_5)$ are the subject of the present Appendix. We focus on the third sheet $\gamma=2$, since the fourth sheet ($\gamma=3=-2$ {\small mod 5}) EFT data can be obtained via a simple conjugation by negating the gauge charges of the chiral multiplets as well as the gauge-gauge CS levels.

On the third sheet of $(A_1,D_5)$ we find a saddle at $(x_1,x_2)=(-1/5,-1/5).$ The two coincident holonomies signal that a larger subgroup of the 4d gauge symmetry containing an $SU(2)$ factor is preserved in 3d. Hence, the EFT is a 3d $\mathcal{N}\!=\!2$ $\mathfrak{su}(2)\oplus \mathfrak{u}(1)$ gauge theory (we will specify the global structure momentarily), as opposed to the case of the first sheet studied in Subsection \ref{subsec:A1D5} where the gauge symmetry was only $U(1)\times U(1)$. In terms of the $U(1)\times U(1)$ subgroup, the gauge-gauge CS levels are
\begin{equation}
    k=\begin{pmatrix}
        \frac{3}{2}&\frac{1}{2}\\
        \frac{1}{2}&\frac{3}{2}
    \end{pmatrix},
\end{equation}
and the matter content is as in Table~\ref{tab:A1D5_3rdSheet_3dEFTcharges}.
\begin{table}[t]
  \centering
  \begin{tabular}{c|cccc}
      \toprule
       & $U(1)_{g_1}\times U(1)_{g_2}$ & $SU(2)_f$ & $\mathfrak{u}(1)_A$ & $\mathfrak{u}(1)_{R}$ \\
    \midrule
      $\Phi_{-1,0}$ & $(-1,0)$ & $\bf 3$ & 0 & $\frac{9}{10}$ \\
      $\Phi_{0,-1}$ & $(0,-1)$ & $\bf 3$ & 0 & $\frac{9}{10}$ \\
      $\Phi_{1,0}$ & $(1,0)$ & $\bf 1$ & $-1$ & $\frac{1}{10}$ \\
      $\Phi_{0,1}$ & $(0,1)$ & $\bf 1$ & $-1$ & $\frac{1}{10}$ \\
      $\Phi_{2,0}$ & $(2,0)$ & $\bf 1$ & 0 & $\frac{1}{5}$ \\
      $\Phi_{0,2}$ & $(0,2)$ & $\bf 1$ & 0 & $\frac{1}{5}$ \\
      $\Phi_{1,1}$ & $(1,1)$ & $\bf 1$ & 0 & $\frac{1}{5}$\\
      \bottomrule
  \end{tabular}
  \caption{Transformation properties of the matter fields under the $U(1)\times U(1)$ subgroup of the gauge symmetry, and under the global and R symmetries in the 3d EFT on the third sheet of $(A_1,D_5)$.}
  \label{tab:A1D5_3rdSheet_3dEFTcharges}
\end{table}
 
The actual $\mathfrak{su}(2)\oplus \mathfrak{u}(1)$ gauge symmetry can be made manifest by performing a redefinition of the gauge fields similar to the one in Subsection~4.2.2 of \cite{ArabiArdehali:2024ysy}
\begin{equation}
    A_1\to A_{\mathfrak{su}(2)}+A_{\mathfrak{u}(1)}\,,\qquad A_2\to -A_{\mathfrak{su}(2)}+A_{\mathfrak{u}(1)}\,,
\end{equation}
where $A_{1,2}$ denote the two $U(1)$ gauge fields. We find a theory with gauge group\footnote{We use conventions in which the CS level is normalized such that for an $SU(2)_k$ theory the classical contribution to the index of the CS interaction is $z^{km}$, where $z$ is the gauge fugacity parametrized such that the character of the fundamental representation is $\chi^{\mathfrak{su}(2)}_{[1]}(z)=z+z^{-1}$ and $m$ is the associated gauge flux. In particular, with this normalization the CS level is quantized such that the effective level $k_{\text{eff}}=k+2\sum_i T({\bf R}_i)$ has to be an even integer, where $T({\bf R}_i)$ is the Dynkin index of the representation ${\bf R}_i$ of $SU(2)$ in which the $i$-th chiral transforms normalized such that for the fundamental $T({\bf 2})=\tfrac{1}{2}$.}
\begin{equation}\label{eq:non_ab_A1D5_3rdSheet}
    \frac{SU(2)_2\times U(1)_4}{\mathbb{Z}_2}\,,
\end{equation}
where the quotient acts as the center of the $SU(2)$ factor and the transformation $\mathrm{e}^{i\pi}$ of the $U(1)$. The matter content is displayed in Table~\ref{tab:A1D5_3rdSheet_3dEFTchargesnew}.
\begin{table}[t]
  \centering
  \begin{tabular}{c|cccc}
      \toprule
       & $\mathfrak{su}(2)\oplus \mathfrak{u}(1)$ & $SU(2)_f$ & $U(1)_A$ & $U(1)_{R}$ \\
    \midrule
      $\Phi_{{\bf 2},-1}$ & $({\bf 2},-1)$ & $\bf 3$ & 0 & $\frac{9}{10}$ \\
      $\Phi_{{\bf 2},1}$ & $({\bf 2},1)$ & $\bf 1$ & $-1$ & $\frac{1}{10}$ \\
      $\Phi_{{\bf 3},2}$ & $({\bf 3},2)$ & $\bf 1$ & 0 & $\frac{1}{5}$ \\
      \bottomrule
  \end{tabular}
  \caption{Transformation properties of the matter fields under the $\mathfrak{su}(2)\times \mathfrak{u}(1)$ gauge symmetry, and under the global and R symmetries in the 3d EFT on the third sheet of $(A_1,D_5)$.}
  \label{tab:A1D5_3rdSheet_3dEFTchargesnew}
\end{table}

The superpotential is
\begin{equation}
    \mathcal{W}=V_{\frac{1}{2},{\frac{1}{2}}}+\mathrm{Tr}\,\epsilon_{ab}\epsilon_{cd}\,\Phi_{{\bf 2},-1}^{ac}\Phi_{{\bf 2},-1}^{bd}\Phi^{}_{{\bf 3},2}\,,
\end{equation}
where the matter part descends from four dimensions \cite{Agarwal:2016pjo}. In particular, the monopole superpotential fixes the BF coupling between the $U(1)$ part of the gauge symmetry and the $U(1)_R$ and $U(1)_A$ symmetries to
\begin{equation}
    k_{gR}=\frac{12}{5}\,,\qquad k_{gA}=-1\,.
\end{equation}
Note also that the only chiral multiplet in Table \ref{tab:A1D5_3rdSheet_3dEFTchargesnew} that has non-zero $\mathfrak{u}(1)_A$ charge is $\Phi_{{\bf 2},1}$, which does not appear in the superpotential. 

As in \cite{ArabiArdehali:2024ysy}, the $\mathbb{Z}_2$ quotient in the gauge group  \eqref{eq:non_ab_A1D5_3rdSheet} is to emphasize that magnetic fluxes of the non-abelian and of the abelian part of the gauge group should sum up to integers, even though they can separately be half-integers. In fact, the monopole involved in the superpotential $V_{\frac{1}{2},{\frac{1}{2}}}$ is the one with the minimal possible values of the gauge magnetic fluxes for such choice of global structure of the gauge group. Moreover, the $\mathbb{Z}_2$ quotient ensures that the theory has no electric 1-form symmetry, as expected from its four-dimensional origin.

The superconformal index of this theory reads
\begin{align}\label{eq:indA1D53dEFT_3rd}
    \mathcal{I}&=1+\chi^{SU(2)_f}_{[2]}(f)\frac{q^\frac{1}{2}}{y}+\left(\frac{1}{y^2}\chi^{SU(2)_f}_{[4]}(f)-1-\chi^{SU(2)_f}_{[2]}(f)\right)q\nonumber\\
    &\qquad+\left(y+\frac{1}{y}+\frac{1}{y^3}\chi^{SU(2)_f}_{[6]}(f)-\frac{1}{y}\chi^{SU(2)_f}_{[4]}(f)\right)q^{\frac{3}{2}}+\cdots\,,
\end{align}
suggesting, similarly to the discussion below \eqref{eq:indA1D53dEFT}, that the theory is SUSY enhancing. We have also checked that it reduces to the expected Hilbert series of $\mathbb{C}^2/\mathbb{Z}_2$ in the Higgs limit and trivial Hilbert series in the Coulomb limit, similarly to what we found on the $\gamma=1$ sheet in \eqref{eq:A1D5_HB_HS} and \eqref{eq:A1D5_CB_HS}. We point out that the indices \eqref{eq:indA1D53dEFT} and \eqref{eq:indA1D53dEFT_3rd} of the 3d EFTs for $(A_1,D_5)$ on the sheets $\gamma=1$ and $\gamma=2$ start differing in the fermionic contributions at order $y^{-1}q^\frac{3}{2}$ (which however do not contribute in the Higgs and Coulomb limits).

The $\mathcal{N}=2$ data appropriate for the A-twist ($s=1$) reads
\begin{equation}
      {\text{\textbf{TQFT}}^{\gamma=2}_{(A_1,D_5)}}:\qquad\frac{{SU}(2)_{2}\times U(1)_{4}}{\mathbb{Z}_2}\ +\ \Phi^{ab\ R_1=1}_{{\bf 2},-1}+\,\Phi^{R_1=1}_{{\bf 2},1}+\,\Phi^{R_1=0}_{{\bf 3},2}\ \ \text{with}\ \  k_{gR_1}=3\,.\label{eq:TQFT_A1D5_gamma=2}
\end{equation}
This presumably corresponds to a non-unitary, non-semisimple TQFT.\\

Finally, note that the $(A_1,D_5)\,\to\,(A_1,A_2)$ flow on the Higgs branch (see \emph{e.g.}~\cite{Buican:2019huq}) implies that this $\mathcal{N}=4$ SCFT has a Higgs branch flow to the Fibonacci unitary TQFT of \cite{ArabiArdehali:2024ysy}. It would be interesting to systematically explore other unitary TQFTs arising from Higgs branch flows of the SCFTs constructed in this work.

\bibliographystyle{JHEP}
\bibliography{references}

\providecommand{\href}[2]{#2}\begingroup\raggedright\begin{thebibliography}{10}

\bibitem{Zamolodchikov:1986db}
A.B.~Zamolodchikov, \emph{{Conformal Symmetry and Multicritical Points in Two-Dimensional Quantum Field Theory. (In Russian)}}, {\emph{Sov. J. Nucl. Phys.} {\bfseries 44} (1986) 529}.

\bibitem{Friedan:1984rv}
D.~Friedan, Z.-a.~Qiu and S.H.~Shenker, \emph{{Superconformal Invariance in Two-Dimensions and the Tricritical Ising Model}}, \href{https://doi.org/10.1016/0370-2693(85)90819-6}{\emph{Phys. Lett. B} {\bfseries 151} (1985) 37}.

\bibitem{Maruyoshi:2016tqk}
K.~Maruyoshi and J.~Song, \emph{{Enhancement of Supersymmetry via Renormalization Group Flow and the Superconformal Index}}, \href{https://doi.org/10.1103/PhysRevLett.118.151602}{\emph{Phys. Rev. Lett.} {\bfseries 118} (2017) 151602} [\href{https://arxiv.org/abs/1606.05632}{{\ttfamily 1606.05632}}].

\bibitem{Gang:2018huc}
D.~Gang and M.~Yamazaki, \emph{{Three-dimensional gauge theories with supersymmetry enhancement}}, \href{https://doi.org/10.1103/PhysRevD.98.121701}{\emph{Phys. Rev. D} {\bfseries 98} (2018) 121701} [\href{https://arxiv.org/abs/1806.07714}{{\ttfamily 1806.07714}}].

\bibitem{ArabiArdehali:2024ysy}
A.~Arabi~Ardehali, M.~Dedushenko, D.~Gang and M.~Litvinov, \emph{{Bridging 4D QFTs and 2D VOAs via 3D high-temperature EFTs}},  \href{https://arxiv.org/abs/2409.18130}{{\ttfamily 2409.18130}}.

\bibitem{Maruyoshi:2016aim}
K.~Maruyoshi and J.~Song, \emph{{$ \mathcal{N}=1 $ deformations and RG flows of $ \mathcal{N}=2 $ SCFTs}}, \href{https://doi.org/10.1007/JHEP02(2017)075}{\emph{JHEP} {\bfseries 02} (2017) 075} [\href{https://arxiv.org/abs/1607.04281}{{\ttfamily 1607.04281}}].

\bibitem{Agarwal:2016pjo}
P.~Agarwal, K.~Maruyoshi and J.~Song, \emph{{$ \mathcal{N} $ =1 Deformations and RG flows of $ \mathcal{N} $ =2 SCFTs, part II: non-principal deformations}}, \href{https://doi.org/10.1007/JHEP12(2016)103}{\emph{JHEP} {\bfseries 12} (2016) 103} [\href{https://arxiv.org/abs/1610.05311}{{\ttfamily 1610.05311}}].

\bibitem{Cecotti:2010fi}
S.~Cecotti, A.~Neitzke and C.~Vafa, \emph{{R-Twisting and 4d/2d Correspondences}},  \href{https://arxiv.org/abs/1006.3435}{{\ttfamily 1006.3435}}.

\bibitem{Xie:2012hs}
D.~Xie, \emph{{General Argyres-Douglas Theory}}, \href{https://doi.org/10.1007/JHEP01(2013)100}{\emph{JHEP} {\bfseries 01} (2013) 100} [\href{https://arxiv.org/abs/1204.2270}{{\ttfamily 1204.2270}}].

\bibitem{Wang:2015mra}
Y.~Wang and D.~Xie, \emph{{Classification of Argyres-Douglas theories from M5 branes}}, \href{https://doi.org/10.1103/PhysRevD.94.065012}{\emph{Phys. Rev. D} {\bfseries 94} (2016) 065012} [\href{https://arxiv.org/abs/1509.00847}{{\ttfamily 1509.00847}}].

\bibitem{Beem:2013sza}
C.~Beem, M.~Lemos, P.~Liendo, W.~Peelaers, L.~Rastelli and B.C.~van Rees, \emph{{Infinite Chiral Symmetry in Four Dimensions}}, \href{https://doi.org/10.1007/s00220-014-2272-x}{\emph{Commun. Math. Phys.} {\bfseries 336} (2015) 1359} [\href{https://arxiv.org/abs/1312.5344}{{\ttfamily 1312.5344}}].

\bibitem{Costello:2018fnz}
K.~Costello and D.~Gaiotto, \emph{{Vertex Operator Algebras and 3d $ \mathcal{N} $ = 4 gauge theories}}, \href{https://doi.org/10.1007/JHEP05(2019)018}{\emph{JHEP} {\bfseries 05} (2019) 018} [\href{https://arxiv.org/abs/1804.06460}{{\ttfamily 1804.06460}}].

\bibitem{Dedushenko:2023cvd}
M.~Dedushenko, \emph{{On the 4d/3d/2d view of the SCFT/VOA correspondence}},  \href{https://arxiv.org/abs/2312.17747}{{\ttfamily 2312.17747}}.

\bibitem{Dedushenko:2018bpp}
M.~Dedushenko, S.~Gukov, H.~Nakajima, D.~Pei and K.~Ye, \emph{{3d TQFTs from Argyres\textendash{}Douglas theories}}, \href{https://doi.org/10.1088/1751-8121/abb481}{\emph{J. Phys. A} {\bfseries 53} (2020) 43LT01} [\href{https://arxiv.org/abs/1809.04638}{{\ttfamily 1809.04638}}].

\bibitem{Aharony:2013dha}
O.~Aharony, S.S.~Razamat, N.~Seiberg and B.~Willett, \emph{{3d dualities from 4d dualities}}, \href{https://doi.org/10.1007/JHEP07(2013)149}{\emph{JHEP} {\bfseries 07} (2013) 149} [\href{https://arxiv.org/abs/1305.3924}{{\ttfamily 1305.3924}}].

\bibitem{Aharony:2013kma}
O.~Aharony, S.S.~Razamat, N.~Seiberg and B.~Willett, \emph{{3$d$ dualities from 4$d$ dualities for orthogonal groups}}, \href{https://doi.org/10.1007/JHEP08(2013)099}{\emph{JHEP} {\bfseries 08} (2013) 099} [\href{https://arxiv.org/abs/1307.0511}{{\ttfamily 1307.0511}}].

\bibitem{ArabiArdehali:2019zac}
A.~Arabi~Ardehali, L.~Cassia and Y.~L\"u, \emph{{From Exact Results to Gauge Dynamics on $\mathbb{R}^3\times S^1$}}, \href{https://doi.org/10.1007/JHEP08(2020)053}{\emph{JHEP} {\bfseries 08} (2020) 053} [\href{https://arxiv.org/abs/1912.02732}{{\ttfamily 1912.02732}}].

\bibitem{Nardoni:2024sos}
E.~Nardoni, M.~Sacchi, O.~Sela, G.~Zafrir and Y.~Zheng, \emph{{Dimensionally reducing generalized symmetries from (3+1)-dimensions}}, \href{https://doi.org/10.1007/JHEP07(2024)110}{\emph{JHEP} {\bfseries 07} (2024) 110} [\href{https://arxiv.org/abs/2403.15995}{{\ttfamily 2403.15995}}].

\bibitem{Gang:2023rei}
D.~Gang, H.~Kim and S.~Stubbs, \emph{{Three-Dimensional Topological Field Theories and Non-Unitary Minimal Models}},  \href{https://arxiv.org/abs/2310.09080}{{\ttfamily 2310.09080}}.

\bibitem{Razamat:2014pta}
S.S.~Razamat and B.~Willett, \emph{{Down the rabbit hole with theories of class $ \mathcal{S} $}}, \href{https://doi.org/10.1007/JHEP10(2014)099}{\emph{JHEP} {\bfseries 10} (2014) 099} [\href{https://arxiv.org/abs/1403.6107}{{\ttfamily 1403.6107}}].

\bibitem{Intriligator:1996ex}
K.A.~Intriligator and N.~Seiberg, \emph{{Mirror symmetry in three-dimensional gauge theories}}, \href{https://doi.org/10.1016/0370-2693(96)01088-X}{\emph{Phys. Lett. B} {\bfseries 387} (1996) 513} [\href{https://arxiv.org/abs/hep-th/9607207}{{\ttfamily hep-th/9607207}}].

\bibitem{Gaiotto:2008ak}
D.~Gaiotto and E.~Witten, \emph{{S-Duality of Boundary Conditions In N=4 Super Yang-Mills Theory}}, \href{https://doi.org/10.4310/ATMP.2009.v13.n3.a5}{\emph{Adv. Theor. Math. Phys.} {\bfseries 13} (2009) 721} [\href{https://arxiv.org/abs/0807.3720}{{\ttfamily 0807.3720}}].

\bibitem{Gaiotto:2024ioj}
D.~Gaiotto and H.~Kim, \emph{{3D TFTs from 4d ${\cal N}=2$ BPS Particles}},  \href{https://arxiv.org/abs/2409.20393}{{\ttfamily 2409.20393}}.

\bibitem{Argyres:1995jj}
P.C.~Argyres and M.R.~Douglas, \emph{{New phenomena in SU(3) supersymmetric gauge theory}}, \href{https://doi.org/10.1016/0550-3213(95)00281-V}{\emph{Nucl. Phys. B} {\bfseries 448} (1995) 93} [\href{https://arxiv.org/abs/hep-th/9505062}{{\ttfamily hep-th/9505062}}].

\bibitem{Eguchi:1996ds}
T.~Eguchi and K.~Hori, \emph{{N=2 superconformal field theories in four-dimensions and A-D-E classification}},  in \emph{{Conference on the Mathematical Beauty of Physics (In Memory of C. Itzykson)}}, pp.~67--82, 7, 1996 [\href{https://arxiv.org/abs/hep-th/9607125}{{\ttfamily hep-th/9607125}}].

\bibitem{Gaiotto:2010jf}
D.~Gaiotto, N.~Seiberg and Y.~Tachikawa, \emph{{Comments on scaling limits of 4d N=2 theories}}, \href{https://doi.org/10.1007/JHEP01(2011)078}{\emph{JHEP} {\bfseries 01} (2011) 078} [\href{https://arxiv.org/abs/1011.4568}{{\ttfamily 1011.4568}}].

\bibitem{Kinney:2005ej}
J.~Kinney, J.M.~Maldacena, S.~Minwalla and S.~Raju, \emph{{An Index for 4 dimensional super conformal theories}}, \href{https://doi.org/10.1007/s00220-007-0258-7}{\emph{Commun. Math. Phys.} {\bfseries 275} (2007) 209} [\href{https://arxiv.org/abs/hep-th/0510251}{{\ttfamily hep-th/0510251}}].

\bibitem{Romelsberger:2005eg}
C.~Romelsberger, \emph{{Counting chiral primaries in N = 1, d=4 superconformal field theories}}, \href{https://doi.org/10.1016/j.nuclphysb.2006.03.037}{\emph{Nucl. Phys. B} {\bfseries 747} (2006) 329} [\href{https://arxiv.org/abs/hep-th/0510060}{{\ttfamily hep-th/0510060}}].

\bibitem{Gadde:2011uv}
A.~Gadde, L.~Rastelli, S.S.~Razamat and W.~Yan, \emph{{Gauge Theories and Macdonald Polynomials}}, \href{https://doi.org/10.1007/s00220-012-1607-8}{\emph{Commun. Math. Phys.} {\bfseries 319} (2013) 147} [\href{https://arxiv.org/abs/1110.3740}{{\ttfamily 1110.3740}}].

\bibitem{Buican:2015hsa}
M.~Buican and T.~Nishinaka, \emph{{Argyres\textendash{}Douglas theories, S$^1$ reductions, and topological symmetries}}, \href{https://doi.org/10.1088/1751-8113/49/4/045401}{\emph{J. Phys. A} {\bfseries 49} (2016) 045401} [\href{https://arxiv.org/abs/1505.06205}{{\ttfamily 1505.06205}}].

\bibitem{Dedushenko:2019mnd}
M.~Dedushenko and Y.~Wang, \emph{{4d/2d \textrightarrow{} 3d/1d: A song of protected operator algebras}}, \href{https://doi.org/10.4310/ATMP.2022.v26.n7.a2}{\emph{Adv. Theor. Math. Phys.} {\bfseries 26} (2022) 2011} [\href{https://arxiv.org/abs/1912.01006}{{\ttfamily 1912.01006}}].

\bibitem{Gadde:2011ik}
A.~Gadde, L.~Rastelli, S.S.~Razamat and W.~Yan, \emph{{The 4d Superconformal Index from q-deformed 2d Yang-Mills}}, \href{https://doi.org/10.1103/PhysRevLett.106.241602}{\emph{Phys. Rev. Lett.} {\bfseries 106} (2011) 241602} [\href{https://arxiv.org/abs/1104.3850}{{\ttfamily 1104.3850}}].

\bibitem{Dolan:2008qi}
F.A.~Dolan and H.~Osborn, \emph{{Applications of the Superconformal Index for Protected Operators and q-Hypergeometric Identities to N=1 Dual Theories}}, \href{https://doi.org/10.1016/j.nuclphysb.2009.01.028}{\emph{Nucl. Phys. B} {\bfseries 818} (2009) 137} [\href{https://arxiv.org/abs/0801.4947}{{\ttfamily 0801.4947}}].

\bibitem{ArabiArdehali:2015ybk}
A.~Arabi~Ardehali, \emph{{High-temperature asymptotics of supersymmetric partition functions}}, \href{https://doi.org/10.1007/JHEP07(2016)025}{\emph{JHEP} {\bfseries 07} (2016) 025} [\href{https://arxiv.org/abs/1512.03376}{{\ttfamily 1512.03376}}].

\bibitem{Ardehali:2021irq}
A.A.~Ardehali and J.~Hong, \emph{{Decomposition of BPS moduli spaces and asymptotics of supersymmetric partition functions}}, \href{https://doi.org/10.1007/JHEP01(2022)062}{\emph{JHEP} {\bfseries 01} (2022) 062} [\href{https://arxiv.org/abs/2110.01538}{{\ttfamily 2110.01538}}].

\bibitem{Cordova:2015nma}
C.~Cordova and S.-H.~Shao, \emph{{Schur Indices, BPS Particles, and Argyres-Douglas Theories}}, \href{https://doi.org/10.1007/JHEP01(2016)040}{\emph{JHEP} {\bfseries 01} (2016) 040} [\href{https://arxiv.org/abs/1506.00265}{{\ttfamily 1506.00265}}].

\bibitem{Buican:2015ina}
M.~Buican and T.~Nishinaka, \emph{{On the superconformal index of Argyres\textendash{}Douglas theories}}, \href{https://doi.org/10.1088/1751-8113/49/1/015401}{\emph{J. Phys. A} {\bfseries 49} (2016) 015401} [\href{https://arxiv.org/abs/1505.05884}{{\ttfamily 1505.05884}}].

\bibitem{Borokhov:2002ib}
V.~Borokhov, A.~Kapustin and X.-k.~Wu, \emph{{Topological disorder operators in three-dimensional conformal field theory}}, \href{https://doi.org/10.1088/1126-6708/2002/11/049}{\emph{JHEP} {\bfseries 11} (2002) 049} [\href{https://arxiv.org/abs/hep-th/0206054}{{\ttfamily hep-th/0206054}}].

\bibitem{Borokhov:2002cg}
V.~Borokhov, A.~Kapustin and X.-k.~Wu, \emph{{Monopole operators and mirror symmetry in three-dimensions}}, \href{https://doi.org/10.1088/1126-6708/2002/12/044}{\emph{JHEP} {\bfseries 12} (2002) 044} [\href{https://arxiv.org/abs/hep-th/0207074}{{\ttfamily hep-th/0207074}}].

\bibitem{Borokhov:2003yu}
V.~Borokhov, \emph{{Monopole operators in three-dimensional N=4 SYM and mirror symmetry}}, \href{https://doi.org/10.1088/1126-6708/2004/03/008}{\emph{JHEP} {\bfseries 03} (2004) 008} [\href{https://arxiv.org/abs/hep-th/0310254}{{\ttfamily hep-th/0310254}}].

\bibitem{Benna:2009xd}
M.K.~Benna, I.R.~Klebanov and T.~Klose, \emph{{Charges of Monopole Operators in Chern-Simons Yang-Mills Theory}}, \href{https://doi.org/10.1007/JHEP01(2010)110}{\emph{JHEP} {\bfseries 01} (2010) 110} [\href{https://arxiv.org/abs/0906.3008}{{\ttfamily 0906.3008}}].

\bibitem{Bashkirov:2010kz}
D.~Bashkirov and A.~Kapustin, \emph{{Supersymmetry enhancement by monopole operators}}, \href{https://doi.org/10.1007/JHEP05(2011)015}{\emph{JHEP} {\bfseries 05} (2011) 015} [\href{https://arxiv.org/abs/1007.4861}{{\ttfamily 1007.4861}}].

\bibitem{Cremonesi:2013lqa}
S.~Cremonesi, A.~Hanany and A.~Zaffaroni, \emph{{Monopole operators and Hilbert series of Coulomb branches of $3d$ $\mathcal{N} = 4$ gauge theories}}, \href{https://doi.org/10.1007/JHEP01(2014)005}{\emph{JHEP} {\bfseries 01} (2014) 005} [\href{https://arxiv.org/abs/1309.2657}{{\ttfamily 1309.2657}}].

\bibitem{Closset:2016arn}
C.~Closset and H.~Kim, \emph{{Comments on twisted indices in 3d supersymmetric gauge theories}}, \href{https://doi.org/10.1007/JHEP08(2016)059}{\emph{JHEP} {\bfseries 08} (2016) 059} [\href{https://arxiv.org/abs/1605.06531}{{\ttfamily 1605.06531}}].

\bibitem{Pasquetti:2019uop}
S.~Pasquetti and M.~Sacchi, \emph{{From 3$d$ dualities to 2$d$ free field correlators and back}}, \href{https://doi.org/10.1007/JHEP11(2019)081}{\emph{JHEP} {\bfseries 11} (2019) 081} [\href{https://arxiv.org/abs/1903.10817}{{\ttfamily 1903.10817}}].

\bibitem{Bhattacharya:2008zy}
J.~Bhattacharya, S.~Bhattacharyya, S.~Minwalla and S.~Raju, \emph{{Indices for Superconformal Field Theories in 3,5 and 6 Dimensions}}, \href{https://doi.org/10.1088/1126-6708/2008/02/064}{\emph{JHEP} {\bfseries 02} (2008) 064} [\href{https://arxiv.org/abs/0801.1435}{{\ttfamily 0801.1435}}].

\bibitem{Kim:2009wb}
S.~Kim, \emph{{The Complete superconformal index for N=6 Chern-Simons theory}}, \href{https://doi.org/10.1016/j.nuclphysb.2009.06.025}{\emph{Nucl. Phys. B} {\bfseries 821} (2009) 241} [\href{https://arxiv.org/abs/0903.4172}{{\ttfamily 0903.4172}}].

\bibitem{Imamura:2011su}
Y.~Imamura and S.~Yokoyama, \emph{{Index for three dimensional superconformal field theories with general R-charge assignments}}, \href{https://doi.org/10.1007/JHEP04(2011)007}{\emph{JHEP} {\bfseries 04} (2011) 007} [\href{https://arxiv.org/abs/1101.0557}{{\ttfamily 1101.0557}}].

\bibitem{Kapustin:2011jm}
A.~Kapustin and B.~Willett, \emph{{Generalized Superconformal Index for Three Dimensional Field Theories}},  \href{https://arxiv.org/abs/1106.2484}{{\ttfamily 1106.2484}}.

\bibitem{Dimofte:2011py}
T.~Dimofte, D.~Gaiotto and S.~Gukov, \emph{{3-Manifolds and 3d Indices}}, \href{https://doi.org/10.4310/ATMP.2013.v17.n5.a3}{\emph{Adv. Theor. Math. Phys.} {\bfseries 17} (2013) 975} [\href{https://arxiv.org/abs/1112.5179}{{\ttfamily 1112.5179}}].

\bibitem{Jafferis:2010un}
D.L.~Jafferis, \emph{{The Exact Superconformal R-Symmetry Extremizes Z}}, \href{https://doi.org/10.1007/JHEP05(2012)159}{\emph{JHEP} {\bfseries 05} (2012) 159} [\href{https://arxiv.org/abs/1012.3210}{{\ttfamily 1012.3210}}].

\bibitem{Beem:2012yn}
C.~Beem and A.~Gadde, \emph{{The $N=1$ superconformal index for class $S$ fixed points}}, \href{https://doi.org/10.1007/JHEP04(2014)036}{\emph{JHEP} {\bfseries 04} (2014) 036} [\href{https://arxiv.org/abs/1212.1467}{{\ttfamily 1212.1467}}].

\bibitem{Razamat:2016gzx}
S.S.~Razamat and G.~Zafrir, \emph{{Exceptionally simple exceptional models}}, \href{https://doi.org/10.1007/JHEP11(2016)061}{\emph{JHEP} {\bfseries 11} (2016) 061} [\href{https://arxiv.org/abs/1609.02089}{{\ttfamily 1609.02089}}].

\bibitem{Evtikhiev:2017heo}
M.~Evtikhiev, \emph{{Studying superconformal symmetry enhancement through indices}}, \href{https://doi.org/10.1007/JHEP04(2018)120}{\emph{JHEP} {\bfseries 04} (2018) 120} [\href{https://arxiv.org/abs/1708.08307}{{\ttfamily 1708.08307}}].

\bibitem{Garozzo:2019ejm}
I.~Garozzo, G.~Lo~Monaco, N.~Mekareeya and M.~Sacchi, \emph{{Supersymmetric Indices of 3d $S$-fold SCFTs}}, \href{https://doi.org/10.1007/JHEP08(2019)008}{\emph{JHEP} {\bfseries 08} (2019) 008} [\href{https://arxiv.org/abs/1905.07183}{{\ttfamily 1905.07183}}].

\bibitem{Beratto:2020qyk}
E.~Beratto, N.~Mekareeya and M.~Sacchi, \emph{{Marginal operators and supersymmetry enhancement in 3d $S$-fold SCFTs}}, \href{https://doi.org/10.1007/JHEP12(2020)017}{\emph{JHEP} {\bfseries 12} (2020) 017} [\href{https://arxiv.org/abs/2009.10123}{{\ttfamily 2009.10123}}].

\bibitem{Comi:2023lfm}
R.~Comi, W.~Harding and N.~Mekareeya, \emph{{Chern-Simons-Trinion theories: One-form symmetries and superconformal indices}}, \href{https://doi.org/10.1007/JHEP09(2023)060}{\emph{JHEP} {\bfseries 09} (2023) 060} [\href{https://arxiv.org/abs/2305.07055}{{\ttfamily 2305.07055}}].

\bibitem{Closset:2020afy}
C.~Closset, S.~Giacomelli, S.~Schafer-Nameki and Y.-N.~Wang, \emph{{5d and 4d SCFTs: Canonical Singularities, Trinions and S-Dualities}}, \href{https://doi.org/10.1007/JHEP05(2021)274}{\emph{JHEP} {\bfseries 05} (2021) 274} [\href{https://arxiv.org/abs/2012.12827}{{\ttfamily 2012.12827}}].

\bibitem{Giacomelli:2020ryy}
S.~Giacomelli, N.~Mekareeya and M.~Sacchi, \emph{{New aspects of Argyres--Douglas theories and their dimensional reduction}}, \href{https://doi.org/10.1007/JHEP03(2021)242}{\emph{JHEP} {\bfseries 03} (2021) 242} [\href{https://arxiv.org/abs/2012.12852}{{\ttfamily 2012.12852}}].

\bibitem{Xie:2021ewm}
D.~Xie, \emph{{3d mirror for Argyres-Douglas theories}},  \href{https://arxiv.org/abs/2107.05258}{{\ttfamily 2107.05258}}.

\bibitem{Beratto:2021xmn}
E.~Beratto, N.~Mekareeya and M.~Sacchi, \emph{{Zero-form and one-form symmetries of the ABJ and related theories}}, \href{https://doi.org/10.1007/JHEP04(2022)126}{\emph{JHEP} {\bfseries 04} (2022) 126} [\href{https://arxiv.org/abs/2112.09531}{{\ttfamily 2112.09531}}].

\bibitem{Carta:2021whq}
F.~Carta, S.~Giacomelli, N.~Mekareeya and A.~Mininno, \emph{{Conformal manifolds and 3d mirrors of Argyres-Douglas theories}}, \href{https://doi.org/10.1007/JHEP08(2021)015}{\emph{JHEP} {\bfseries 08} (2021) 015} [\href{https://arxiv.org/abs/2105.08064}{{\ttfamily 2105.08064}}].

\bibitem{Benvenuti:2017bpg}
S.~Benvenuti and S.~Giacomelli, \emph{{Lagrangians for generalized Argyres-Douglas theories}}, \href{https://doi.org/10.1007/JHEP10(2017)106}{\emph{JHEP} {\bfseries 10} (2017) 106} [\href{https://arxiv.org/abs/1707.05113}{{\ttfamily 1707.05113}}].

\bibitem{Benvenuti:2010pq}
S.~Benvenuti, A.~Hanany and N.~Mekareeya, \emph{{The Hilbert Series of the One Instanton Moduli Space}}, \href{https://doi.org/10.1007/JHEP06(2010)100}{\emph{JHEP} {\bfseries 06} (2010) 100} [\href{https://arxiv.org/abs/1005.3026}{{\ttfamily 1005.3026}}].

\bibitem{Xie:2016uqq}
D.~Xie and S.-T.~Yau, \emph{{New N = 2 dualities}},  \href{https://arxiv.org/abs/1602.03529}{{\ttfamily 1602.03529}}.

\bibitem{Beratto:2020wmn}
E.~Beratto, S.~Giacomelli, N.~Mekareeya and M.~Sacchi, \emph{{3d mirrors of the circle reduction of twisted A$_{2N}$ theories of class S}}, \href{https://doi.org/10.1007/JHEP09(2020)161}{\emph{JHEP} {\bfseries 09} (2020) 161} [\href{https://arxiv.org/abs/2007.05019}{{\ttfamily 2007.05019}}].

\bibitem{Beem:2017ooy}
C.~Beem and L.~Rastelli, \emph{{Vertex operator algebras, Higgs branches, and modular differential equations}}, \href{https://doi.org/10.1007/JHEP08(2018)114}{\emph{JHEP} {\bfseries 08} (2018) 114} [\href{https://arxiv.org/abs/1707.07679}{{\ttfamily 1707.07679}}].

\bibitem{Closset:2019hyt}
C.~Closset and H.~Kim, \emph{{Three-dimensional $\mathcal{N}= 2$ supersymmetric gauge theories and partition functions on Seifert manifolds: A review}}, \href{https://doi.org/10.1142/S0217751X19300114}{\emph{Int. J. Mod. Phys. A} {\bfseries 34} (2019) 1930011} [\href{https://arxiv.org/abs/1908.08875}{{\ttfamily 1908.08875}}].

\bibitem{Nekrasov:2014xaa}
N.A.~Nekrasov and S.L.~Shatashvili, \emph{{Bethe/Gauge correspondence on curved spaces}}, \href{https://doi.org/10.1007/JHEP01(2015)100}{\emph{JHEP} {\bfseries 01} (2015) 100} [\href{https://arxiv.org/abs/1405.6046}{{\ttfamily 1405.6046}}].

\bibitem{Closset:2018ghr}
C.~Closset, H.~Kim and B.~Willett, \emph{{Seifert fibering operators in 3d $\mathcal{N}=2$ theories}}, \href{https://doi.org/10.1007/JHEP11(2018)004}{\emph{JHEP} {\bfseries 11} (2018) 004} [\href{https://arxiv.org/abs/1807.02328}{{\ttfamily 1807.02328}}].

\bibitem{Cho:2020ljj}
G.Y.~Cho, D.~Gang and H.-C.~Kim, \emph{{M-theoretic Genesis of Topological Phases}}, \href{https://doi.org/10.1007/JHEP11(2020)115}{\emph{JHEP} {\bfseries 11} (2020) 115} [\href{https://arxiv.org/abs/2007.01532}{{\ttfamily 2007.01532}}].

\bibitem{Gang:2021hrd}
D.~Gang, S.~Kim, K.~Lee, M.~Shim and M.~Yamazaki, \emph{{Non-unitary TQFTs from 3D $ \mathcal{N} $ = 4 rank 0 SCFTs}}, \href{https://doi.org/10.1007/JHEP08(2021)158}{\emph{JHEP} {\bfseries 08} (2021) 158} [\href{https://arxiv.org/abs/2103.09283}{{\ttfamily 2103.09283}}].

\bibitem{Cordova:2016uwk}
C.~Cordova, D.~Gaiotto and S.-H.~Shao, \emph{{Infrared Computations of Defect Schur Indices}}, \href{https://doi.org/10.1007/JHEP11(2016)106}{\emph{JHEP} {\bfseries 11} (2016) 106} [\href{https://arxiv.org/abs/1606.08429}{{\ttfamily 1606.08429}}].

\bibitem{Neitzke:2017cxz}
A.~Neitzke and F.~Yan, \emph{{Line defect Schur indices, Verlinde algebras and $U(1)_r$ fixed points}}, \href{https://doi.org/10.1007/JHEP11(2017)035}{\emph{JHEP} {\bfseries 11} (2017) 035} [\href{https://arxiv.org/abs/1708.05323}{{\ttfamily 1708.05323}}].

\bibitem{Witten:1988hf}
E.~Witten, \emph{{Quantum Field Theory and the Jones Polynomial}}, \href{https://doi.org/10.1007/BF01217730}{\emph{Commun. Math. Phys.} {\bfseries 121} (1989) 351}.

\bibitem{Creutzig:2012sd}
T.~Creutzig and D.~Ridout, \emph{{Modular Data and Verlinde Formulae for Fractional Level WZW Models I}}, \href{https://doi.org/10.1016/j.nuclphysb.2012.07.018}{\emph{Nucl. Phys. B} {\bfseries 865} (2012) 83} [\href{https://arxiv.org/abs/1205.6513}{{\ttfamily 1205.6513}}].

\bibitem{Creutzig:2013yca}
T.~Creutzig and D.~Ridout, \emph{{Modular Data and Verlinde Formulae for Fractional Level WZW Models II}}, \href{https://doi.org/10.1016/j.nuclphysb.2013.07.008}{\emph{Nucl. Phys. B} {\bfseries 875} (2013) 423} [\href{https://arxiv.org/abs/1306.4388}{{\ttfamily 1306.4388}}].

\bibitem{Dimofte:2019zzj}
T.~Dimofte, N.~Garner, M.~Geracie and J.~Hilburn, \emph{{Mirror symmetry and line operators}}, \href{https://doi.org/10.1007/JHEP02(2020)075}{\emph{JHEP} {\bfseries 02} (2020) 075} [\href{https://arxiv.org/abs/1908.00013}{{\ttfamily 1908.00013}}].

\bibitem{Creutzig:2021ext}
T.~Creutzig, T.~Dimofte, N.~Garner and N.~Geer, \emph{{A QFT for non-semisimple TQFT}}, \href{https://doi.org/10.4310/ATMP.2024.v28.n1.a4}{\emph{Adv. Theor. Math. Phys.} {\bfseries 28} (2024) 161} [\href{https://arxiv.org/abs/2112.01559}{{\ttfamily 2112.01559}}].

\bibitem{Garner:2023pmt}
N.~Garner and W.~Niu, \emph{{Line Operators in $U(1|1)$ Chern-Simons Theory}},  \href{https://arxiv.org/abs/2304.05414}{{\ttfamily 2304.05414}}.

\bibitem{Garner:2024yin}
N.~Garner, N.~Geer and M.B.~Young, \emph{{B-twisted Gaiotto-Witten theory and topological quantum field theory}},  \href{https://arxiv.org/abs/2401.16192}{{\ttfamily 2401.16192}}.

\bibitem{Creutzig:2021seo}
T.~Creutzig, D.~Ridout and M.~Rupert, \emph{{A Kazhdan\textendash{}Lusztig Correspondence for $L_{-\frac{3}{2}}(\mathfrak {sl}_3)$}}, \href{https://doi.org/10.1007/s00220-022-04602-8}{\emph{Commun. Math. Phys.} {\bfseries 400} (2023) 639} [\href{https://arxiv.org/abs/2112.13167}{{\ttfamily 2112.13167}}].

\bibitem{Buican:2019huq}
M.~Buican and Z.~Laczko, \emph{{Rationalizing CFTs and Anyonic Imprints on Higgs Branches}}, \href{https://doi.org/10.1007/JHEP03(2019)025}{\emph{JHEP} {\bfseries 03} (2019) 025} [\href{https://arxiv.org/abs/1901.07591}{{\ttfamily 1901.07591}}].

\end{thebibliography}\endgroup

\end{document}